\documentclass[11pt,reqno]{amsart}
\usepackage[dvips]{color}
\usepackage{amsmath}
\usepackage{amsxtra}
\usepackage{amscd}
\usepackage{amsthm}
\usepackage{amsfonts}
\usepackage{amssymb}
\usepackage{eucal}
\usepackage{epsfig}
\usepackage{graphics}
\usepackage[matrix,arrow]{xy}
\usepackage{url}

\input xy
\xyoption{all}

\theoremstyle{plain}
\newtheorem{theorem}{Theorem}[section]

\newtheorem{thm}[theorem]{Theorem}

\newtheorem{conj}[theorem]{Conjecture}

\numberwithin{equation}{section}

\newcommand{\R}{\mathbb{R}}
\newcommand{\C}{\mathbb{C}}

\newcommand{\Z}{\mathbb{Z}}

\newcommand{\g}{\mathfrak{g}}

\newcommand{\nc}{\newcommand}
\nc{\on}{\operatorname}
\nc{\la}{\lambda}
\nc{\wh}{\widehat}
\nc{\wt}{\widetilde}
\nc{\sw}{{\mathfrak s}{\mathfrak l}}
\nc{\ghat}{\wh{\g}}
\nc{\hhat}{\wh{\h}}
\nc{\mc}{\mathcal}
\nc{\bi}{\bibitem}
\nc{\pa}{\partial}
\nc{\ppart}{(\!(t)\!)}
\nc{\pparl}{(\!(\la)\!)}
\nc{\zpart}{(\!(z)\!)}
\nc{\n}{{\mathfrak n}}
\nc{\ol}{\overline}
\nc{\mb}{\mathbf}
\nc{\bb}{{\mathfrak b}}
\nc{\su}{\wh\sw_2}
\nc{\h}{{\mathfrak h}}
\nc{\can}{\on{can}}
\nc{\ntil}{\wt{\n}}
\nc{\pone}{{\mathbb P}^1}
\nc{\bs}{\backslash}
\nc{\al}{\alpha}
\nc{\gt}{{\mathfrak g}'}
\nc{\ds}{\displaystyle}
\nc{\Bun}{\on{Bun}}
\nc{\Gr}{\on{Gr}}

\def\neg{\negthinspace}

\nc{\ka}{\kappa}
\def\lg{{}^L\neg\g}
\def\hlg{\wh{\lg}}
\def\LG{{}^L\neg G}
\nc{\lka}{{}^L\neg\ka}
\nc{\LP}{{}^L\neg P}
\nc{\LQ}{{}^L\neg Q}
\nc{\lkt}{{}^L\neg\wt\ka}
\nc{\Hom}{\on{Hom}}
\nc{\OO}{\mathcal O}
\nc{\Loc}{\on{Loc}}
\nc{\sto}{\!\!\shortto\!\!}
\nc{\hsl}{\widehat{\mathfrak sl}}

\DeclareMathOperator{\Vir}{Vir}
\DeclareMathOperator{\sVir}{sVir}

\DeclareMathOperator{\td}{d}
\DeclareMathOperator{\tsu}{su}
\DeclareMathOperator{\tso}{so}
\DeclareMathOperator{\DS}{DS}
\DeclareMathOperator{\Ff}{\bigwedge_{\R}}
\DeclareMathOperator{\KL}{KL}

\theoremstyle{definition}
\newtheorem{rem}{Remark}[section]

\usepackage{stmaryrd}
\usepackage{trimclip}

\makeatletter
\DeclareRobustCommand{\shortto}{%
  \mathrel{\mathpalette\short@to\relax}%
}

\newcommand{\short@to}[2]{%
  \mkern2mu
  \clipbox{{.3\width} 0 0 0}{$\m@th#1\vphantom{+}{\shortrightarrow}$}%
  }
\makeatother

\begin{document}

\title[Quantum Langlands dualities]{Quantum Langlands dualities of
  boundary conditions, $D$-modules, and conformal blocks}

\author{Edward Frenkel}

\address{Department of Mathematics, University of California,
  Berkeley, CA 94720, USA}

\author{Davide Gaiotto}

\address{Perimeter Institute for Theoretical Physics, Waterloo,
  Ontario, N2L 2Y5, Canada}

\begin{abstract}
  We review and extend the vertex algebra framework linking gauge
  theory constructions and a quantum deformation of the Geometric
  Langlands Program. The relevant vertex algebras are associated to
  junctions of two boundary conditions in a 4d gauge theory and can be
  constructed from the basic ones by following certain standard
  procedures. Conformal blocks of modules over these vertex algebras
  give rise to twisted $D$-modules on the moduli stacks of $G$-bundles
  on Riemann surfaces which have applications to the Langlands
  Program. In particular, we construct a series of vertex algebras for
  every simple Lie group $G$ which we expect to yield $D$-module
  kernels of various quantum Geometric Langlands dualities. We pay
  particular attention to the full duality group of gauge theory,
  which enables us to extend the standard qGL duality to a larger
  duality groupoid. We also discuss various subtleties related to the
  spin and gerbe structures and present a detailed analysis for the
  $U(1)$ and $SU(2)$ gauge theories.
\end{abstract}

\maketitle

\setcounter{tocdepth}{1}
\tableofcontents

\section{Introduction}

A geometric version of the Langlands Program originated in the 1980s
in the works of Deligne, Drinfeld, and Laumon. In the early 1990s,
Beilinson and Drinfeld discovered its deep connections to 2d conformal
field theory and representation theory of affine Kac--Moody algebras at
the critical level \cite{BD}. This led them to the insight that
deforming the Kac--Moody level, one should be able to obtain a quantum
deformation of the Geometric Langlands correspondence (see \cite{St}
for an early formulation). Today, this is usually stated as a
conjectural equivalence between derived categories of twisted
$D$-modules on the moduli stacks $\Bun_G$ and $\Bun_{\LG}$ of bundles
on a compact Riemann surface $X$ for a pair of Langlands dual
simple\footnote{This can be generalized to reductive Lie groups. For
  abelian $G$, it has been proved in \cite{PR}.} Lie groups $G$ and
$\LG$:
\begin{equation}    \label{qGL Sm}
S_m: \quad D_\ka(\Bun_G) \longleftrightarrow
D_{-1/m \ka}(\Bun_{\LG})
\end{equation}
where $m$ is the lacing number of $G$ and we equip $X$ with a spin
structure\footnote{Without a choice of spin structure the statement is
  modified in an interesting way, which we will review later on.} (it
is expected that $S_m^2$ is the identity in the categorical
sense). Here $\ka$ is a twisting parameter for $D$-modules
corresponding to the level of the affine Kac--Moody algebra $\ghat$
shifted by the critical value. The equivalence \eqref{qGL Sm} is
expected to hold for irrational $\ka$ (i.e. $\ka \in \C \bs {\mathbb
  Q}$) and satisfy some natural compatibilities. For rational $\ka$
some modifications may have to be applied to the above categories for
it to hold.

The ordinary (non-quantum) Geometric Langlands correspondence appears
in the $\ka \to 0$ limit of \eqref{qGL Sm}. In this limit $D_{-1/m
  \ka}(\Bun_{\LG})$ becomes a suitably modified version of the derived
category of quasicoherent sheaves on $\on{Loc}_{\LG}$, the moduli
stack of flat $\LG$-bundles on $X$ (see \cite{AG} for a precise
conjecture).

In the past twenty years, a great effort has been made to search for
general methods and structures that could shed light on how and why
quantum Geometric Langlands (qGL) duality \eqref{qGL Sm} comes about
(see \cite{Frenkel1,KW,Kapustin,JT,gaitsW,gaitsQ,Ga1,Ga2,CG,AFO}
for a partial list of references).

It turns out that a lot of valuable information can be obtained from
the study of dualities in 4d supersymmetric gauge theories. In their
groundbreaking work \cite{KW}, Kapustin and Witten related quantum qGL
duality \eqref{qGL Sm} to the $S$-duality of twisted 4d gauge theories
and mirror symmetry of Hitchin moduli spaces (see also
\cite{Kapustin,Witten}). In their approach, $S$-duality is manifested
as an equivalence of certain categories of branes on the Hitchin
moduli spaces for $G$ and $\LG$, which they have linked to categories
of $D$-modules. On the other hand, Witten and one of the authors
undertook in \cite{GW1,GW2} an in-depth study of boundary conditions
in 4d gauge theory and their behavior under quantum dualities. This
dramatically expanded the class of 4d gauge theory data that one could
employ to gain further insights into the qGL dualities.

In this paper, we build on these results, as well as the recent works
\cite{Ga1,Ga2,GR,CG}, to present a systematic study of the quantum
Geometric Langlands dualities in the framework of boundary conditions
in 4d gauge theory and the corresponding junction vertex
algebras. Here we list the main ingredients of our approach:

\bigskip

\begin{enumerate}

\item We heed a lesson of 4d gauge theory and consider, instead of a
  single duality $S_m$ sending $\ka \mapsto -1/m\ka$ and $G \mapsto
  \LG$, the full group (more precisely, a groupoid) of quantum
  dualities. It combines $S_m$ with the dualities $T^n$ sending $\ka
  \to \ka+n, n \in n(G) \cdot \Z$ and $G \mapsto G$, which correspond
  to the equivalences
\begin{equation}    \label{qGL T}
\quad D_\ka(\Bun_G)  \longleftrightarrow  D_{\ka+n}(\Bun_G)  
\end{equation}
given by taking the tensor product with a power of a ``minimal'' line
bundle ${\mc L}_G$ on $\Bun_G$ (corresponding to level $n(G)$ defined
in Section \ref{symmetries}). The functors $S_m$ and $T^n$ are then
expected to generate a categorical action of a certain groupoid ${\mc
  G}^G_\ka$ on the categories of twisted $D$-modules on $\Bun_G$ (at
least, for irrational $\ka$). Thus, for any $g \in {\mc G}^G_\ka$ we
should have functors
\begin{equation}    \label{gen qGL}
{\mc E}^{G,g}_\ka: D_\ka(\Bun_G) \to D_{g(\ka)}(\Bun_{g(G)}),
\end{equation}
satisfying the relations in the groupoid ${\mc G}^G_\ka$, which
ultimately boil down to relations between the elements
\begin{equation}
S_m = \begin{pmatrix}
0 & -1 \\
m & 0
\end{pmatrix} \qquad \qquad T = \begin{pmatrix}
1 & 1 \\
0 & 1
\end{pmatrix}  
\end{equation}
in the group $PGL_2(\Z)$. The study of all of these functors together
(rather than $S_m$ of \eqref{qGL Sm} alone) allows for a much greater
flexibility and leads to a number of non-trivial
consequences.\footnote{For a non-simply connected group $G$ only
  powers of $T$ divisible by a positive integer $n(G)$ are allowed as
  quantum dualities. However, gauge theory predicts that an extension
  of the duality groupoid ${\mc G}^G_\ka$, generated by $S_m$ and $T$,
  acts on the more general categories of gerbe-twisted $D$-modules on
  $\Bun_{G'}$ where $G'$ is in the isogeny class of $G$ or $\LG$ but
  does not necessarily coincide with either of them. The vertex
  algebra technology provides information about these new dualities as
  well, and about further refinements which occur when the choice of
  spin structure on $X$ is removed.}


\bigskip

\item We start with the basic boundary conditions of the twisted 4d
  gauge theory labeled by $(G,\ka)$: Dirichlet, Neumann, and Nahm, and
  consider their images under the action of the groupoid ${\mc
    G}^G_\ka$. Thus, we obtain a big collection of boundary conditions
  $B$. To each of them corresponds a ribbon category ${\mc
    C}^G_\ka(B)$ of line defects (or a spin-ribbon category if we
  choose a spin structure on the underlying manifold; we could also
  view them as chiral categories). In general, this is a derived, or a
  DG, category. We describe these categories for the basic boundary
  conditions explicitly (see Section \ref{ex bc}). We then assign to a
  boundary condition $g(B)$ with label $(g(G),g(\ka))$, where $B$ is
  basic and $g \in {\mc G}^G_\ka$, the same category ${\mc
    C}^G_\ka(B)$. All equivalences of categories that this assignment
  entails are assumed to hold (in particular, this way we naturally
  get the statement that ${\mc C}^G_\ka(\on{Neumann})$ is equivalent
  to ${\mc C}^{\LG}_{1/m\ka}(\on{Nahm})$, which is related to a
  conjecture of Gaitsgory and Lurie \cite{gaitsW,gaitsQ}).

\bigskip

\item For each Riemann surface $X$ and a point $x \in X$, there is a
  {\em compactification functor} from the category of line defects
  ${\mc C}^G_\ka(B)$ to $D_\ka(\Bun_G)$. Thus, we can go from line
  defects of boundary conditions directly to objects of the derived
  category $D_\ka(\Bun_G)$ (in what follows we will refer to them
  simply as twisted $D$-modules on $\Bun_G$), bypassing the categories
  of branes. These functors have multi-point generalizations.

\bigskip

\item To each junction of boundary conditions $B_1 \to B_2$ in the 4d
  theory labeled by $(G,\ka)$ one associates a vertex algebra
  $V^G_\ka(B_1 \sto B_2)$ and a functor
$$
{\mc C}^G_\ka(B_1)^\vee
  \boxtimes {\mc C}^G_\ka(B_2) \to V^G_\ka(B_1 \sto B_2)\on{-mod}.
$$
We argue that the compactification functor of (iii) can be constructed
rigorously as a localization functor for the vertex algebra
$V^G_\ka(D^G_{0,1} \sto B)$, where $D^G_{0,1} \to B$ is any
non-degenerate junction from the Dirichlet boundary condition
$D^G_{0,1}$ to $B$.

The notion of localization functor is familiar from representation
theory of vertex algebras and 2d conformal field theory (see
e.g. \cite{F:review}). It assigns to a module over a vertex algebra
with affine Kac--Moody symmetry its sheaf of coinvariants twisted by
various $G$-bundles (these are the dual spaces to the spaces of
conformal blocks). This sheaf is naturally a twisted $D$-module on
$\Bun_G$.

\bigskip

\item We conjecture that morphisms between the $D$-modules associated
  to objects in ${\mc C}^G_\ka(B_1)$ and ${\mc C}^G_\ka(B_2)$ can be
  identified with the spaces of coinvariants of the corresponding
  modules over the vertex algebra $V^G_\ka(B_1 \sto B_2)$ for a
  non-degenerate junction $B_1 \to B_2$ (for such junctions it follows
  therefore that the spaces of coinvariants of $V^G_\ka(B_1 \sto
  B_2)$-modules should be independent of the specifics of the junction
  $B_1 \to B_2$).

\bigskip

\item For irrational $\ka$, we conjecture that a kernel of the
  equivalence ${\mc E}^{G,g}_\ka$ given by \eqref{gen qGL} can be
  constructed as the sheaf of coinvariants of the vertex algebra
$$
V^G_\ka(D_{0,1} \sto \; g(D_{0,1}))
$$
associated to a non-degenerate junction $D_{0,1} \to g(D_{0,1})$,
where $g(D_{0,1})$ is the boundary condition in 4d gauge obtained by
applying the duality $g \in {\mc G}^G_\ka$ to the Dirichlet boundary
condition $D^G_{0,1}$. Unlike $D^G_{0,1}$, these dual boundary
conditions are notoriously difficult to describe directly (see
\cite{GW1,GW2}). However, as we explain below, there are various
tricks which enable us to construct explicitly various vertex algebras
$V^G_\ka(D_{0,1} \sto g(D_{0,1}))$ without knowing what $g(D_{0,1})$
is. In fact, in this paper we construct two families of such vertex
algebras with the favorable property that all of their conformal
dimensions (apart from the vacuum) are strictly positive and the
graded components corresponding to all conformal dimensions are
finite-dimensional.

\end{enumerate}

\bigskip

Although 4d gauge theory is a powerful motivator, many of these
results and conjectures can be understood directly and rigorously in
terms of the junction vertex algebras and their properties. Thus, the
junction vertex algebras enable us to translate subtle and complicated
structures of 4d gauge theory into a simpler, mathematically
rigorous world of vertex algebras, their modules, conformal blocks and
localization functors. In this world, the structures relevant to the
qGL dualities can be expressed in terms of a kind of lego game, in
which building blocks are labeled by the basic boundary conditions and
their images under various dualities (a precise nomenclature is set up
in Section \ref{nomen}).

To each junction between these blocks and a label $(G,\ka)$, which
identifies the bulk 4d gauge theory in which the junction is
implemented, we assign a vertex algebra. (We want to emphasize that a
given pair of boundary conditions may have many different junctions;
the corresponding vertex algebras may well be non-isomorphic.) There
are several basic junctions for which we know the corresponding vertex
algebra at the outset. For instance, a junction from Dirichlet to
Neumann gives us an affine Kac--Moody algebra, a junction from Nahm to
Neumann gives us the corresponding ${\mc W}$-algebra, and so on. And
then there are two standard moves which enable us to produce new
junctions and new vertex algebras.

The first move is {\em composition}: we can compose two junctions $B_1
\to B_2$ and $B_2 \to B_3$ with label $(G,\ka)$ to produce a new
junction $B_1 \to B_3$ with the same label. Furthermore, for many
junctions, we can construct the corresponding vertex algebra
explicitly. For irrational $\ka$, it is an extension of the tensor
product of the vertex algebra associated to $B_1 \to B_2$ and the
vertex algebra $B_2 \to B_3$ by a specific family of modules.

The second move is {\em duality}: the vertex algebras arising in
different duality frames of a given junction should be the same. In
other words, for each element $g$ of the duality groupoid ${\mc
  G}^G_\ka$, the vertex algebra assigned to a junction $B_1 \to B_2$
with a label $(G,\ka)$ should be isomorphic to the vertex algebra
associated to its duality image $g(B_1) \to g(B_2)$ with the label
$(g(G),g(\ka))$. If both junction vertex algebras can be constructed
explicitly, we obtain an isomorphism between them, which could be
non-trivial (such as the duality of ${\mc W}$-algebras
\cite{FF}). More often than not, however, only one of these vertex
algebras is known {\em a priori}, and then this move gives us a way to
define the other vertex algebra.

Thus, we obtain ``junction calculus'' with two standard moves at our
disposal: composition of two junctions produces another junction, and
changing the duality frame of a junction gives rise to an equivalent
junction. Iterating these moves, we obtain a vast array of junctions,
and hence the corresponding vertex algebras, many examples of which
are presented below.

Since all these structures arise from the 4d gauge theory, we expect
that this junction calculus satisfies a kind of bootstrap
consistency. In other words, whenever we obtain the same junction by
means of different sequences of the standard moves applied to
basic junctions, the corresponding vertex algebras should be
isomorphic. It is interesting to ask what sort of mathematical
structure this represents and what are the minimal requirements for
its consistency (as a useful analogy, consider the Kirby calculus). We
do not address this question in the present paper, but hope to return
to it in a future work. Here we focus on various applications of the
rich framework provided by this junction calculus.

\bigskip

The first application has already been mentioned in (vi) above: we can
construct junctions $D^G_{0,1} \to g(D^G_{0,1})$ from Dirichlet to its
$g$-dual boundary condition. The corresponding vertex algebra has two
commuting affine Kac--Moody algebra symmetries: $\ghat$ and $\wh{\g'}$,
where $\g'$ is the $g$-dual of $\g$ (i.e. it is $\g$ or $\lg$
depending on what $g$ is). We conjecture that the sheaf of
coinvariants of this vertex algebra on $\Bun_G \times \Bun_{g(G)}$ is
a kernel of the qGL equivalence ${\mc E}^{G,g}_\ka$ corresponding to
$g$ (see formula \eqref{gen qGL}).

Our construction of the kernels is automatically compatible with the
expected characteristic property of the functor $S_m$; namely, that it
should relate two important families of $D$-modules:
\begin{equation}
S_m: {\mc D}_\ka^{x_i, \lambda_i} \longleftrightarrow \Psi^{x_i,
  \lambda_i}_{-1/m \ka}, \qquad x_i \in X, \quad \la_i \in P^+(G)
\end{equation}
which are local modifications of the sheaf ${\mc D}_\ka$ of
$\ka$-twisted differential operators on $\Bun_G$ and the Whittaker
sheaf $\Psi_{-1/m \ka}$ on $\Bun_{\LG}$ defining a geometric
analogue of the Whittaker functional on the space of automorphic
functions (the Whittaker sheaf may also be viewed as a quantization
the structure sheaf of the oper manifold, which appears in the limit
$\ka \to 0$).

\medskip

The second application is that we can express morphisms between
various $D$-modules ${\mc F}_1$ and ${\mc F}_2$ on $\Bun_G$ in terms
of conformal blocks of a suitable vertex algebra. The most basic
example is when ${\mc F}_1$ is the $D$-module of $\delta$-functions
supported at a point ${\mc P} \in \Bun_G$ and ${\mc F}_2$ is in the
image of the compactification functor corresponding to a boundary
condition $B$. In this case, $\on{Hom}({\mc F}_1,{\mc F}_2)$ should be
isomorphic to the space of ${\mc P}$-twisted coinvariants of a module
over the vertex algebra obtained from the junction from Dirichlet to
$B$.


For example, the fibers of the $D$-modules ${\mc D}_\ka^{x_i,
  \lambda_i}$ at points of $\Bun_G$ are the spaces of coinvariants of
the tensor product of the corresponding Weyl modules over the
Kac--Moody vertex algebra $V_\ka(\mathfrak{g})$.

As another example, the vector space
\begin{equation}
\Hom(\Psi^{x_i, \mu^\vee_i}_{\ka}, {\mc D}_\ka^{x_i, \lambda_i})
\end{equation}
is expected to be isomorphic to the space of coinvariants of the ${\mc
  W}$-algebra ${\mc W}_\ka(\g)$ obtained by the quantum
Drinfeld--Sokolov reduction of $V_\ka(\mathfrak{g})$.

A pair of boundary conditions, $B_1$ and $B_2$ may well have different
junctions $B_1 \to B_2$ which give rise to non-isomorphic vertex
algebras. However, we conjecture that the corresponding spaces of
coinvariants of modules over these vertex algebras are isomorphic to
each other for all non-degenerate junctions. Some of these
isomorphisms may be quite non-trivial.

\medskip

The third application is that we can construct many interesting
$D$-modules that are qGL dual to each other. For example, take $G =
GL(n)$ or semisimple and self-dual with $m=1$ (this means that $G$ is
a product of $E_8$ factors). Then the entire group $PSL_2(\Z)$ is
realized by functors ${\mc E}^{G,g}_\ka$, and we have a relation
$(ST)^3=1$.  Furthermore, the functor $T$ leaves $\Psi^{x_i,
  \mu^\vee_i}_{\ka}$ invariant, which implies that $STS$ must send
${\mc D}_\ka^{x_i, \lambda_i}$ to ${\mc D}_{STS(\ka)}^{x_i,
  \lambda_i}$. But $STS = T^{-1} S T^{-1}$. We thus learn (almost for
free!) that under the qGL duality $S$,
\begin{equation}    \label{selfdual}
{\mc L}^{-1}_G \otimes {\mc D}_{\ka+1}^{x_i, \lambda_i} \in
D_\ka(\Bun_G) \quad \longleftrightarrow
\quad {\mc L}_G \otimes {\mc D}_{-\ka^{-1}-1}^{x_i,
  \lambda_i} \in D_{-1/\ka}(\Bun_G).
\end{equation}
Many non-trivial and perhaps even surprising statements of this kind
can be obtained this way.



Finally, we expect that it is possible to obtain a kernel of the
Geometric Langlands duality proper as a carefully defined critical
level limit $\ka \to 0$ of the kernels constructed using the junction
calculus (see Section \ref{hecke} for a brief discussion). We leave
the details to a future work.

\bigskip

The paper is organized as follows. In Section \ref{overview} we give a
brief overview of the subject, summarizing the links between boundary
conditions in 4d gauge theories, vertex algebras, and $D$-modules on
$\Bun_G$. In Section \ref{bc} we focus on a specific class of 4d gauge
theories, the GL twisted $N=4$ supersymmetric theories defined in
\cite{KW}. For these theories, we give concrete examples of boundary
conditions and categories associated to them, junction vertex
algebras, compactification functors, and the action of the duality
groupoid. We also formulate conjectures linking conformal blocks of
the junction vertex algebras and morphisms between $D$-modules on
$\Bun_G$ obtained via the compactification functors. In Section
\ref{qGL dual} we explain how the action of the duality groupoid on
the categories of line defects associated to boundary conditions gives
rise to qGL dualities between categories of twisted $D$-modules on
$\Bun_G$ and $\Bun_{\LG}$. In Section \ref{fam ker} we present an
explicit construction of a family of junction vertex algebras
$X_{p,q}(G)$ for an arbitrary simple Lie group $G$, whose sheaves of
coinvariants we expect to give rise to kernels of specific qGL
dualities. For positive $p$ and $q$ with either of them greater than
$1$, these vertex algebras have the favorable property that conformal
dimensions of all fields other than the vacuum are strictly positive.

Section \ref{stalks} starts with a detailed discussion of the links
between the $D$-modules and the branes associated to the basic
boundary conditions. We then explain how the compactification functor
(with values in twisted $D$-modules on $\Bun_G$) comes about from the
point of view of 4d gauge theory and from the point of view of the
theory of vertex algebras. In particular, we show that under some
natural assumptions it can be constructed as the localization functor
for the corresponding junction vertex algebra (up to tensoring with a
line bundle on $\Bun_G$). In Sections \ref{U1} and \ref{SU2} we
discuss in more detail various examples of the categories of line
defects and junction vertex algebras arising from 4d gauge theories
with gauge groups $U(1)$ and $SU(2)/SO(3)$, respectively, paying
special attention to the spin and gerbe subtleties. These subtleties
for general groups are discussed in Section \ref{sec:topo}. Finally,
in Section \ref{other}, we produce more examples of kernel vertex
algebras for general simple Lie groups. Section \ref{future} lists
some open questions and directions for future research.

\bigskip

\noindent {\bf Acknowledgements.} E.F. thanks Joerg Teschner and
Edward Witten for valuable discussions.  D.G. thanks Alexander
Braverman, Kevin Costello, Thomas Creutzig, Dennis Gaitsgory and
especially Philsang Yoo for valuable discussions and explanations.

E.F. was supported by NSF grant DMS-1601934. D.G. is supported by the
NSERC Discovery Grant program and by the Perimeter Institute for
Theoretical Physics. Research at Perimeter Institute is supported by
the Government of Canada through Industry Canada and by the Province
of Ontario through the Ministry of Research and Innovation.

\section{Overview of the gauge theory setup}    \label{overview}

Here we summarize the links between boundary conditions in 4d
supersymmetric gauge theories, vertex algebras, and $D$-modules on the
moduli stack $\Bun_G$ of $G$-bundles on a curve $X$.  We also discuss
the relations between dualities of gauge theories, quantum Langlands
dualities, and various operations on vertex algebras. The purpose of
this section is to give a bird's-eye view of the
subject. Therefore, we only give brief descriptions of some these
objects and emphasize the links between them. More details are given
in the subsequent sections. We also refer the reader to the earlier
works \cite{GW1,GW2,Ga1,Ga2,GR,CG}. 

Useful mathematical groundwork on boundary conditions for 4d
supersymmetric gauge theories is presented in \cite{EY,BY}. Some
important links between boundary conditions and the Geometric
Langlands Program have been found in (yet unpublished) work by Yoo
\cite{Y:Perimeter}.  That includes, in particular, the
characterization (reviewed below) of the categories of boundary line
defects at Neumann and Nahm boundary conditions and their junction
local operator algebras. The characterization of the category of line
defects at Dirichlet boundary conditions was developed by Yoo and one
of the authors in the course of an ongoing project on the local
Geometric Langlands Program \cite{GY}.

Several of the statements below have a rather natural TFT
interpretation. Despite that, we warn the reader that our setup
differs in an important way from the standard TFT setup, where one may
describe a 4d topological theory as some sort of 3-category, with
objects associated to topological 3d boundary conditions, morphisms
associated to topological 2d junctions, 2-morphisms associated to 1d
junctions of junctions, etc.

Topologically twisted 4d gauge theory is equipped with topological 3d
boundary conditions, which are the main actors in our setup. On the
other hand, it admits no topological 2d junctions between generic
pairs of boundary conditions. Instead, it admits {\it holomorphic} 2d
junctions, which support interesting vertex algebras of local
operators. These vertex algebras are our main computational
tools.\footnote{The resulting structure is a higher analogue of the
  structure which arises naturally in the study of 3d Chern--Simons
  theory, which is a topological field theory but typically admits no
  topological 2d boundary conditions. Instead, it admits holomorphic
  2d boundary conditions, supporting rational vertex algebras. }

In this paper we do not employ the full toolbox available to us in
topologically twisted 4d gauge theory. In particular, this toolbox
includes topological 3d interfaces (domain walls) and topological 2d
surface defects. These tools are invaluable for uncovering other
aspects of qGL dualities: the study of interfaces will enable us to
construct functors between the categories of $D$-modules on $\Bun_G$
and $\Bun_H$, generalizing the geometric Langlands functoriality (see
Sect. 4 of \cite{F:AMS}); the study of surface defects will allow us
to introduce ramification into the picture (generalizing \cite{GW}),
and to set up a proper framework for the local qGL dualities
respectively. We leave these topics for future work.

Our main objective in this paper is to understand the {\em
  compactification functor} associated to a (possibly decorated)
Riemann surface $X$. Intuitively, this is the map from structures in
the 4d gauge theory ${\mc T}$ to structures in the effective 2d
topological theory ${\mc T}[X]$ which describes the compactification
of ${\mc T}$ on $X$.

In the first approximation, this 2d theory is a twisted sigma model
whose target as the moduli space ${\mc M}_H(G)$ of solutions of
Hitchin's equations \cite{Hit}. However, this description is
incomplete and inadequate for understanding important objects and
phenomena in this 2d theory, such as boundary
conditions. Mathematically, this can be expressed as saying that in
the first approximation, the category of 1d boundary conditions in the
2d theory ${\mc T}[X]$ is described by some category of branes on
${\mc M}_H(G)$, but this description omits important boundary
conditions. A better choice is the category of twisted $D$-modules on
$\Bun_G$, and for irrational values of the coupling constant $\ka$ (which
corresponds to the twisting parameter) we do expect this to be the
correct answer. But for rational values of $\ka$ there are subtle
differences between the category of 1d boundary conditions in ${\mc
  T}[X]$ and the category of $\ka$-twisted $D$-modules on $\Bun_G$;
this can be seen in the fact that some of the $D$-modules may be
unphysical because of various issues with the 2d theory (see Remark
\ref{rational} below).

In this paper we will mostly sidestep these concerns. Our objective is
not to identify specific categories of $\ka$-twisted $D$-modules on
$\Bun_G$ for which qGL duality would be an equivalence for all values
of $\ka$. Rather, we want to find an explicit and practical
description of the compactification functor mapping decorated boundary
conditions of the 4d gauge theory and various junctions between them
to twisted $D$-modules on $\Bun_G$ and morphisms between them.

Once we achieve that, the general machine of 4d gauge theory dualities
will then allow us to construct specific collections of qGL dual
twisted $D$-modules with matching properties as well as a variety of
duality ``kernels''. We expect that studying these data further will
ultimately allow us to define the qGL dualities precisely and in full
generality.

See Table \ref{tab:one} for a brief summary of the role played by
objects in various dimensions.

\begin{table}[h]
\begin{center}
\begin{tabular}{|c||c||c|}
\hline
{\em Gauge Theory} & {\em Vertex Algebra} & {\em Compactification on $X$}  \\
\hline \hline
4d bulk theory ${\mc T}_G^\ka$ & ? & ``$D_\ka(\Bun_G)$''  \\ 
\hline
3d boundary & Spin ribbon(chiral)  & Functor  \\
condition $B$ &  category ${\mc C}({\mc T},B)$ & ${\mc C}({\mc T},B)
\to D_\ka(\Bun_G)$ \\
\hline
2d junction & Vertex algebra with & Conformal blocks  \\
$B_1 \to B_2$ & ${\mc C}^\vee_1\boxtimes {\mc C}_2$ action & map to
morphisms \\
\hline
\end{tabular} 
\end{center}
\vspace*{5mm}
\caption{A brief summary of the relations between gauge theory, vertex algebra and quantum Geometric Langlands 
structures.}\label{tab:one}
\end{table}
 
\subsection{Categories of Boundary Lines}

The starting point is the following:

\bigskip

$\bullet$ {\em Each boundary condition $B$ in a 4d topologically
  twisted supersymmetric gauge theory $T$ gives rise to a ribbon
  category (as defined e.g. in \cite{EGNO}) ${\mc C}({\mc T},B)$ of
  ``boundary line defects''.}

\bigskip

In a more careful treatment, we should replace ribbon categories with
chiral categories (see e.g.  \cite{Raskin-chiral}), which are better
suited for algebraic-geometric considerations. Furthermore, in general
${\mc C}({\mc T},B)$ should be a derived, or a DG category. We will
mostly ignore this issue because in the examples we consider below
(corresponding to the irrational level $\ka$) we can work with the
abelian categories.

Upon compactification of $B$ on a Riemann surface $X$, possibly
decorated by line defects at points $x_i \in X$, we expect to obtain
an object in the category of 1d boundary conditions for the 2d theory
${\mc T}[X]$.  This map should give rise to a functor from ${\mc
  C}({\mc T},B) \boxtimes {\mc C}({\mc T},B) \boxtimes \cdots$, with
different factors corresponding to different points $x_i$, to that
category, which is compatible with braiding and fusion.

Without loss of generality, we can focus on a single point $x \in X$
and denote the corresponding compactification functor from ${\mc
  C}({\mc T},B)$ by $F_{\mc T}^B$.

\subsection{Vertex algebra at a junction}    \label{va j}

Given a 4d bulk theory ${\mc T}$ and two 3d boundary conditions $B_1$
and $B_2$, we can look for 2d junctions interpolating from one
boundary condition to the other. We denote such a junction as $J_{12}:
B_1 \to B_2$.  In general, the same pair of boundary conditions may
admit a variety of distinct junctions, with microscopic definitions
which may involve various auxiliary holomorphic 2d degrees of freedom.

Most of the time, we will suppress the specific choice of junction
$J_{12}$ in our notation, unless we want to draw special attention to
it.

To these data, we expect to associate a {\em vertex algebra} $V({\mc
  T},B_1\sto B_2)$ of local operators together with a functor
$$
F_{{\mc T},B_1\, \sto \; B_2}: {\mc C}({\mc T},B_1)^\vee \boxtimes {\mc
  C}(B_2) \to V({\mc T},B_1\sto B_2)\on{-mod},
$$
where ${\mc C}({\mc T},B_1)^\vee$ is the dual category to ${\mc
  C}({\mc T},B_1)$.\footnote{In special situations, the bulk 4d theory
  may admit non-trivial category ${\mc S}({\mc T})$ of bulk line
  defects, with functors to the categories of boundary lines.  If
  that's the case, the product should taken over this category. Also,
  it might be necessary to consider here the dual of the derived, or
  DG, categories.}
  
\begin{figure}
\center
\includegraphics[width=0.7\textwidth]{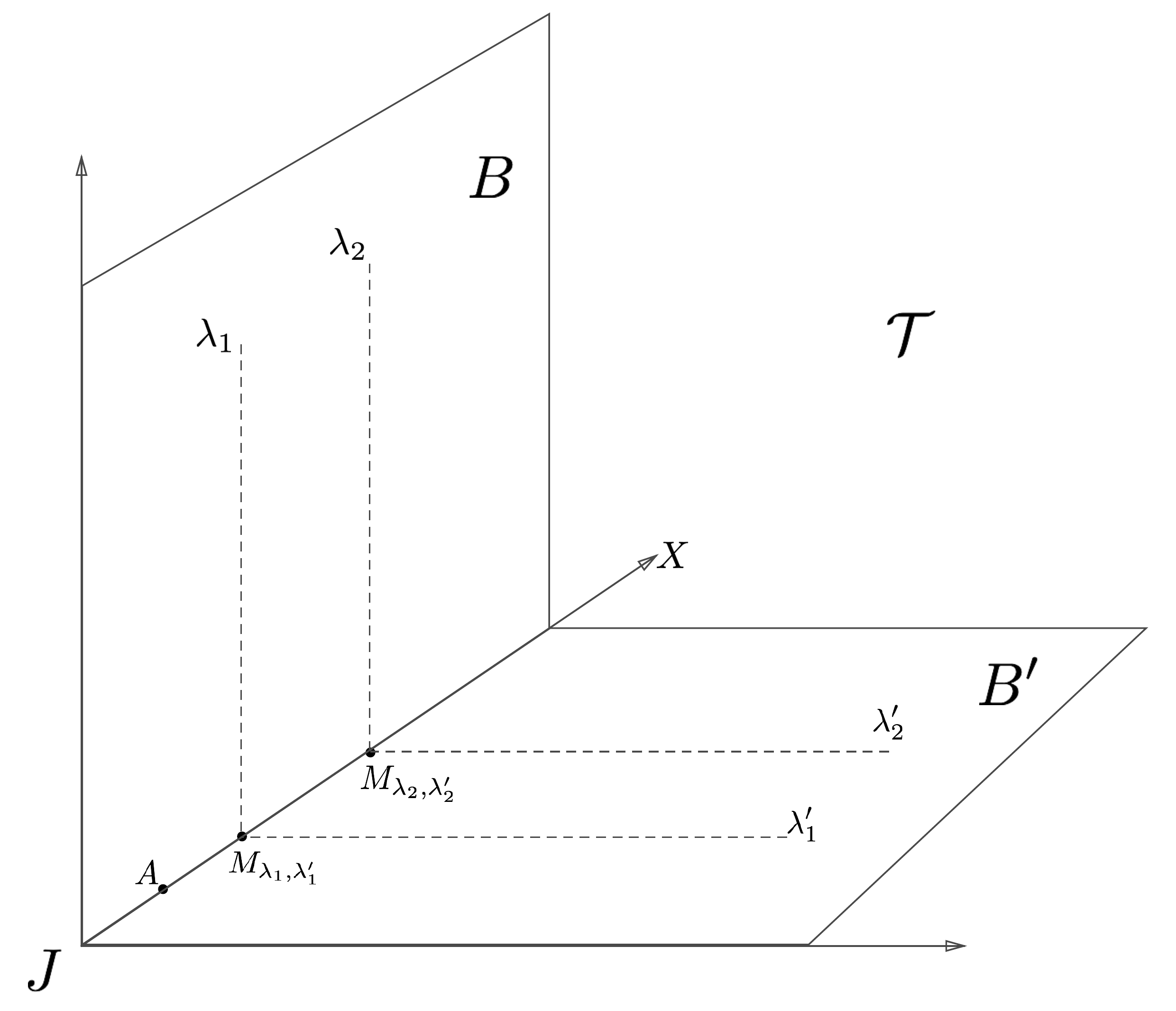}
\caption{The main physical actors: a 4d theory, 3d boundary conditions
  $B$ and $B'$, boundary line defects
$\lambda_i$ and $\lambda'_i$ placed at $x_i \in X$, a 2d junction $J$,
the vertex algebra $A$ of junction local operators,
the $A$-modules $M_{\lambda_i, \lambda_i'}$ of local operators at the
endpoints of boundary line defects at the junction. We drew $B$ and
$B'$ as
orthogonal in the topological plane transverse to the junction, but
any angle is possible.} \label{fig1}
\end{figure}

Physically, the functor maps boundary lines to the spaces of local
operators supported at points where the lines end at the junction.
These naturally form a module for the vertex algebra $V({\mc
  T},B_1\sto B_2)$ of local operators supported at generic points of
the junction.  See Figure \ref{fig1} for an illustration of these
ideas.

This functor is not an equivalence in general, but we expect it to be
fully faithful in many interesting cases. If is it fully faithful, we
will call the junction {\em non-degenerate}.

\subsection{Conformal blocks}    \label{cb}

It is natural to consider conformal blocks (more precisely, their dual
spaces called spaces of coinvariants, see \cite{FB,BD1} and Section
\ref{stalks} below) of a junction vertex algebra $V({\mc T},B_1\sto
B_2)$, possibly involving the $V({\mc T},B_1\sto B_2)$-modules
associated to the corresponding ribbon categories.

We will get such a space of conformal blocks for any choice of curve
$X$, possibly decorated by modules placed at one or more points $x_i
\in X$. By construction, the operation of taking conformal blocks (or
coinvariants) gives a functor from the category of $N$-tuples of
$V({\mc T},B_1\sto B_2)$-modules to vector spaces, where $N$ is the
number of points. In what follows, we restrict ourselves to the case
$N=1$, but a generalization to $N>1$ is straightforward.

What interests us the most is that there is a direct link between
morphisms between the images of the compactification functor and
conformal blocks of the corresponding $V^G_\ka(B_1\sto B_2)$-module.

More precisely, let ${\mc A}_1 \in \on{Ob}({\mc C}({\mc T},B_1))$,
${\mc A}_2 \in \on{Ob}({\mc C}({\mc T},B_2))$. Then on the one hand,
these objects can be sent by the compactification functors to objects
$\ol{\mc A}_1 = F_{\mc T}^{B_1}({\mc A_1})$ and $\ol{\mc A}_2 = F_{\mc
  T}^{B_2}({\mc A_2})$ of the category of 1d boundary conditions for
${\mc T}[X]$.
  
On the other hand, $F_{{\mc T},B_1 \, \sto \; B_2}({\mc A}_1^\vee
\otimes{\mc A}_2)$ is a $V(B_1\sto B_2)$-module.

\begin{conj}    \label{Hom and conf}
Suppose that the functor $F_{{\mc T},B_1\, \sto \; B_2}$ is fully
faithful. Then $\on{Hom}(\ol{\mc A}_1,\ol{\mc A}_2)$ is
isomorphic to the space of coinvariants of the $V(B_1\sto B_2)$-module
$F_{{\mc T},B_1 \, \sto \; B_2}({\mc A}_1^\vee \otimes {\mc A}_2)$.
\end{conj}

This is the main tool we will employ to reconstruct the 
compactification functor.

We also have the following

\begin{conj}    \label{meta}
  Suppose that the functor $F_{{\mc T},B_1\, \sto \; B_2}$ is fully
  faithful. Then the category of $V(B_1\sto B_2)$-modules and
  the corresponding spaces of coinvariants (and conformal blocks)
  depend only on $B_1$ and $B_2$ and not on the junction data between
  them.
\end{conj}

In Sections \ref{conf blocks} and \ref{comp and loc} we will explain
the significance of these conjectures.

The physical interpretation of these conjectures is
straightforward. It involves the space of states for the theory on a
space manifold of the form $ [0,1] \times X$, with boundary conditions
$B_1$ and $B_2$ at the endpoints of the segment.

By definition, the space of states computes the $\Hom$ in the category
of boundary conditions in ${\mc T}[X]$. It can also be interpreted as
the space of states of the 3d TFT resulting from compactification of
${\mc T}$ on the segment.

\begin{figure}
\center
\includegraphics[width=0.2\textwidth]{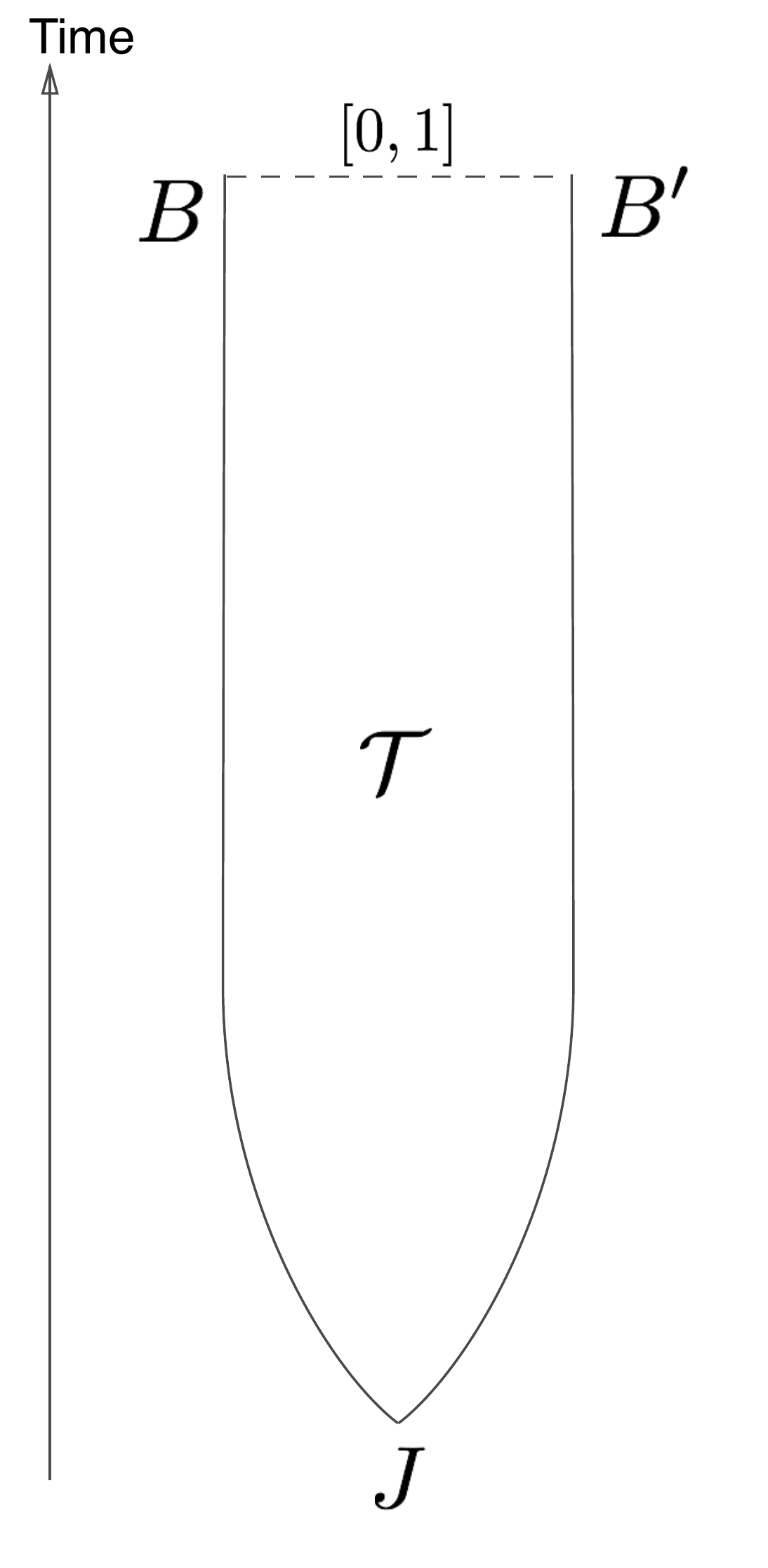}
\caption{The space-time justification for relating conformal blocks of
  junction vertex algebras
and the Hilbert space of states on $[0,1] \times X$ with $B$ and $B'$
boundary conditions: the junction creates
states in the the Hilbert space. The $X$ holomorphic direction is not
depicted.} \label{fig2}
\end{figure}

The 2d junction descends to a boundary condition for such a TFT.  Our
conjectures can then be seen as a variant of the standard relation
between the space of states of a 3d TFT and the space of conformal
blocks of its boundary vertex algebras. See Figure \ref{fig2} for an
illustration of this setup.

\begin{rem} We can illustrate such a relation in further detail for
  the simplest situation namely, a 3d TFT $T[C]$ described by some
  modular tensor category $C$ and a 2d rational vertex algebra
  $A$. This is not an example which occurs in our setup (except when
  describing some useful auxiliary degrees of freedom in later
  sections) but it is nevertheless instructive.

  Any rational vertex algebra $A$ gives rise to a modular tensor
  category $A\mathrm{-mod}$. That means $A$ can always be found at a
  boundary for the 3d TFT $T[A\mathrm{-mod}]$ defined by
  $A\mathrm{-mod}$.  The space of states of $T[A\mathrm{-mod}]$ on a
  Riemann surface $X$ coincides with the space of conformal blocks of
  $A$ on $X$.

  However, the vertex algebra $A$ can also be found as the
  algebra of boundary local operators at boundaries of other 3d TFTs.
  The MTC $A\mathrm{-mod}$ has a universal property: any boundary
  condition for $T[C]$ supporting $A$ can be factored as the
  composition of the canonical boundary condition for the
  $T[A\mathrm{-mod}]$ 3d TFT and a topological interface from $T[C]$
  to $T[A\mathrm{-mod}]$.

  In particular, the functor $C \to A\mathrm{-mod}$ from lines to
  local operators can be reinterpreted as, or factored through, the
  functor which describes the action of such an interface on the lines
  of the 3d TFTs. The interface, though, also controls the relation
  between the spaces of states for the two 3d TFTs on any Riemann
  surface.

  Hence in this basic setup one can derive sharp statements relating
  the properties of a functor $C \to A\mathrm{-mod}$ and the
  relation between the spaces of states of $T[C]$ and conformal blocks
  of $A$.\qed
\end{rem}

\subsection{Boundary conditions with global symmetries}

A final ingredient is the observation that certain boundary conditions
are equipped with a non-trivial group of {\it global symmetries},
which is defined independently of the gauge symmetry of the bulk
theory.

Whenever boundary condition $B$ has a non-trivial global symmetry
group $H$, the compactification functor can be modified by coupling
the system to an $H$-bundle on $X$.  This gives families of 1d
boundary conditions for ${\mc T}[X]$ parametrized by $\Bun_H$.

Furthermore, junctions of the form $B \to B'$ or $B' \to B$ which
preserve the $H$ symmetry will support vertex algebras with an
$H$-symmetry implemented by an $\wh{\mathfrak{h}}$ Kac--Moody
subalgebra. As a consequence, spaces of coinvariants can be promoted
to twisted $D$-modules on $\Bun_H$.

Morphisms between these $D$-modules can be used to give a concrete
realization for the spaces of morphisms between the families of 1d
boundary conditions parametrized by $\Bun_H$.

Each 4d gauge theory with gauge group $G$ comes equipped with a
special family of ``Dirichlet'' boundary conditions with global
symmetry $G$. These will be key to giving a precise formulation of the
compactification functors with values in the categories of twisted
$D$-modules on $\Bun_G$ (see Section \ref{bc to D}).

\subsection{Composition of junctions}    \label{comp junctions}

Consider now three boundary conditions $B_1$, $B_2$ and $B_3$ in a
bulk theory ${\mc T}$, with pairwise junctions $J_{12}$ and
$J_{23}$. On physical grounds, we then expect to have a new junction
$J_{12} \circ J_{23}$ from $B_1$ to $B_3$.

The corresponding vertex algebra $V({\mc T},B_1\sto B_3)$ in which we
take as junction data the composition of the data $J_{12}$ and
$J_{23}$ is expected to include the tensor product vertex algebra
$$
V({\mc T},B_1\sto B_2) \otimes V({\mc T},B_2\sto B_3).
$$
Furthermore, conjecturally, $V({\mc T},B_1\sto B_3)$ can be built as an extension of the latter by
a (super)algebra object ${\mathcal A}_{13}$ of the category
$$
\left[V({\mc T},B_1\sto B_2) \otimes V({\mc
    T},B_2\sto B_3) \right]\on{-mod}
$$
which is the image of $\on{Id}^\vee_1 \times \mathrm{Diag} \times
\on{Id}_3$ under $F_{{\mc T},B_1\, \sto \; B_2} \boxtimes F_{{\mc
    T},B_2\, \sto \; B_3}$. Here $\on{Id}^\vee_1$ and $\on{Id}_3$
denote the identity objects in ${\mc C}({\mc T},B_1)^\vee$ and ${\mc
  C}({\mc T},B_3)$, respectively, while $\mathrm{Diag}$ is a
``diagonal object'' in ${\mc C}({\mc T},B_2) \boxtimes {\mc C}({\mc
  T},B_2)^\vee$. This object is easy to define if ${\mc C}({\mc
  T},B_2)$ is semisimple as an abelian category. In general, the
construction of $\mathrm{Diag}$ requires special care.

The composition of the obvious map
$${\mc C}({\mc T},B_1)^\vee \boxtimes {\mc C}({\mc T},B_3) \to {\mc
  C}({\mc T},B_1)^\vee \boxtimes \mathrm{Diag} \boxtimes {\mc C}({\mc
  T},B_3)$$
and $F_{{\mc T},B_1 \, \sto \; B_2} \boxtimes F_{{\mc T},B_2 \, \sto
  \; B_3}$ gives a functor to ${\mathcal A}_{13}\on{-mod}$ and thus to
$V({\mc T},B_1\sto B_3)\on{-mod}$, as needed.

\begin{figure}
\center
\includegraphics[width=0.7\textwidth]{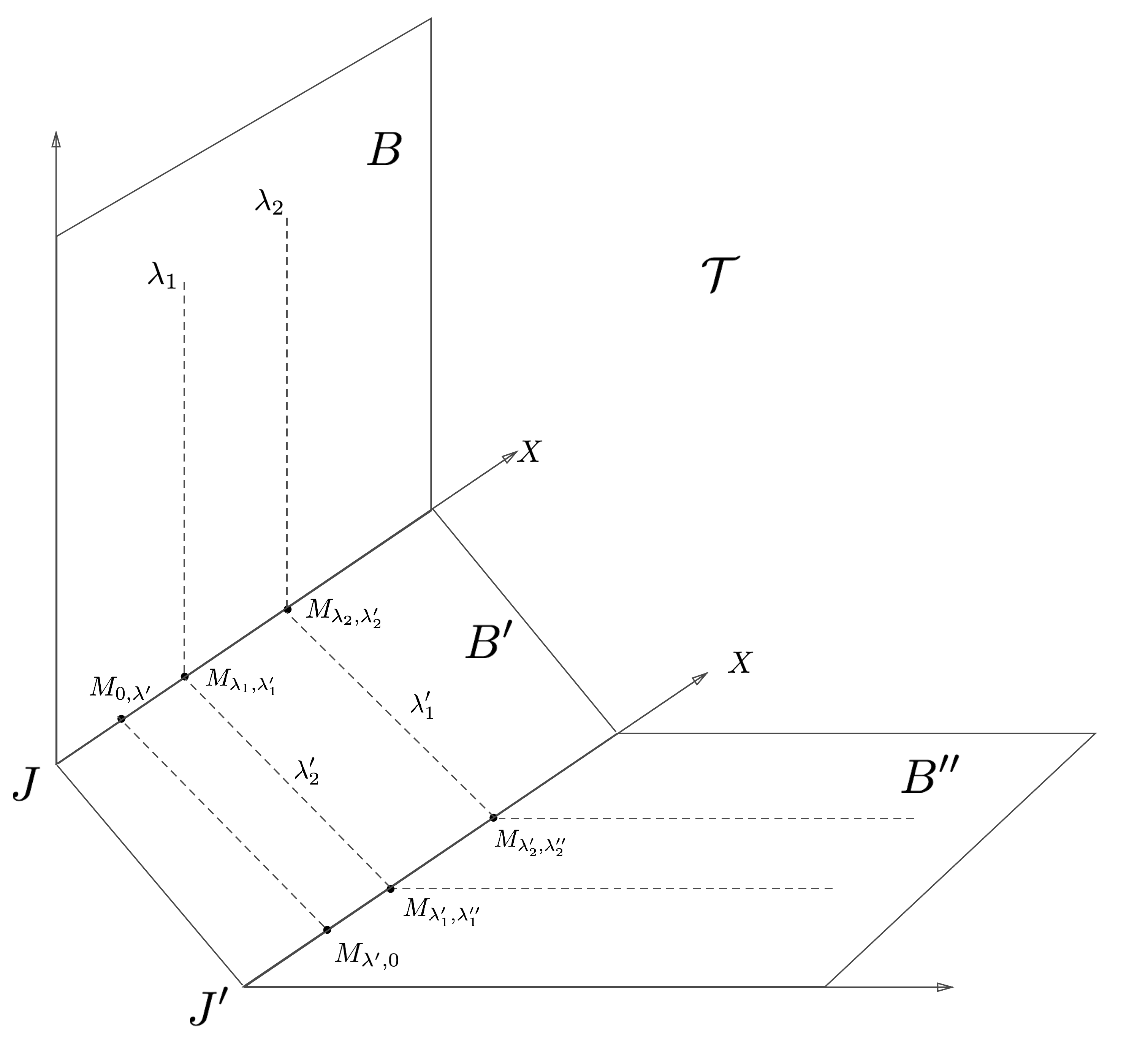}
\caption{Composition of junctions $J: B' \to B$ and $J': B'' \to
  B'$. The local operators for the composite junction
can arise from finite segments of line defects in the intermediate
boundary. Thus the new junction vertex algebra is an extension of
$A \times A'$ by a combination of product of modules $M_{0,\lambda'}
\times M_{\lambda',0}$. Modules for the
new vertex algebra are similarly built from combinations of the form
$M_{\lambda,\lambda'} \times M_{\lambda',\lambda''}$.
}\label{fig3}
\end{figure}

See Figure \ref{fig3} for the physical explanation of this prescription. 

In the rest of the paper, we will discuss several such compositions. A
particularly important application of these compositions is to produce
junctions between boundary conditions which do not simultaneously
admit weakly-coupled descriptions, in a sense we will explain momentarily. 

\subsection{Dualities}

A duality is an equivalence between different definitions of the same
4d theory. The quantum Geometric Langlands dualities are expected to
relate the different definitions (in other words, different duality
frames) of the category of 1d boundary conditions for ${\mc T}[X]$.

Each definition of a theory will come with some collection of ``weakly
coupled'' boundary conditions and junctions, which can be defined
microscopically in terms of the fields of the theory and possibly
other auxiliary fields. These are the boundary conditions and
junctions for which direct calculations of the categories of line
defects and vertex algebras are usually possible.

A duality rarely identifies two weakly coupled boundary
conditions. More often, it maps weakly coupled boundary conditions to
other ``strongly coupled'' ones for which direct calculations are
not feasible.

We will often face the problem to define and compute properties of
junctions between boundary conditions which are not simultaneously
weakly coupled.  For example, in order to study the qGL dualities, we
need to employ duality images of the Dirichlet boundary conditions.

Our solution to the problem is simple: we find a chain of boundary
conditions such that each consecutive pair is simultaneously weakly
coupled in {\it some} duality frame of the 4d theory; find and compute
the properties of the corresponding junctions; and then compose these
junctions into a single junction between the first and the last
boundary conditions in the chain.

\subsection{Spin subtleties}

We should mention here an important subtlety which will be recurring
in our analysis. {\em A priori}, the constructions we employ involve
theories and defects which are spin-topological or spin-holomorphic:
they are defined only on manifolds equipped with a spin structure. In
many situations, though, we can refine the constructions so that the
corresponding objects become truly topological or holomorphic.

Working with ``spin'' constructions is somewhat simpler. For example,
the dualities of gauge theories are well understood in the spin case,
while the non-spin refinement of the dualities is only known in some
examples.\footnote{As the physical gauge theory contains spinors, the
  theory is initially defined only on spin manifolds and the duality
  group(oid) has been studied in detail only on such manifolds. The
  topologically twisted theory, however, has no spinors and can be
  readily defined on a manifold without a spin structure. But then the
  duality group(oid) needs to be modified in ways that are only
  partly understood in the existing literature.

As the spinors in the physical theory transform in the
fundamental representation of the $SU(4)_R\simeq Spin(6)_R$ R-symmetry
group, while fields of integral spin transform in the representations
of $SO(6)_R$, it should be possible to couple the theory to
$\mathrm{Spin}_{SU(4)_R}$-bundles, i.e. bundles on a four-manifold $M$
with the structure group $(Spin(4) \times SU(4))/\Z_2$ (which is a
$\Z_2$-extension of the structure group $SO(4) \times (SU(4)/\Z_2) =
SO(4) \times SO(6)$), such that the corresponding $SO(4) \times
(SU(4)/\Z_2)$-bundle has the frame bundle of $M$ along the first
factor. The latter choice is probably better in preparation
for a topological twist. It suggests that the spin-refined duality
action may already be probed within the physical theory.}

The most immediate consequence of working with spin constructions is
that boundary line defects form spin-ribbon categories, a variant of
ribbon categories in which the topological twisting of an object is
defined only up to an integral multiple of $\pi$ (rather than $2\pi$).
Spin constructions also employ spin-vertex algebras, i.e. vertex
algebras in which some of the fields have half-integral spin.
Conformal blocks of modules over the spin-vertex algebras are only
well-defined on a complex curve equipped with a specific choice of
spin structure (i.e. a square root of the canonical line bundle).

If a spin-vertex algebra can be coupled to $H$-bundles and $H$ has a
$\Z_2$ central subgroup which acts as $-1$ on the fields of
half-integral spin, then instead of choosing a spin structure, we can
work with $\mathrm{Spin}_H$ bundles rather than $G$-bundles (we could
say we have a $\mathrm{Spin}_H$ vertex algebra) in this
case). Conformal blocks then become dependent on $\mathrm{Spin}_H$
bundles, and they give rise to twisted $D$-modules on the stack
$\Bun_{\mathrm{Spin}_H}$ of such bundles. See Section \ref{gerbes} for
more details, including the definition of $\mathrm{Spin}_H$
bundles.

Finally, notice that ``spin'' and ``super'' are distinct notions. Our
vertex algebras are allowed to be vertex superalgebras, in the sense
of including odd (Grassmann) fields. But these fields do not
necessarily have half-integral spin, and conversely, fields of
half-integral spin may or may not be odd.

\section{Boundary conditions, dualities, and junction vertex
  algebras}    \label{bc}

In this section we discuss specific boundary conditions and junctions
between them. Though they exist universally in all twisted 4d gauge
theories, we will focus on the GL twisted 4d gauge theory, as defined
in \cite{KW}, with a compact connected gauge group $G_c$, whose
complexification is denoted by $G$ (this is a connected reductive Lie
group over $\C$), and the topological coupling constant which is
denoted by $\Psi$ in \cite{KW}, but which we will denote by
$\ka$.\footnote{There is in fact a separate coupling for each simple
  factor of the Lie algebra of $G$, together with a matrix of couplings for Abelian factors.} Upon compactification
of the 4d gauge theory on a Riemann surface $X$, we naturally obtain a
functor from the category of line defects associated with a given
boundary condition to the category $D_\ka(\Bun_G)$ of twisted
$D$-modules on the moduli stacks $\Bun_G$ of $G$-bundles on $X$. We
will call it the compactification functor. In subsequent sections, we
will see how these functors, together with the action of quantum
dualities on the boundary conditions and junctions between them,
naturally lead us to valuable insights into the quantum Geometric
Langlands theory.

In fact, there is a larger class of theories that we need to consider
in order to fully explore the duality groups of gauge theories. These
theories include extra ``discrete theta angles''. For now, we set the
discrete theta angles to $0$, but we will come back to the more
general theories in later sections.

The parameter $\ka$ of gauge theory may be viewed as the level of the
corresponding affine Kac--Moody algebra $\ghat$ (the affinization of
the Lie algebra $\g$ of $G$) shifted by the critical value, or as the
twisting parameter for $D$-modules on the affine Grassmannian $\Gr_G$
and on $\Bun_G$, also shifted by the critical value (corresponding to
the properly defined square root of the canonical line bundle). We
will denote this bulk theory by ${\mc T}^G_\ka$.

The properties of the theory ${\mc T}^G_\ka$ are rather uniform as a
function of $\ka$, as long as $\ka$ is not rational. A variety of new
phenomena occurs for rational values of $\ka$, such as the existence
of a non-trivial category ${\mc S}({\mc T}^G_\ka)$
of bulk line defects. For simplicity, unless specified otherwise, we
will assume throughout this paper that $\ka$ is irrational (i.e. $\ka
\in \C \bs {\mathbb Q}$).

\subsection{Examples of boundary conditions}    \label{ex bc}

In the examples below, we consider basic boundary conditions in the
theory ${\mc T}^G_\ka$ and the ribbon categories corresponding
to them, which we will denote by ${\mc C}^G_\ka(B)$.

\bigskip

\begin{enumerate}

\item {\em The Dirichlet boundary condition}. The corresponding
  category ${\mc C}^G_\ka(B)$ is the category $D_{-\ka}(\Gr_G)$ of
  right $(-\kappa)$-twisted $D$-modules on the affine Grassmannian
  $\Gr_G = G\zpart/G[[z]]$. Recall that $\Gr_G$ is an ind-scheme, a
  union of its closed subschemes of finite type (which are proper
  finite-dimensional complex algebraic varieties). By definition, the
  support of each object of $D_{-\ka}(\Gr_G)$ should belong to one of
  those subschemes (see e.g. \cite{FG1} for a precisely definition).

\bigskip

\item {\em The Neumann boundary condition}. The corresponding category
  ${\mc C}^G_\ka(B)$ is the Kazhdan--Lusztig category $\KL_\ka(G)$
  \cite{KL} of finitely generated $\ghat_\ka$-modules $M$ on which the
  action of the Lie subalgebra $\g[[z]] \subset \ghat_\ka$ is locally
  finite and can be exponentiated to an action of $G[[z]]$
  (equivalently, $M$ decomposes into a direct sum of
  finite-dimensional modules over the constant Lie subalgebra $\g$
  which can be exponentiated to an action of $G$, and for any vector
  $v \in M$, there exists $n>0$ such that $\g \otimes t^n\C[[t]] \cdot
  v = 0$).

\bigskip

\item {\em The principal Nahm boundary condition}. The corresponding
  category ${\mc C}^G_\ka(B)$ is the Whittaker category
  $\on{Whit}_{-\ka}(G)$, of $(-\kappa)$-twisted $D$-modules on $\Gr_G$
  which are $N\zpart$-equivariant with respect to a fixed
  non-degenerate character $\psi$ (corresponding to the principal
  element appearing in the Nahm boundary condition). Here $N$ is the
  unipotent subgroup of $G$.\footnote{The category
    $\on{Whit}_{-\ka}(G)$ is an example of a category that requires,
    for some $G$, a choice of spin structure, i.e. a specific choice
    of a square root of the canonical line bundle on a formal disc (or
    on a curve $X$). Indeed, one needs to make this choice in order to
    define a non-degenerate character on the group $N\zpart$ in a
    coordinate-independent fashion.} See \cite{FGV,Gaitsgory} for the
  precise definition. The chiral category structure is defined in
  \cite{Beraldo,Raskin}.

\bigskip

\item {\em Other Nahm boundary conditions.} There is a more general
  family of boundary conditions labeled by the extra data of a
  conjugacy class of an embedding $\rho$ of $\mathfrak{sl}_2$ into
  $\g$ (or equivalently, a non-zero nilpotent conjugacy class in $\g$).
  The corresponding category $\on{Whit}^\rho_{-\ka}(G)$ can be defined
  in a similar manner as for the principal Nahm boundary condition.
\end{enumerate}

\bigskip

In addition, we expect that each of these categories is equipped with
a functor:

\medskip

\begin{itemize}

\item
For each boundary condition $B$ in the theory ${\mc
  T}^G_\ka$ and every compact Riemann surface (smooth projective
curve) $X$ with a point $x \in X$, there is a {\em compactification
  functor} $F^{G,B}_\ka$ from the category ${\mc C}^G_\ka(B)$ to the
category $D_\ka(\Bun_G)$ of left $\ka$-twisted $D$-modules on
$\Bun_G$. In general, this is a functor between derived, or DG,
categories. This functor has a multi-point generalization compatible
with fusion and braiding in ${\mc C}^G_\ka(B)$.

\medskip

\item
The image $F^{G,B}_\ka(I)$ of the identity object $I$ in ${\mc C}^G_\ka(B)$ 
does not depend on the point $x$. We call the map from boundary
conditions to twisted $D$-modules on $\Bun_G$,
$$
B \mapsto F^{G,B}_\ka(I)
$$
the {\em compactification map}.

\end{itemize}

\medskip

In Section \ref{stalks}, under some natural assumptions, we will
describe the compactification functor as a localization functor for a
certain vertex algebra with affine Kac--Moody symmetry. We will also
discuss the connection between the corresponding twisted $D$-modules
on $\Bun_G$ and branes on the Hitchin moduli space ${\mc M}_H(G)$
associated to $X$ constructed in \cite{KW}. Here we will only mention
that the boundary conditions available at general $\ka$ are
non-trivial {\it deformations} of standard half-BPS boundary
conditions of the physical theory. Half-BPS boundary conditions can be
employed either at $\ka=0$ or at $\ka=\infty$, but only a few can be
deformed to other values of $\kappa$. Neumann are deformed from
$\ka=\infty$, while Dirichlet and Nahm are deformed from $\ka=0$.

The compactification functor can be interpreted as the composition of
two functors: a compactification of boundary conditions to branes and
a functor from branes to twisted $D$-modules. The brane interpretation
is useful and instructive, but we stress that the compactification
functors can be defined directly, i.e. bypassing the categories of
branes on ${\mc M}_H(G)$ introduced in \cite{KW}. See Sections
\ref{Dmod branes} and \ref{branes vs D} for more details.

This simplifies some aspects of the construction. For example, as
explained in \cite{KW}, the sigma model description is simplest for
real values of $\ka$, but for other values of $\ka$ one needs to
consider generalized complex structures on ${\mc M}_H(G)$. On the
other hand, the categories of $\ka$-twisted $D$-modules on $\Bun_G$
behave in a more uniform way relative to $\ka$.

\subsection{Examples of the compactification functors}    \label{ex
  fun}

We now describe examples of the compactification functors, as well as
the images of the standard objects in ${\mc C}^G_\ka(B)$ under these
functors. We also comment on the corresponding branes on the Hitchin
moduli space ${\mc M}_H(G,X)$.

\medskip

\begin{enumerate}

\item {\em The Dirichlet boundary condition}. The compactification
  functor is the suitably defined push-forward corresponding to the
  surjective map
$$
\Gr_G \to \Bun_G = G(X\bs x) \bs \Gr_G.
$$
Note that under this functor right $(-\ka)$-twisted $D$-modules on
$\Gr_G$ are mapped to left $\ka$-twisted $D$-modules on $\Bun_G$.

The standard objects in the category $D_{-\ka}(\Gr_G)$ are
``$\delta$-function'' (right) $(-\ka)$-twisted $D$-modules supported
at points of $\Gr_G$ (see below), which we denote by $\delta^{-\ka}_p,
p \in \Gr_G$. The corresponding $D$-module on $\Bun_G$ is the
``$\delta$-function'' (left) $\ka$-twisted $D$-module $\delta^\ka_{\mc
  P}$, where ${\mc P}$ is the image of $p \in \Gr_G$ in $\Bun_G$. (If
${\mc P}$ has non-trivial automorphisms, then the category of twisted
$D$-modules supported at ${\mc P}$ is non-trivial, but we ignore this
for now. See the discussion at the end of this subsection, as well as
Remark \ref{p and P}.)

For irrational $\kappa$, the $D$-module $\delta^\ka_{\mc P}$ is a
counterpart of the $(A,B,A)$ brane ${\mc F}'_{\mc P}$. These are
supported on the fibers of the maps from an open subset of ${\mc
  M}_H(G,X)$, seen as a moduli space of flat $G$-bundles, to ${\mc
  M}(G,X)$, the moduli space of semi-stable $G$-bundles on $X$ (see
Section \ref{Dmod branes} for more details).

These branes are deformations of the $(B,A,A)$ branes $F_{\mc P}$, the
fibers of the projection of ${\mc M}_H(G,X)$, seen as a moduli space
of Higgs bundles on $X$, to ${\mc M}(G,X)$.  These originate from
the undeformed Dirichlet boundary conditions at $\kappa=0$.

\bigskip

\item {\em The Neumann boundary condition}. The compactification
  functor is the localization functor $\Delta_\ka$ (see \cite{FB},
  Ch. 18, \cite{F:review} and Section \ref{stalks} below).

  The standard objects in $\KL_\ka(G)$ are the Weyl modules ${\mathbb
    V}_{\la,\ka}$, where $\la \in P^+(G)$, the set of dominant
  integral weight of $G$ (i.e. highest weights of irreducible
  representations of $G$). (Note that for $\ka \not\in {\mathbb Q}$,
  $\KL_\ka(G)$ is a semisimple category with each object isomorphic
  to the direct sum of finitely many Weyl modules ${\mathbb
    V}_{\la,\ka}$.) The corresponding (left) twisted $D$-module
  $\Delta_\ka({\mathbb V}_{\la,\ka}) = {\mc D}^{x,\la}_\ka$ on
  $\Bun_G$ is the tensor product ${\mc D}_\ka \otimes_{\mc O} {\mc
    V}_{x,\la}$, where ${\mc D}_\ka$ is the sheaf of $\ka$-twisted
  differential operators on $\Bun_G$ and ${\mc V}_{x,\la}$ is the
  sheaf of sections of the tautological vector bundle on $\Bun_G$
  corresponding to the finite-dimensional representation $V_\la$ of
  $G$ and the point $x$ of the curve $X$.
  
In particular, for the vacuum module ${\mathbb V}_{0,\ka}$ we have
$\Delta_\ka({\mathbb V}_{0,\ka}) = {\mc D}_\ka$. This is the image of
the compactification map.

The brane on ${\mc M}_H(G,X)$ corresponding to ${\mc D}_\ka$ for
irrational $\ka$ is the canonical coisotropic brane ${\mc B}_{\on{c.c.}}$
(see Section \ref{Dmod branes}).  This is a deformation away from
$\kappa=\infty$ of the space-filling $(B,B,B)$ brane $\wt{\mc B}$
associated to the undeformed Neumann boundary conditions.

\bigskip

\item {\em The (principal) Nahm boundary condition}. Consider for
  simplicity the case that $G$ is simple. Then the Whittaker category
  $\on{Whit}_{-\ka}(G)$ can be realized as a direct limit of the
  categories of Whittaker sheaves (left $\ka$-twisted $D$-modules) on
  the moduli stacks $\ol\Bun_N^{{\mc F}_{T_0}(\mu^\vee \cdot x)}$ of
  $B$-bundles on $X$, where $\mu^\vee \in \LP^+$, the set of dominant
  integral coweights of $G$ (see \cite{FGV}). For each of these
  categories, we take the functor of direct image corresponding to the
  natural maps $j_{x,\mu^\vee}: \ol\Bun_N^{{\mc F}_{T_0}(\mu^\vee
    \cdot x)} \to \Bun_G$. The compactification functor from
  $\on{Whit}_{-\ka}(G)$ to the category of $\ka$-twisted $D$-modules
  on $\Bun_G$ is naturally ``glued'' from these.

The standard objects of $\on{Whit}_{-\ka}(G)$ are the Whittaker
sheaves $\Psi^{x,\mu^\vee}_\ka, \mu^\vee \in \LP^+$. In particular,
the ${\mc D}_\ka$-module corresponding to $\Psi^{x,0}_\ka$ is
constructed as follows. Let ${\mc F}_T^0$ be a $T$-bundle on $X$
(where $T$ is the Cartan subgroup of $G$) corresponding to the
canonical line bundle $K_X$ on $X$ and a
cocharacter $\check\rho$. Note that we may need to choose a square
root of the canonical line bundle on $X$ to make sense of this
$T$-bundle.\footnote{This is an example of how we may be forced to
  make a choice of a spin structure on $X$. Once we make such a
  choice, there is a parameter for ${\mc F}_T^0$: an element of
  $H^1(X,Z(G))$. For simplicity, we will assume here that $G$ is of
  adjoint type so that its center $Z(G)$ is trivial.} Let
$\ol\Bun_N^{{\mc     F}_T^0}$ be the corresponding moduli stack and
$j_{x,0}$ the corresponding map $\ol\Bun_N^{{\mc F}_T^0} \to
\Bun_G$. It is equipped
with a natural map $ev: \Bun_N^{{\mc F}_T^0} \to {\mathbb A}^1$, the
affine line. Let ${\mc E}$ be the $D$-module on ${\mathbb A}^1$
generated by the exponential function. Then, $\Psi^{x,0}_\ka$ is a
generalization to arbitrary $\ka$ of $j_{x,0!}(ev^*({\mc E}))$. The
construction of the other ${\mc D}_\ka$-modules
$\Psi^{x,\mu^\vee}_\ka$ is similar (for $\ka=0$, they can be obtained
by applying the Hecke functors to $\Psi^{x,0}_0$). See
\cite{FGV,Gaitsgory,Beraldo,Raskin} for more details.

For $\ka \neq 0$, the $(A,B,A)$ brane corresponding to
$\Psi_\ka^{0,x}$ is the brane of opers. This is a deformation away
from $\ka=0$ of the $(B,A,A)$-brane of ``classical opers'' (also known
as the Hitchin section) associated to undeformed Nahm boundary
conditions. See Section \ref{Dmod branes} for more details.

\end{enumerate}

\bigskip

\begin{table}[h]
\begin{center}
\begin{tabular}{|c||c||c|}
\hline
{\em Boundary Condition} & {\em Category of Lines} & {\em
  Compactification Functor}  \\
\hline \hline
Dirichlet ($D_{0,1}$) & $D_{-\ka}(\Gr_G)$  & Push-Forward $\Gr_G \to \Bun_G$  \\
\hline
Neumann ($N_{1,0}$) & Kazhdan--Lusztig $\KL_\ka$ & Localization
functor $\Delta_\ka$ \\
\hline
Principal Nahm ($N_{0,1}$) & Whittaker $\on{Whit}_{-\ka}(G)$ & Direct image  \\
\hline
\end{tabular} 
\end{center}
\vspace*{5mm}
\caption{The basic boundary conditions that exist universally for all
  $G$}\label{tab:two}
\end{table}

A few comments on the definition of the sheaf of
``$\delta$-functions'' supported at a point. Let $Z$ be a smooth
algebraic variety over $\C$ (or an ind-scheme such as $\Gr_G$) and $p$
a point of $Z$. By definition,
$$
\delta_p = i_{p!}(\C_p) = {\mc D}_Z \underset{{\mc O}_Z}\otimes \C_p
$$
is the (left) $D$-module push-forward of the constant sheaf at a
point (viewed as a $D$-module at that point) under the embedding
$i_p: \on{pt} \to Z$. Concretely, for any open subset $U \subset M$
containing $p$, we have
$$
\delta_p(U) = {\mc D}(U) \underset{{\mc O}(U)}{\otimes} \C_p,
$$
where ${\mc O}(U)$ is the ring of functions on $U$ acting on $\C_p$ by
evaluation at $p$. If we choose local coordinates $x_1,,\ldots,x_n$ at
$p$, and choose vector fields $\pa_1,\ldots,\pa_n$ on $U$ such that
the value of $\pa_i$ at $p$ is $\pa/\pa x_i$, then
$$
\delta_p(U) \simeq \C[\pa_i]_{i=1,\ldots,n}.
$$
The definition of the sheaf of $\delta$-functions supported at $p$ in
the categories of twisted left and right $D$-modules is similar.

If we deal with a smooth stack (such as $\Bun_G$) rather than a
variety, then a point ${\mc P}$ may well come with a non-trivial group
$\on{Aut}({\mc P})$ of automorphisms (in the case of a $G$-bundle
${\mc P}$, this is the group of global automorphisms of ${\mc P}$,
which is a subgroup of $G$). In this case, instead of a single
$\delta$-function $D$-module supported at ${\mc P}$, we have a
category of $D$-modules supported at ${\mc P}$ which is equivalent to
the category of representations of $\on{Aut}({\mc P})$. In this case,
we denote by $\delta_{\mc P}$ the $D$-module corresponding to the
regular representation of $\on{Aut}({\mc P})$ (i.e. the space of
functions on $\on{Aut}({\mc P})$). For further comments about these
$D$-modules, see Remark \ref{p and P} at the end of Section \ref{comp
  and loc}.

\subsection{Quantum dualities}    \label{symmetries}

Supersymmetric quantum gauge theories in 4d possess important
dualities generalizing the electromagnetic duality. Each duality
relates various attributes of two gauge theories, which usually have
different coupling constants and often different groups, $G$ and its
Langlands dual $\LG$. Upon compactification of the 4d theory on a
Riemann surface $X$, these dualities give rise to interesting
equivalences of categories, which can be viewed as quantum versions of
the geometric Langlands correspondence. Tracing these equivalences to
the ``first principles'' of 4d gauge theory yields unexpected insights
into the quantum geometric Langlands theory, which would be difficult
to realize otherwise.

Gauge theory for a general compact Lie group can be decomposed into
gauge theories for the simple and abelian factors, up to some
topological subtleties which however do not affect the
non-perturbative dynamics behind quantum duality symmetries.

We will discuss the properties of general gauge theories in Section
\ref{sec:topo}. For now, we will focus on a connected simple Lie group
$G$ over $\C$. Denote by $m$ the {\em lacing number} of $G$ -- the
maximal number of edges connecting two vertices of the Dynkin diagram
of $G$.  Thus, $m$ is $1$ for simply-laced groups, and is equal to $2$
or $3$ for non-simply laced simple Lie groups.

There are two basic types of dualities: the orientation reversal
symmetry $R$ and the dualities corresponding to elements of a
subgroup of $PGL_2(\Z)$, which depends on the bulk theory. We start by
outlining their action on bulk theories ${\mc T}^G_\ka$ introduced
above.

\medskip

\begin{itemize}

\item $R$ preserves the group $G$ but sends $\ka$ to $-\ka$: ${\mc
  T}^G_\ka \mapsto {\mc T}^G_{-\ka}$.

\medskip

\item For a given $G$, only a particular subgroup of $PGL_2(\Z)$ acts
  by duality transformations. Furthermore, these transformations in
  general change not only the parameter $\ka$ but the group $G$ as
  well, so in fact it is better to view them as forming a {\em
    groupoid} rather than a group.

\end{itemize}

\medskip

Let us describe this groupoid, which we will denote by ${\mc G}^G_\ka$.

The group $PGL_2(\Z)$ naturally acts on the projective line, which
gives rise to the standard action on $\ka$ viewed as a point on the
projective line with respect to a particular coordinate:
$$
\ka \mapsto \frac{a\ka+b}{c\ka+d}.
$$
Now introduce the following elements of $PGL_2(\Z)$:
$$
S_m = \begin{pmatrix}
0 & -1 \\
m & 0
\end{pmatrix} \qquad \ka \mapsto \check\ka = - 1/m\ka
$$
and
$$
T = \begin{pmatrix}
1 & 1 \\
0 & 1
\end{pmatrix} \qquad \ka \mapsto \ka+1.
$$

Denote by $n(G)$ the minimum positive coefficient of a well-defined
Chern--Simons action for $G_c$ \cite{DW} (equivalently, the level of
the affine Kac--Moody algebra corresponding to the minimal line bundle
${\mc L} _G$ on the moduli stack of $G$-bundles on a curve
X).\footnote{Here is the first place where spin subtleties occur: for
  some $G$, the minimal Chern--Simons action is only defined on spin
  manifolds unless we replace $n(G)$ by $2n(G)$. Equivalently, for
  those groups we need to pick the square root of the canonical line
  bundle on $X$ in order to define the line bundle ${\mc L}_G$;
  without such a choice, we can only define ${\mc L}_G^{\otimes 2}$.}
For example, $n(G)=1$ if $G$ is simply-connected simple Lie
group. Likewise, we denote by $n(\LG)$ the corresponding number for
$\LG$. Let ${\mc S}_G$ (resp., ${\mc S}_{\LG}$) be the subgroup of
$PSL_2(\Z)$ generated by $T^{n(G)}$ and $S_m T^{n(\LG)} S_m$ (resp.,
$T^{n(\LG)}$ and $S_m T^{n(G)} S_m$).


We define the groupoid ${\mc G}^G_\ka$ as the category with the
objects (which we will sometimes refer to as nodes) labeled by pairs
$$
(G,\ka'), \quad \ka' \in {\mc S}_G \cdot \ka, \qquad \on{and} \qquad
(\LG,\ka'), \quad \ka' \in {\mc S}_{\LG} \cdot \check\ka,
$$
where $\check\ka=-1/m\ka$. We have morphisms (arrows) from $(G,\ka')$
to $(G,\ka'')$ for each element $g$ in ${\mc S}_G$ such that $\ka'' =
g(\ka')$; from $(G,\ka')$ to $(\LG,\ka'')$ for each element $g$ in
${\mc S}_{\LG} S_m$ or $S_m{\mc S}_G$ such that $\ka'' = g(\ka')$; and
similarly for the morphisms starting from nodes labeled by $\LG$.


\bigskip

Now we consider the action of these symmetries on the boundary
conditions discussed above and the corresponding categories.

The symmetry $R$ sends the bulk theory ${\mc T}^G_\ka$ with a boundary
condition $B$ to the bulk theory ${\mc T}^G_{-\ka}$ with a boundary
condition that we denote by $R(B)$. The resulting ribbon category
should be dual to the original one, so that we have an equivalence
\begin{equation}    \label{dual}
{\mc C}^G_{-\ka}(R(B)) \simeq {\mc C}^G_\ka(B)^\vee
\end{equation}
In general, we need to consider here the dual of the derived, of a DG,
category. However, for irrational values of $\ka$, it appears that it
is sufficient to work with the abelian categories.

On the other hand, the action of the duality groupoid ${\mc G}^G_\ka$
should take the ribbon categories to equivalent ones:
$$
{\mc C}^{g(G)}_{g(\ka)}(g(B)) \simeq {\mc C}^G_\ka(B),
$$
for every $g \in {\mc S}_G$ (in this case, $g(G)=G$) and $g \in S_m
\cdot {\mc S}_G$ (in this case, $g(G)=\LG$).

\subsection{Examples of the action of quantum dualities  on boundary
  conditions}

Now we look at how $R$ and the duality groupoid ${\mc G}^G_\ka$ act on
the basic boundary conditions.

\bigskip

\begin{enumerate}

\item {\em The action of $R$} preserves the Neumann, Dirichlet, and
  Nahm boundary conditions. In fact, for irrational $\ka$ the
  equivalences \eqref{dual} hold in these cases at the level of
  abelian categoies.

  For example, for irrational $\ka$ the dual category to
  $\on{KL}_\ka(G)$ is $\on{KL}_{-\ka}(G)$. This can be seen from the
  Kazhdan--Lusztig equivalence between the category $\on{KL}_\ka(G)$
  with irrational $\ka$ and the category $U_q(\g)$-mod of
  finite-dimensional $U_q$-modules, where $q=e^{\pi i/\ka}$. As
  explained in Example 2.10.14 of \cite{EGNO}, the dual category to
  the latter is equivalent to the opposite category, which is known to
  be equivalent to $U_{q^{-1}}(\g)$-mod and hence to
  $\on{KL}_{-\ka}(G)$ via the Kazhdan--Lusztig equivalence.

\bigskip

\item {\em The action of $S_m$} exchanges the Neumann and principal
  Nahm boundary conditions. The corresponding categories are therefore
  expected to be equivalent:
\begin{equation}    \label{KLWhit}
\KL_\ka(G) \simeq \on{Whit}_{1/m\ka}(\LG).
\end{equation}
This is essentially the statement of a conjecture of Gaitsgory and
Lurie, proved by Gaitsgory for irrational $\ka$
\cite{gaitsW,gaitsQ}.\footnote{This is an example of an equivalence
  that for some groups requires a choice of a spin structure (or a
  square root of the canonical line bundle), because the category
  $\KL_\ka(G)$ does not require such a choice, but the category
  $\on{Whit}_{1/m\ka}(\LG)$ does require it for some $G$.}

\bigskip

\item {\em The $S_m$-dual of the Dirichlet boundary condition} is much
  more complicated if $G$ is non-abelian (see \cite{GW1,GW2,
    Ga1,Ga2}). (If $G$ is abelian, there is no difference between the
  Dirichlet and Nahm boundary conditions, and therefore the dual of
  Dirichlet is Neumann.) From the perspective of boundary conditions in
  4d gauge theory, this is the reason why it is difficult to construct
  the quantum geometric Langlands correspondence. All we can say at
  the outset is that the corresponding category should be equivalent
  to the category of right $1/m\ka$-twisted $D$-modules on
  $\Gr_{\LG}$. Similar considerations apply to the intermediate Nahm
  boundary conditions.

\bigskip

\item {\em The action of $T^p, p \in n(G) \cdot \Z$}. If $B$ is any of
  the Nahm boundary conditions or a Dirichlet boundary condition, then
  $T^p(B) = B$, but for the Neumann boundary conditions, $T^p(B) \neq
  B$ for $p \in \Z$. In any case, we expect the equivalences
$$
{\mc C}^G_{\ka+p}(T^p(B)) \simeq {\mc C}^G_\ka(B),
$$
In other words,
\begin{equation}    \label{Tp}
{\mc C}^G_\ka(T^p(B)) \simeq {\mc C}^G_{\ka-p}(B).
\end{equation}
In particular, the equivalences
\begin{equation}    \label{WhitWhit}
D_\ka(\Gr) \simeq D_{\ka+p}(\Gr), \qquad \on{Whit}_{\ka}(G) \simeq
\on{Whit}_{\ka+p}(G), \qquad \forall p \in n(G) \cdot \Z
\end{equation}
are given by tensoring with the appropriate line bundle on $\Gr_G$.
\end{enumerate}

\bigskip

In Section \ref{qGL dual} we conjecture that for irrational $\ka$ the
groupoid ${\mc G}^G_\ka$ acts by equivalences on the categories of
twisted $D$-modules on $\Bun_G$ and $\Bun_{\LG}$ (the situation
becomes more subtle for rational values of $\ka$).

\subsection{Nomenclature for boundary conditions}    \label{nomen}

We will consider boundary conditions in bulk theories of type ${\mc
  T}^G_\ka$ discussed above and therefore for now we will keep $G$ in
our notation for boundary conditions. But since all boundary
conditions we consider are defined uniformly for all values of
$\kappa$, we will not keep $\kappa$. Hence we use the notation
$B^G$. Starting from every a boundary condition $B^G$, we can generate
a whole family of boundary conditions: $R(B^G)$ and $g(B^G)$ with $g
\in {\mc S}_G$ and $g(B^{\LG})$ with $g \in S_m {\mc S}_{G}$ by the
action of the duality groupoid.
This suggests the following nomenclature (it is similar but not
identical to the one adopted in \cite{CG}).

For a given bulk theory ${\mc T}^G_\ka$ we will use the following
notation:

\medskip

\begin{enumerate}

\item $N^G_{0,1}$ for the (principal) Nahm
boundary condition;

\medskip

\item $N^G_{1,0}$ for the Neumann boundary condition;

\medskip

\item $D^G_{0,1}$ for the Dirichlet boundary condition;

\medskip

\item $N^{G,\rho}_{0,1}$ for the $\rho$-Nahm boundary condition.
\end{enumerate}

\bigskip

Next, we denote by $N^G_{g \circ (0,1)}$ and $N^G_{g \circ (1,0)}$
(modulo the identification $N^G_{ap,aq} = N^G_{p,q}$) the images of
these boundary conditions under $g \in {\mc S}_G$, with $g$ acting on
the pairs $(p,q)$ according to its standard action on the column
vectors ${q \choose p}$ (note the switch of $p$ and $q$).

Thus, every element in ${\mc G}^G_\ka$ sends
$$
N^G_{p,q} \mapsto N^{g(G)}_{g \circ (p,q)},
$$
where $g(G)$ is either $G$ or $\LG$. At the same time, each $g$ acts
on $\ka$ in the standard way.

Further, for the orientation reversal $R$ we define
$$
R: N^G_{p,q} \mapsto N^G_{p,-q}
$$
In this case, $R({\mc T}^G_\ka) = {\mc T}^G_{-\ka}$, and so $R(\ka) =
-\ka$.

We can give a similar definition of the families $D^G_{p,q}$ and
$N^{G, \rho}_{p,q}$ starting from Dirichlet or $\rho$-Nahm boundary
conditions in the bulk gauge theories ${\mc T}^G_\ka$ and ${\mc
  T}^{\LG}_\ka$. For example, $D^G_{0,1}$ denotes the Dirichlet
boundary condition in ${\mc T}^G_\ka$, while $D^G_{1,0}$ denotes the
$S_m$-image of the Dirichlet boundary condition $D^{\LG}_{0,1}$ in
${\mc T}^{\LG}_\ka$. Also, $R$ maps $D^G_{p,q}$ to $D^G_{p,-q}$, etc.

Using $R$ together with the groupoid of duality transformations, we
obtain various non-trivial identifications between boundary conditions
$N^G_{p,q}$ (beyond $N^G_{ap,aq} = N^G_{p,q}$). Each of them is
expected to give rise to an equivalence between the corresponding
ribbon categories of line defects. Some of these equivalence are
highly non-trivial. For instance, we have
\begin{equation}    \label{identS}
({\mc T}^{\LG}_{-1/m\ka},N^{\LG}_{0,1}) =
({\mc T}^G_\ka,N^G_{1,0}), \qquad ({\mc T}^{\LG}_{-1/m\ka},N^{\LG}_{1,0}) =
({\mc T}^G_\ka,N^G_{0,1}).
\end{equation}
The corresponding equivalences of categories are given by formula
\eqref{KLWhit} and its analogue in which we reverse $G$ and $\LG$,
$\ka$ and $1/m\ka$. We will encounter more equivalences of this nature
below.

\subsection{Extended families of boundary conditions}

With some care, the duality groupoid can be extended to a larger
groupoid involving all transformations generated by $T$ and $S_m$. 
The nodes of the extended groupoid involve gauge theories 
with gauge algebra $\mathfrak{g}$ or ${}^L\mathfrak{g}$
but gauge group which may differ from $G$ or $\LG$.
They are furthermore modified by certain ``discrete theta angles'' \cite{AST}. 
It is possible to define well-behaved analogues
of Dirichlet, Neumann, Nahm boundary conditions in these generalized
theories, but it requires some extra structures, which we will
describe in section \ref{sec:topo}.

As a consequence, one may accordingly extend the range of pairs $(p,q)$ 
for which the above families of boundary conditions are
well-defined. The compactification functors should map the categories
associated to these boundary conditions to the categories of
$D$-modules on $\Bun_G$ twisted by appropriate gerbes.

\subsection{Vertex algebras at a junction}    \label{va junction}

As explained in Section \ref{va j}, we can attach a vertex algebra to
a junction of two boundary conditions $B_1$ and $B_2$ in a bulk theory
${\mc T}$, possibly with some extra data attached to the junction. We
now focus on the case of the bulk theory ${\mc T}={\mc T}^G_\ka$, as
above, and denote the corresponding junction vertex algebra $V({\mc
  T}^G_\ka,B_1\sto B_2)$ by $V^G_\ka(B_1\sto B_2)$. Then we have the
corresponding functor $F_{{\mc T},B_1\, \sto \; B_2}$, which we now
denote by $F^G_{\ka,B_1\; \sto \; B_2}$. Thus,
\begin{equation}    \label{box}
F^G_{\ka,B_1\; \sto \; B_2}: {\mc C}^G_\ka(B_1)^\vee \boxtimes {\mc
  C}^G_\ka(B_2) \to V^G_\ka(B_1\sto B_2)\on{-mod}.
\end{equation}

This functor should satisfy the following conditions:

First,
\begin{equation}    \label{II}
F^G_{\ka,B_1\; \sto \; B_2}(I \otimes I) = V^G_\ka(B_1\sto B_2),
\end{equation}
where $I$ denotes the identity object.

Second, if we apply any duality symmetry $g \in {\mc G}^G_\ka$ as
defined in Section \ref{symmetries} to all data, then the vertex
algebra should be unchanged:
\begin{equation}    \label{g junction}
V^G_\ka(B_1\sto B_2) \simeq V^{g(G)}_{g(\ka)}(g(B_1)\sto g(B_2)),
\end{equation}
and we should have a commutative diagram
\begin{equation}    \label{compat}
\begin{CD}
{\mc C}^G_\ka(B_1)^\vee \boxtimes {\mc
  C}^G_\ka(B_2) @>{F^G_{\ka,B_1\; \sto \; B_2}}>>
V^G_\ka(B_1\sto B_2)\on{-mod} \\
@VVV @VVV \\
{\mc C}^{g(G)}_{g(\ka)}(g(B_1))^\vee \boxtimes {\mc
  C}^{g(G)}_{g(\ka)}(g(B_2))  @>{F^{g(G)}_{g(\ka),g(B_1)\; \sto \; g(B_2)}}>>
V^{g(G)}_{g(\ka)}(g(B_1)\sto g(B_2))\on{-mod}
\end{CD}
\end{equation}

Third, under the orientation reversal $R$, we should have an
isomorphism of vertex algebras
\begin{equation}   \label{R junction}
V^G_\ka(B_1\sto B_2) \simeq V^G_{-\ka}(R(B_2)\sto R(B_1)),
\end{equation}
compatible with the functors $F^G_{\ka,B_1\; \sto \; B_2}$ and
$F^G_{-\ka,R(B_2)\; \sto \; R(B_1)}$ and the equivalences
$$
{\mc C}^G_\ka(B_i)^\vee \simeq {\mc C}^G_{-\ka}(R(B_i)).
$$

Let us consider some basic examples. More examples will be presented
in later sections.

\bigskip

\begin{enumerate}

\item Let $B_1=D^G_{0,1}$ and $B_2=N^G_{1,0}$ be the Dirichlet and Neumann
  boundary conditions, respectively. There is a standard junction
  between them described in \cite{CG} such that
  $V^G_\ka(B_1\sto B_2)$ is the affine Kac--Moody vertex algebra
  $V_\ka(\g)$.

In this case, we have
$$
{\mc C}^G_\ka(D^G_{0,1})^\vee \boxtimes {\mc C}^G_\ka(N^G_{1,0}) =
\left(D_{-\ka}(\Gr_G)\right)^\vee \boxtimes \KL_\ka(G)
\simeq D_\ka(\Gr_G) \boxtimes \KL_\ka(G),
$$
and there is indeed a functor from the latter category to the category
of $V_\ka(\g)$-modules (i.e., $\ghat$-modules of level
$\kappa$). Namely, we view $\KL_\ka(G)$ as a subcategory of
$\ghat_\ka\on{-mod}$ and use the functor
\begin{equation}    \label{star}
{\mc F} \in D_\ka(\Gr_G), M \in \KL_\ka(G) \mapsto {\mc
  F} \star M,
\end{equation}
where $\star$ denotes the categorical convolution functor (see
\cite{BD,FG}).

Beilinson has conjectured (see the Introduction of \cite{Bei},
especially Remark (ii)) that the category of $\ghat$-modules of any
level $\kappa$ is expected to ``fiber'' over the stack of flat
$\LG$-bundles on the punctured disc. This conjecture suggests that the
essential image of ${\mc C}^G_\ka(D^G_{0,1})^\vee \boxtimes {\mc
  C}^G_\ka(N^G_{1,0})$ under the functor $F^G_{\ka,D^G_{0,1}\; \sto
\; N^G_{1,0}}$ is the subcategory of $\ghat$-modules of level $\kappa$
supported on the formal neighborhood of the trivial flat bundle.

\bigskip

\item Let $B_1=N^G_{0,1}$ and $B_2=N^G_{1,0}$ be the Nahm and Neumann
  boundary conditions, respectively. Then there is a standard junction
  between them described in \cite{NW,GR,CG} such that $V^G_\ka(B_1\sto B_2)$
  is the ${\mc W}$-algebra ${\mc W}_\ka(\g)$.

The corresponding functor
$$
F^G_{\ka,N^G_{0,1}\; \sto \; N^G_{1,0}}: \on{Whit}_{\ka}(G) \boxtimes
\KL_\ka(G) \to {\mc W}_\ka(\g)\on{-mod}
$$
is constructed as follows. Recall \cite{FF,FKW} (see Ch. 15 of
\cite{FB} for a survey) that the Drinfeld--Sokolov reduction functor
$H^{\g}_{\on{DS}}$ is defined as the semi-infinite cohomology of the
nilpotent Lie subalgebra $\n_+\zpart$ of $\ghat$ twisted by a
non-degenerate character $\psi$ of $\n_+\zpart$ which takes non-zero
values on the $(-1)$st Fourier coefficients of the generating
currents. This functor, applied to $\KL_\ka(G)$, defines the
restriction of $F^G_{\ka,N^G_{0,1}\; \sto \; N^G_{1,0}}$ to $I
\boxtimes \KL_\ka(G))$, where $I=\Psi^{x,0}_\ka$.

For irrational $\ka$, it follows from the results of Arakawa
\cite{Arakawa,Arakawa1} that the functor $H^{\g}_{\on{DS}}$ is exact,
and its essential image in the category ${\mc W}_\ka(\g)$-mod is a
semi-simple subcategory with the simple modules $M_{(\la,0),\ka} =
H^{\g}_{\on{DS}}({\mathbb V}_{\la,\ka})$. See also \cite{Raskin:w}
for some general results about this functor.

In order to incorporate the category $\on{Whit}_{\ka}(G)$, observe
that for any dominant integral coweight $\mu^\vee$ of $G$, we can
twist the character $\psi$ by the element $\mu^\vee(z) \in
H(\!(z)\!)$. Let us denote the quantum Drinfeld--Sokolov reduction
functor with respect to this twisted character by
$H^{\g}_{\on{DS},\mu^\vee}$. This functor has been previously studied
in \cite{FG,CG}.

For any object $M$ of $KL_\ka(G)$, the functor $F^G_{\ka,N^G_{0,1}\;
  \sto \; N^G_{1,0}}$ sends the object $\Psi^{x,\mu^\vee}_\ka \otimes
M$ of $\on{Whit}_{\ka}(G) \boxtimes \KL_\ka(G)$ to the ${\mc
  W}_\ka(\g)$-module $H^{\g}_{\on{DS},\mu^\vee}(M)$. This can be
generalized to arbitrary objects of $\on{Whit}_{\ka}(G)$. Thus,
$\on{Whit}_{\ka}(G)$ can be viewed as the category that controls the
data of the Drinfeld--Sokolov reduction.

It is natural to conjecture that for irrational $\ka$ the essential
image of $F^G_{\ka,N^G_{0,1}\; \sto \; N^G_{1,0}}$ is a semi-simple
subcategory of ${\mc W}_\ka(\g)$-mod with simple modules
\begin{equation}    \label{Mka}
M_{(\la,\mu^\vee),\ka} = H^{\g}_{\on{DS},\mu^\vee}({\mathbb
  V}_{\la,\ka}) = F^G_{\ka,N^G_{0,1}\; \sto \;
N^G_{1,0}}(\Psi^{x,\mu^\vee}_\ka \otimes {\mathbb
  V}_{\la,\ka}).
\end{equation}

According to our conventions, the duality $S_m$ acts as follows:
$$
S_m(N^G_{0,1}) = N^{\LG}_{1,0}, \qquad S_m(N^G_{1,0}) = N^{\LG}_{0,1}
$$
Hence $RS_m$ should send the junction $N^G_{0,1}\to N^G_{1,0}$ to
$N^{\LG}_{0,1}\sto N^{\LG}_{1,0}$, leading to the Feigin-Frenkel duality
\begin{equation}    \label{dualW}
{\mc W}_\ka(\g) \simeq {\mc W}_{1/m\ka}(\lg).
\end{equation}
Moreover, it follows from the diagram \eqref{compat} that the
subcategories of ${\mc W}_\ka(\g)$-mod and ${\mc W}_{1/m\ka}(\lg)$-mod
that are the essential images of the functors $F^G_{\ka,N^G_{0,1}\;
  \sto \; N^G_{1,0}}$ and $F^{\LG}_{1/m\ka,N^{\LG}_{0,1}\; \sto \;
  N^{\LG}_{1,0}}$ should be equivalent. In particular, under the
duality \eqref{dualW} we expect to have an isomorphism
\begin{equation}
M^{\g}_{(\la,\mu^\vee),\ka} \simeq M^{\lg}_{(\mu^\vee,\la),1/m\ka} \, .
\end{equation}
(see Conjecture 1.5 of \cite{CG}).\footnote{This has been proved in
  \cite{AF}.}

Note that if this is the case, then the Whittaker category
$\on{Whit}_{\ka}(G)$ can be realized for irrational $\ka$ as a
semi-simple subcategory ${\mc W}_\ka(\g)_m$-mod of ${\mc
  W}_\ka(\g)$-mod, which has as simple objects the modules
$$
M_{(0,\mu^\vee),\ka} =
H^{\g}_{\on{DS},\mu^\vee}({\mathbb V}_{0,\ka})
$$
with ``magnetic'' highest weights $(0,\mu^\vee), \mu^\vee \in
\LP^+$. The equivalence \eqref{KLWhit} between $\on{Whit}_{\ka}(G)$
and $KL_{1/m\ka}(\LG)$ can therefore be rephrased as the statement
that the category ${\mc W}_\ka(\g)_m$-mod is equivalent, as a ribbon
category, to $KL_{1/m\ka}(\LG)$. This statement (and its extension to
negative rational $\ka$) is essentially Conjecture 6.3 of
\cite{AFO}.

\bigskip

\item More generally, if $B$ is any boundary condition, then we expect
  that the vertex algebra
$$
V^G_\ka(D^G_{0,1}\sto B)
$$
contains the affine vertex algebra $V_{\ka'}(\g)$, where $\ka' - \ka
\in n(G) \cdot \Z$, as a subalgebra. \footnote{This expectation is
  conditional on the existence of a physical junction $D^G_{0,1}\sto
  B$ which preserves the global $G$ symmetry associated to the
  Dirichlet boundary condition. If such a junction exists, then by
  Noether's theorem we obtain the corresponding conserved currents
  which generate the affine vertex algebra at the junction.}

We cannot determine $\ka'$ on general grounds, because we can always modify the junction data
$J_{12}$ by adding some holomorphic vertex algebra, and this operation
may shift the level by a multiple of $n(G)$.

Furthermore, we expect that
$$
V^G_\ka(N^G_{0,1}\sto B)
$$
can be obtained by the quantum Drinfeld--Sokolov reduction of this
affine subalgebra, and
$$
V^G_\ka(N^G_{1,\ka-\ka'}\sto B)
$$
can be obtained by the BRST reduction of
$$
V^G_\ka(D^G_{0,1}\sto B) \otimes V_{-\ka'}(\g)
$$
(with respect to the diagonal action of $\ghat$, with the level equal
to twice the critical level). \footnote{These expectations are motivated from the existence of physical operations which convert a 
$D^G_{0,1}$ boundary condition to either $N^G_{0,1}$ or $N^G_{1,\ka-\ka'}$. For example, an $N^G_{1,\ka-\ka'}$
boundary condition can be obtained by ``gauging the $G$ global symmetry of $D^G_{0,1}$''. 
The effect of these operations on the junction vertex algebra is encoded by the above BRST reductions. }

From the last two constructions, we can obtain many examples of vertex
algebras, some of which we consider below.

These statements are consistent with the categorical statements: The
averaging functor $D_\ka(\Gr_G) \to \on{Whit}_{\ka}(G)$ gives rise to
a functor
$$
D_\ka(\Gr_G) \boxtimes {\mc C}^G_\ka(B) \to \on{Whit}_{\ka}(G)
\boxtimes {\mc C}^G_\ka(B).
$$
The composition of natural functors $D_\ka(\Gr_G) \to D_{\ka'}(\Gr_G)
\to \KL_{-\ka'}(G)$ (where the first functor is tensoring with a line
bundle on $\Gr_G$ and the second functor is pulling back to $G\zpart$
and taking $G[[z]]$-invariants with respect to the right action) and
the equivalence $\KL_{-\ka'}(G) = {\mc C}^G_{-\ka'}(N^G_{1,0}) \simeq
{\mc C}^G_{-\ka}(N^G_{1,\ka'-\ka})$ gives a functor
$$
D_\ka(\Gr_G) \boxtimes {\mc C}^G_\ka(B) \to {\mc
  C}^G_{-\ka}(N^G_{1,\ka'-\ka}) \boxtimes {\mc C}^G_\ka(B).
$$

\bigskip

\item Similar statements hold for general Nahm boundary conditions
  corresponding to an embedding $\rho: \mathfrak{sl}_2 \to \g$. In
  this case, the vertex algebra is the ${\mc W}$-algebra obtained by
  the generalized quantum Drinfeld--Sokolov reduction of $V_\ka(\g)$
  associated to the embedding $\rho$ \cite{KRW}.
\end{enumerate}

\subsection{Conformal blocks and compactification
  functors}    \label{conf blocks}

It is natural to consider the spaces of conformal blocks and
coinvariants of a junction vertex algebra $V^G_\ka(B_1\sto B_2)$.

Recall from Section \ref{cb} that we expect the coinvariants of
$V^G_\ka(B_1\sto B_2)$ to be closely related to the Hom's of the
images of line defects in $B_1$ and $B_2$ under the compactification
functors.

More precisely, let ${\mc A}_1 \in \on{Ob}({\mc C}^G_\ka(B_1)$, ${\mc
  A}_2 \in \on{Ob}({\mc C}^G_\ka(B_2))$. The compactification functors
send them to twisted $D$-modules $\ol{\mc A}_1 = F^{G,B_1}_\ka({\mc
  A_1})$ and $\ol{\mc A}_2 = F^{G,B_2}_\ka({\mc A_2})$ on $\Bun_G$.

On the other hand, we have a $V^G_\ka(B_1\sto B_2)$-module
$F^G_{\ka,B_1\; \sto \; B_2}({\mc A}_1^\vee \otimes {\mc A}_2)$ which
we will denote simply by ${\mc A}_1^\vee \otimes {\mc A}_2$. Then the
statement of Conjecture \ref{Hom and conf} becomes

\begin{conj}    \label{Hom and conf1}
Suppose that the functor $F^G_{\ka,B_1\; \sto \; B_2}$ is fully
faithful. Then $\on{Hom}(\ol{\mc A}_1,\ol{\mc A}_2)$ is
isomorphic to the space of coinvariants of the $V^G_\ka(B_1\sto
B_2)$-module ${\mc A}_1^\vee \otimes {\mc A}_2$.
\end{conj}

An important special case arises when $B_1$ is the Dirichlet boundary
condition $D^G_{0,1}$ whose global symmetry group is the group
$G$.\footnote{More generally, the global symmetry group of the general
  Nahm pole boundary conditions $N^{G,\rho}_{1,0}$ is the centralizer
  of $\rho$.} Then we can choose as $\ol{\mc A}_1$ the
$\delta$-function $D$-module $\delta_{\mc P}$ supported at ${\mc P}
\in \Bun_G$. Thus, given ${\mc A} \in \on{Ob}({\mc C}^G_\ka(B))$, we
obtain a family of vector spaces $\on{Hom}(\delta_{\mc P},\ol{\mc
  A})$. According to conjecture \ref{Hom and conf1}, they should be
isomorphic to the spaces of coinvariants of the $V(B_1\sto
B_2)$-module corresponding to $\delta_p \otimes {\mc A}$.

We note that important special cases of this isomorphism were studied
(in the language of branes) in \cite{BT} as an expression of a
relation between the Kapustin--Witten and Beilinson--Drinfeld
approaches to the Geometric Langlands correspondence (in this regard,
see also Section \ref{hecke}).

In fact, as we explain in Section \ref{comp and loc}, we expect a
stronger statement to be true. Indeed, on the one hand,
$\on{Hom}(\delta_{\mc P},\ol{\mc A})$ is a fiber of the $D$-module
$\ol{\mc A} = F^{G,B}_\ka({\mc A})$ at ${\mc P} \in \Bun_G$. On the
other hand, recall that we expect the vertex algebra
$V^G_\ka(D^G_{0,1} \sto B)$ to contain an affine Kac--Moody vertex
subalgebra $V_{\ka'}(\g)$ of level $\ka'$ such that $\ka-\ka' = p
\cdot n(G)$, where $p$ is an integer. This allows us to couple
coinvariants of $V^G_\ka(D^G_{0,1} \sto B)$-modules to $G$-bundles on
$X$. Mathematically, this means that for any $V^G_\ka(D^G_{0,1} \sto
B)$-module, we can define its sheaf of coinvariants, which is a
$\ka'$-twisted $D$-module on $\Bun_G$. This $D$-module can then be
mapped to a $\ka$-twisted $D$-module by tensoring it with the $p$th
power of the minimal line bundle ${\mc L}_G$, an infinite order
generator of the Picard group of $\Bun_G$.

This enables us to {\em define}, under some natural assumptions, the
compactification functor $F^{G,B}_\ka$ in a mathematically rigorous
way in terms of the localization functor for the vertex algebra
$V^G_\ka(D^G_{0,1} \sto B_2)$, i.e. the functor assigning to a module
over this vertex algebra its sheaf of coinvariants on $\Bun_G$ (see
Section \ref{comp and loc}).

\bigskip

Another interesting case to consider is that of $B_1 = N^G_{0,1}$ and
$B_2 = N^G_{1,0}$. Then the vertex algebra $V^G_\ka(N^G_{0,1} \sto
N^G_{1,0})$ is the ${\mc W}$-algebra ${\mc W}_\ka(\g)$. If we take
${\mc A}_1 = \Psi^{x,\mu^\vee}_\ka$ and ${\mc A}_2 = {\mathbb
  V}_{\la,\ka}$, then the corresponding ${\mc W}_\ka(\g)$-module is
$M_{(\la,\mu^\vee),\ka}$ which appeared in formula
\eqref{Mka}. Conjecture \ref{Hom and conf1} then implies that there is
an isomorphism between $\on{Hom}(\Psi^{x,\mu^\vee}_\ka,{\mc
  D}^{x,\la}_\ka)$ and the space of coinvariants of the ${\mc
  W}_\ka(\g)$-module $M_{(\la,\mu^\vee),\ka}$. In the case $\mu^\vee =
0$ and $\la=0$, a closely related isomorphism (in the language of
branes) was proposed in \cite{NW} as a possible interpretation of the
AGT conjecture \cite{AGT}.

\begin{rem}
  This statement has a multi-point generalization: an isomorphism
  between the space $\on{Hom}(\Psi^{x_i,\mu^\vee_i}_\ka,{\mc
    D}^{x_i,\la_i}_\ka)$ and the space of coinvariants of the modules
  $M_{(\la_i,\mu^\vee_i),\ka}$ inserted at the points $x_i \in
  X$. Consider the case $\g=\sw_2$, $X=\pone$, and suppose that the
  set of points $x_i$ is a disjoint union of two subsets; for points
  of one of subset we have $\la_i=0$ and $\mu^\vee_i$ is the
  fundamental coweight, and for points of the other subset we have
  arbitrary $\la_i$ but $\mu^\vee_i=0$. Then, if we allow the
  points of the first kind to vary, we can view this isomorphism as a
  reformulation of the quantum separation of variables linking
  conformal blocks of $\su$ and the Virasoro algebra
  \cite{Stoyan,JT,FGT} (at the critical level, it becomes the
  separation of variables of the corresponding Gaudin model
  \cite{Frenkel1}). Thus, the above isomorphism (with points
  corresponding to dominant integral weights or to the fundamental
  coweights) could be viewed as a generalization of the quantum
  separation of variables to Lie algebras of higher rank. We leave the
  details to a future work.\qed
\end{rem}

Replacing $N^G_{1,0}$ by another boundary condition $B$, we obtain a
generalization of the above isomorphism, which expresses coinvariants
of modules over $V^G_\ka(N^G_{0,1} \sto B)$ as Hom's between the
corresponding $D$-modules on $\Bun_G$.

Indeed, recall from item (iii) of Section \ref{va junction} that if
the junction $N^G_{0,1} \to B$ is obtained from a junction $D^G_{0,1}
\to B$, then we expect the vertex algebra $V^G_\ka(N^G_{0,1} \sto B)$
to be the quantum Drinfeld--Sokolov reduction of $V^G_\ka(D^G_{0,1}
\sto B)$. Now, given an object ${\mc A}$ of the category ${\mc
  C}^G_\ka(B)$, we obtain a family of ``magnetic'' $V^G_\ka(N^G_{0,1}
\sto B)$-modules $H^{\g}_{\on{DS},\mu^\vee}({\mc A})$. Applying
Conjecture \ref{Hom and conf1}, we obtain that the space of
coinvariants of the $V^G_\ka(N^G_{0,1} \sto B)$-module
$H^{\g}_{\on{DS},\mu^\vee}({\mc A})$ is isomorphic to
$\on{Hom}(\Psi^{x,\mu^\vee}_\ka,F^{G,B}_\ka({\mc A}))$.

\section{Quantum Langlands dualities of twisted
  $D$-modules}    \label{qGL dual}

It is important to realize that using the quantum dualities, we can
generalize the compactification functor to a whole family of functors
from the category ${\mc C}^G_\ka(B)$ to various categories of twisted
$D$-modules on $\Bun_G$ and $\Bun_{\LG}$. This will naturally lead us
to postulate the existence of certain functors between the categories
$D_\ka(\Bun_G)$. We will conjecture that for irrational $\ka$ these give
rise to a plethora equivalences of categories, generalizing the
quantum geometric Langlands correspondence \eqref{qGL Sm}.

\subsection{A family of dual compactification functors}

Recall the duality groupoid ${\mc G}^G_\ka$ introduced in Section
\ref{symmetries}. Let us observe that a duality does not affect the
group of global symmetries of a boundary condition. In particular, for
any $g \in {\mc G}^G_\ka$, the boundary condition $D^G_{p,q} =
g^{-1}(D^{g(G)}_{0,1})$ should have the same global symmetry group as
the Dirichlet boundary condition $D^{g(G)}_{0,1}$, which is the group
$g(G)$. Therefore, arguing as above, for any junction $D^G_{p,q} \to
B$ in the bulk theory ${\mc T}^G_\ka$, where $D^G_{p,q} =
g^{-1}(D^{g(G)}_{0,1})$, we obtain a functor from the category ${\mc
  C}^G_\ka(B)$ to the category of twisted $D$-modules on
$\Bun_{g(G)}$.

To determine the twisting parameter, observe that the junction
$D^G_{p,q} \to B$ in the bulk theory ${\mc T}^G_\ka$ can be seen in
another duality frame as the junction $D^{g(G)}_{0,1} \to g(B)$ in the
bulk theory ${\mc T}^{g(G)}_{g(\ka)}$, which gives rise to a functor
(in general, of derived categories)
\begin{equation}    \label{FgG}
F^{g(G),g(B)}_{g(\ka)}: {\mc C}^{g(G)}_{g(\ka)}(g(B)) \to
D_{g(\ka)}(\Bun_{g(G)}).
\end{equation}
But the category ${\mc C}^{g(G)}_{g(\ka)}(g(B))$ should be equivalent
to ${\mc C}^G_\ka(B)$, so $F^{g(G),g(B)}_{g(\ka)}$ gives rise to a
functor
\begin{equation}    \label{gFG}
_g F^{G,B}_\ka: {\mc C}^G_\ka(B) \to D_{g(\ka)}(\Bun_{g(G)}).
\end{equation}
The functors $_g F^{G,B}_\ka$ are generalizations of the
compactification functor $F^{G,B}_\ka$ (which corresponds to $g$ being
the identity). They correspond to junctions $D^G_{p,q} \to B$ in the
same sense in which the compactification functor $F^{G,B}_\ka$
corresponds to a junction $D^G_{0,1} \to B$. Note that in both cases
there could be multiple junctions between these boundary conditions,
but the corresponding functors $_g F^{G,B}_\ka$ should be equivalent
to each other according to Conjecture \ref{meta} (see Conjecture
\ref{meta1} for a more precise formulation).

Once the functors $_g F^{G,B}_\ka$ labeled by $g$ in the groupoid
${\mc G}^G_\ka$ have been defined, it is natural to suppose that there
is also a collection of {\em duality functors} ${\mc E}^{G,g}_\ka$,
labeled by the same groupoid, acting between the corresponding
categories of twisted $D$-modules on $\Bun_G$ and $\Bun_{\LG}$ and
intertwining the functors $_g F^{G,B}_\ka$ for all boundary $B$
conditions. This leads to a vast generalization of the idea of quantum
Geometric Langlands duality \eqref{qGL Sm}, which from the 4d gauge
theory point of view is just one of many duality functors; namely, the
functor ${\mc E}^{G,S_m}_\ka$ attached to the duality $S_m$.

Putting aside the subtleties related to the spin structures, which
will be addressed in later sections, we expect that for irrational $\ka$
these functors ${\mc E}^{G,g}_\ka$ are in fact equivalences between
categories of twisted $D$-modules on $\Bun_G$ and $\Bun_{\LG}$. In
other words, we conjecture that for each $g \in {\mc G}^G_\ka$ there
is an equivalence ${\mc E}^{G,g}_\ka$ between $D_\ka(\Bun_G)$ and
$D_{g(\ka)}(\Bun_{g(G)})$ which intertwines the equivalences ${\mc
  C}^G_\ka(B) \simeq {\mc C}^{g(G)}_{g(\ka)}(g(B))$ via the
corresponding compactification functors, for all boundary conditions
$B$.

Before stating the conjecture, it is useful to recall the subgroup
${\mc S}_G$ of $PGL_2(\Z)$ generated by $T^{n(G)}$ and $S_m T^{n(\LG)}
S_m$. Each element $g \in {\mc S}_G$ gives rise to a particular
element (arrow) of the groupoid ${\mc G}^G_\ka$ connecting $(G,\ka)$
to $(G,g(\ka))$. In this case, we write $g(G)=G$. Similarly, each
element $g \in S_m \cdot {\mc S}_G$ gives rise to the element of ${\mc
  G}^G_\ka$ connecting $(G,\ka)$ to $(\LG,g(\ka))$. In case case, we
write $g(G)=\LG$. Likewise, for $g \in {\mc S}_{\LG}$ and $g \in S_m
\cdot {\mc S}_{\LG}$. When no confusion can arise, we use the same
notation $g$ for the corresponding elements of the groupoid ${\mc
  G}^G_\ka$.

\begin{conj}    \label{qGL functors}
Let $\ka$ be irrational. Then for each $g \in {\mc G}^G_\ka$, connecting
the nodes $(G,\ka)$ and $(g(G),g(\ka))$, there is a functor
$$
{\mc E}^{G,g}_\ka: D_\ka(\Bun_G) \to
  D_{g(\ka)}(\Bun_{g(G)})
$$
that fits in a commutative diagram
\begin{equation}    \label{diagram}
\begin{CD}
{\mc C}^G_{\ka}(B) @>>> {\mc C}^{g(G)}_{g(\ka)}(g(B)) \\
@V{F^{G,B}_\ka}VV @VV{F^{g(G),g(B)}_{g(\ka)}}V \\
D_\ka(\Bun_G) @>{{\mc E}^{G,g}_\ka}>> D_{g(\ka)}(\Bun_{g(G)})
\end{CD}
\end{equation}
and these functors combine into a categorical action of the groupoid
${\mc G}^G_\ka$ on the categories $D_\ka(\Bun_G)$.

Furthermore, if $g=T^n, n = p \cdot n(G), p \cdot \Z$, then the
functor
$$
{\mc E}^{G,T^n}_\ka: D_{\ka}(\Bun_G) \to D_{\ka+n}(\Bun_G)
$$
is given by the formula
$$
{\mc F} \mapsto {\mc
  F} \otimes {\mc L}^{\otimes m}_G,
$$
where ${\mc L}_G$ is a minimal line bundle on $\Bun_G$. In particular,
the functor ${\mc E}^{G,1}_\ka$ is the identity functor.
\end{conj}

A categorical action of the groupoid ${\mc G}^G_\ka$ means that for
any pair $g,h \in {\mc G}^G_\ka$, we have an isomorphism of functors
$$
i_{g,h}: {\mc E}^{h(G),g}_{h(\ka)} \circ {\mc E}^{G,h}_\ka
\simeq {\mc E}^{G,gh}_\ka
$$
and these isomorphisms satisfy a cocycle condition for every triple of
elements of ${\mc G}^G_\ka$:
$$
i_{g,hk} \circ i_{h,k} = i_{g,h} \circ i_{gh,k}.
$$

This implies that the functors ${\mc E}^{G,g}_\ka$ satisfy (in the
categorical sense) the relations satisfied by the corresponding
elements in the groupouid ${\mc G}^G_\ka$ (all of these relations
boil down to some relations in $PGL_2(\Z)$). This leads to various
non-trivial statements.

For example, Conjecture \ref{qGL functors} states that $T^{n(G)} \in
{\mc S}_G$ gives rise to an equivalence between the categories of
twisted $\ka$- and $(\ka+n(G))$-twisted $D$-modules on $\Bun_G$
obtained by tensoring a twisted $D$-module with the minimal line
bundle ${\mc L}_G$ on $\Bun_G$, and likewise for $T^{n(\LG)}$ and
$\Bun_{\LG}$. On the other hand, the duality transformation $S_m \in
PGL_2(\Z)$ should give rise to the equivalence \eqref{qGL Sm}:
$$
D_{\ka}(\Bun_G) \simeq D_{-1/m\ka}(\Bun_{\LG}).
$$
Therefore these equivalences should satisfy whatever relations are
satisfied by $T^{n(G)}, T^{n(\LG)}$ and $S_m$ in the group $PGL_2(\Z)$.

This is one of the reasons we find it more fruitful to look at the
{\em entire} groupoid ${\mc G}^G_\ka$ (rather than the specific
equivalence $S_m$) as the collection of {\em quantum Geometric
  Langlands dualities} (qGL dualities for short). Another reason is
that, as we show below, using the method of constructing vertex
algebras via compositions of junctions, we can conjecturally construct
the {\em kernels} of many of these qGL dualities directly. However,
the vertex algebra corresponding to the duality $S_m$ may not be the
optimal one (it may have unbounded conformal dimensions and other
unfavorable features). Therefore, we may be better off constructing
other qGL dualities first. Using these qGL dualities and the
equivalences $T^{n(G)}$ and $T^{n(\LG)}$, we can then construct the
qGL duality $S_m$ as well.

\subsection{Kernels}

Let us discuss these kernels in more detail. Let $g$ be an arrow
between the nodes $(G,\ka)$ and $(g(G),g(\ka))$ in the groupoid ${\mc
  G}^G_\ka$. Then, according to Conjecture \ref{qGL functors}, there
should be a functor
$$
{\mc E}^{g(G),g}_{\ka,g(\ka)}: D_\ka(\Bun_G) \to
D_{g(\ka)}(\Bun_{g(G)}).
$$
We hope to realize this functor as the correspondence induced by a
kernel
$$
{\mc F}^{G,g}_\ka \in D_{-\ka,g(\ka)}(\Bun_G \times \Bun_{g(G)})
$$
which is obtained by applying a localization functor to the vertex
algebra
$$
V^{g(G)}_{g(\ka)}(D^G_{0,1}\sto g(D^G_{0,1})).
$$
corresponding to a junction $D^G_{0,1}\to g(D^G_{0,1})$.

Indeed, according to our general conjectures, this vertex algebra
should have commuting actions of the affine Kac--Moody algebras $\ghat$
of level $-\ka$ and $\wh{\g'}$ (where $\g'$ is the Lie algebra of
$g(G)$) of level $g(\ka)$. Therefore we can apply the localization
functor with respect to both actions, which yields a twisted
$D$-module in the category $D_{-\ka,g(\ka)}(\Bun_G \times
\Bun_{g(G)})$.

According to our general conjectures, it should be independent of the
junction data of $D^G_{0,1}\to g(D^G_{0,1})$ and should coincide with
the image of the identity object
$$
I \otimes I \in {\mc
  C}^G_\ka(D^G_{0,1})^\vee \boxtimes {\mc
  C}^{g(D^G_{0,1})}_\ka(g(D^G_{0,1}))
$$
under the compactification functor of this category, viewed as a
category arising in the 4d gauge theory with the semisimple group $G
\times g(G)$ and the coupling $(-\ka,g(\ka))$. In the next subsection
we will explain a general strategy for constructing these kernels. We
will give a number of examples in the later sections; in particular,
two families of kernel vertex algebras in Sections \ref{fam ker} and
\ref{other}.

\medskip

The following heuristic point of view on the kernels ${\mc
  F}^{G,g(G)}_\ka$ might be useful: let's think of the functors
$F^{g(G),g(B)}_{g(\ka)}$ as kind of bookkeeping devices, giving us
``coordinate representations'' of objects of the category ${\mc
  C}^{g(G)}_{g(\ka)(g(B)} \simeq {\mc C}^G_\ka(B)$. Indeed, given an
object $A$ of ${\mc C}^G_\ka(B)$, we obtain an object $g(A)$ of the
category ${\mc C}^{g(G)}_{g(\ka)}(g(B))$, and the functor
$F^{g(G),g(B)}_\ka$ then assigns to $g(A)$ a $g(\ka)$-twisted
$D$-module on $\Bun_{g(G)}$. Let's think of the collection of its
fibers (to simplify notation, we denote $F^{g(G),g(B)}_\ka(g(A))$
simply by $g(A)$ here)
$$
A \mapsto \on{Hom}(\delta_{\mc P},g(A)), \qquad {\mc P} \in
\Bun_{g(G)}
$$
as a categorical version of the set of ``$g$-coordinates'' of $A$,
relative to a specific ``basis'' $\{ \delta_{\mc P} \}_{{\mc P} \in
  \Bun_{g(G)}}$ in the category $D_{g(\ka)}(\Bun_{g(G)})$.

Now let's compare these ``$g$-coordinates'' to the ``coordinates'' of
$A$ relative to the standard basis $\{ \delta_{{\mc P}'} \}_{{\mc P}'
  \in \Bun_G}$ in the category $D_\ka(\Bun_G)$:
$$
A \mapsto \on{Hom}(\delta_{{\mc P}'},A), \qquad {\mc P}' \in \Bun_{G}.
$$
As in linear algebra, converting one ``coordinate
representation'' into another requires a categorical version of
``change of basis matrix'' which we can think of as a $D$-module on
$\Bun_G \times \Bun_{g(G)}$ with fibers
$$
\on{Hom}(\delta_{\mc P},g(\delta_{{\mc P}'})), \qquad
({\mc P}',{\mc P}),
$$
with the Hom's taken in the category $D_{g(\ka)}(\Bun_{g(G)}$.

Heuristically, for each ${\mc P} \in \Bun_{g(G)}$,
\begin{align*}
\on{Hom}(\delta_{\mc P},g(A)) &= \underset{{\mc P}' \in \Bun_G}\int
\on{Hom}(\delta_{\mc P},g(\delta_{{\mc P}'})) \otimes
\on{Hom}(g(\delta_{{\mc P}'}),g(A)) \\ & = \underset{{\mc P}' \in
  \Bun_G}\int
\on{Hom}(\delta_{\mc P},g(\delta_{{\mc P}'})) \otimes
\on{Hom}(\delta_{{\mc P}'},A),
\end{align*}
where the first factor represents the kernel ${\mc F}^{G,g}_\ka$.

The upshot of this discussion is that the kernel of the qGL duality
$g$ should be a $D$-module on $\Bun_G \times \Bun_{g(G)}$ whose fibers
at $({\mc P}',{\mc P})$ are $\on{Hom}(\delta_{\mc P},g(\delta_{{\mc
    P}'}))$. It should also be clear from the above discussion (see
also Section \ref{comp and loc} below) that this $D$-module can be
obtained by applying the localization functor to the vertex algebra
$V^{g(G)}_{g(\ka)}(D^G_{0,1}\sto g(D^G_{0,1}))$.

We expect that all of this works out nicely for irrational $\ka$. For
rational $\ka$, some complications arise. We no longer expect that our
functors yield equivalences between the corresponding categories of
twisted $D$-modules. Rather, we expect such equivalences between
certain ``tempered'' subcategories \cite{AG}. The reason is that from
the 4d gauge theory point of view \cite{GW1,GW2}, the non-tempered
part is accommodated by additional fields, denoted by $\sigma,
\ol\sigma$ in \cite{KW}, which become relevant at rational values of
$\ka$ as additional degrees of freedom \cite{EY2}. They play a role
similar to the role Arthur's $SL_2$ plays in the classical Langlands
correspondence (\cite{F:bourbaki}, Sect. 6.2). See Remark
\ref{rational} below for more details.

Another issue arising at rational $\ka$ is that there are two derived
categories of $\ka$-twisted $D$-modules, dual to each other
\cite{DG}. In order to extend our setup to rational $\ka$ one should
probably consider one of them for positive rational values of $\ka$
and the other for negative ones.\footnote{We thank P. Yoo for useful
  comments on this issue.}




\subsection{Extension of the compactification functor}

The above construction is a special case of a general phenomenon:
Whenever a boundary condition $B$ has a global symmetry group $H$, the
compactification functor $F^{G,B}_\ka$ can be extended to a functor
$$
F^{G,B,H}_\ka: {\mc C}^G_\ka(B) \to D_{\ka,-\ka_H}(\Bun_G \times
\Bun_{H,(x)}).
$$
Here $\Bun_{H,(x)}$ denotes the moduli stack of $H$-bundles on $X$
equipped with a trivialization on the formal disc around the point
$x$. Such a trivialization is necessary in general in order to insert
a line defect at $x$. However, specific line defects may only depend
on a trivialization on the $n$th formal neighbourhood of $x$ for some
integer $n\geq 0$, in which case the image of the functor can be
well-defined on the corresponding quotient of $\Bun_{H,(x)}$ by a
congruence subgroup of $H[[z]]$. It should be possible to express this
statements as the requirement that the functor $F^{G,B,H}_\ka$ should
intertwine the appropriate actions of the loop group of $H$ on the
source and target categories.

In particular, if we take the identify object $I$ in the category
${\mc C}^G_\ka(B)$, then we do not insert any line defect at all, and
therefore the corresponding twisted $D$-module $F^{G,B,H}_\ka(I)$ is
actually well-defined in $D_{\ka,-\ka_H}(\Bun_G \times \Bun_H)$. Thus,
$F^{G,B,H}_\ka(I)$ can be viewed as an extension of the
compactification map, taking values in $D_{\ka,-\ka_H}(\Bun_G \times
\Bun_H)$. We can obtain the functor $F^{G,B,H}_\ka(I)$ by applying the
localization functor to the vertex algebra $V^G_\ka(D^G_{0,1}\sto B)$
(for appropriate choices of junctions). Indeed, on general grounds we
expect this vertex algebra to contain a Kac--Moody subalgebra
$V_{\ka'}(\mathfrak{g})\times V_{-\ka'_H}(\mathfrak{h})$, with $\ka'$
and $\ka'_H$ could in general differ from $\ka$ by integral multiples
of $n(G)$ and $n(H)$. Applying the localization functor to
$V^G_\ka(D^G_{0,1}\sto B)$ with respect to this Kac--Moody subalgebra
and tensoring the resulting $D$-modules by a line bundle if necessary,
we obtain the desired $D$-module in $D_{\ka,\ka_H}(\Bun_G \times \Bun_H)$.

As an example, suppose that $B$ is also the Dirichlet boundary
condition. Then $H=G$. For irrational $\ka$, the compactification map
should produce the diagonal $D$-module in $D_{\ka,-\ka}(\Bun_G \times
\Bun_G)$, and the compactification functor should be the natural
functor
$$
D_{-\ka}(\Gr_G) \to D_{-\ka,\ka}(\Bun_G \times \Bun_{G,(x)})
$$
assigning to a compactly supported $D$-module on $\Gr_G$ a kernel on
$\Bun_G \times \Bun_{G,(x)}$. This is a categorical version of a
construction of the kernel of a convolution operator that is familiar
from the theory of automorphic representations. The kernel of the
convolution with a $G[[z]]$-invariant compactly supported object (a
function or a $D$-module) on $\Gr_G$ is well-defined on $\Bun_G \times
\Bun_G$, but the convolution with a general compactly supported object
on $\Gr_G$ gives rise to a kernel on $\Bun_G \times \Bun_{G,(x)}$.

Now let $B$ be any duality image $g(D^G_{0,1})$ of the Dirichlet
boundary condition discussed above. Then we get an extended
compactification functor
$$
D_{-g(\ka)}(\Gr_{g(G)}) \to D_{-\ka,g(\ka)}(\Bun_G
\times \Bun_{g(G),(x)}).
$$
Applying it to the identity object of the category
$D_{-g(\ka)}(\Gr_{g(G)})$, we obtain a $D$-module on
$D_{-\ka,g(\ka)}(\Bun_G \times \Bun_{g(G)})$ which is nothing but the
kernel we discussed above. It can also be viewed as the qGL image of
the diagonal $D$-module in $\Bun_G \times \Bun_G$ under the qGL
duality $g$ applied along the second factor. Just as the categorical
integral transform induced by the diagonal gives rise to the identity
functor, integral transforms induced by these kernels give rise to
functors taking twisted $D$-modules on $\Bun_G$ to their images on
$\Bun_{g(G)}$ under the qGL duality $g$.

Thus, we see that junctions between $D^G_{0,1}$ and $g(D^G_{0,1})$ are
crucial for constructing qGL dualities. Namely, we expect the sheaves
of coinvariants of the corresponding vertex algebras give rise to the
qGL kernels. This begs the question: How to build junctions between
$D^G_{0,1}$ and $g(D^G_{0,1})$?

\subsection{Building duality kernels}    \label{building}

A particularly powerful method for constructing junctions between
$D^G_{0,1}$ and its dual $g(D^G_{0,1})$ is using compositions of
junctions \cite{CG}. More generally, this method can allow us to produce many
interesting junctions between boundary conditions which do not
simultaneously admit weakly-coupled descriptions, such as $D^G_{0,1}$
and $D^G_{p,q}$.

In general, we will always compose junctions along boundary conditions
of type $N^G_{p',q'}$, whose category of boundary lines is semisimple
for irrational $\ka$. This considerably simplifies the construction of
composite vertex algebras. We leave for future work calculations
involving composition along other types of boundary conditions, as
well as the limits of the objects we construct to rational values of
$\ka$.

Thus, our basic strategy is to ``resolve'' the junction between
$D^G_{0,1}$ and $D^G_{p,q}$ by a composition of junctions of the form
\begin{equation}
  D^G_{p_0=0,q_0=1} \to N^G_{p_1,q_1} \to \cdots \to
  N^G_{p_{n-1},q_{n-1}} \to D^G_{p_n=p,q_n=q},
\end{equation}
so that each the vertex algebra corresponding to each intermediate
junction is a Kac--Moody or ${\mc W}$-algebra.

The resulting vertex algebra $V({\mc T},D^G_{0,1}\sto D^G_{p,q})$ is
then an extension of an algebra which contains two Kac--Moody
subalgebras at appropriate levels, together with a sequence of ${\mc
  W}$-algebras in-between. The extension involves diagonal objects in
the products of two categories of the form $\KL_{\ka_i}(G)$ or
$\KL_{\ka_i}(\LG)$ for appropriate intermediate levels.

As an important example, in which we don't have to deal with the
subtleties related to the center and the spin structures, consider the
case of the simple Lie group $G$ of type $E_8$, which is Langlands
self-dual; it is simply-laced, simply-connected, and has trivial
center. Then we can interpolate between $D^{G}_{0,1}$ and
$D^{G}_{1,0}$ by $N^{G}_{1,-1}$.
\begin{itemize}
\item We have $T^{-1}(D^{G}_{0,1}) = D^{G}_{0,1}$ and
  $T^{-1}(N^G_{1,0}) = N^G_{1,-1}$. Therefore, if we apply $T^{-1}$ to
  the standard junction $D^{G}_{0,1} \to N^{G}_{1,0}$, we obtain the
  junction $D^{G}_{0,1} \to N^{G}_{1,-1}$, and according to
  formula \eqref{Tp}, the latter supports $V_{\ka+1}(\g)$.
\item We have $RS(D^{G}_{1,0})=D^G_{0,1}$ and $RS(N^{G}_{1,-1}) =
  N^G_{1,-1}$. Therefore, if we apply $RS$ to the junction
  $D^{G}_{0,1} \to N^{G}_{1,-1}$ (or equivalently, $RST^{-1}$ to the
  standard junction $D^{G}_{0,1} \to N^{G}_{1,0}$), we obtain the
  junction $N^{G}_{1,-1} \to D^G_{1,0}$ (recall that the junction
  arrow is reversed under $R$, see formula \eqref{R junction}). The
  latter junction supports $V_{\ka^{-1}+1}(\g)$, because if
  $g=RST^{-1}$, then $g^{-1}(\ka) = TSR(\ka) = \ka^{-1} +1$ (see
  formula \eqref{g junction}).
\item The vertex algebra corresponding to the composite junction
  $D^{G}_{0,1} \to D^G_{1,0}$ is thus the following extension of
  $V_{\ka+1}(\g) \otimes V_{\ka^{-1} +1}(\g)$:
$$
V^G_\ka(D^{G}_{0,1} \sto D^G_{1,0}) = \bigoplus_{\la \in P^+(G)}
{\mathbb V}_{\la,\ka+1} \otimes {\mathbb V}_{\la,\ka^{-1}+1}.
$$
\end{itemize}

Conjecturally, if we apply the localization functor to this composite
junction vertex algebra, we obtain a twisted $D$-module on $\Bun_G
\times \Bun_G$ with the twists $\ka+1$ along the first factor and
$\ka^{-1}+1$ along the second factor, which should be the kernel of
the functor of qGL duality ${\mc E}^{G,TST}_{-\ka-1}$ corresponding to
$TST$ from $D_{-\ka-1}(\Bun_G)$ to $D_{\ka^{-1}+1}(\Bun_G)$. We will
discuss a generalization of this junction vertex algebra, and the
corresponding kernel, in Section \ref{other}.

As a more basic example, consider the composition of basic junctions
from $D^G_{0,1}$ to $N^G_{1,0}$ and from $N^G_{1,0}$ to
$D^G_{0,1}$. This composition gives the vertex algebra of (twisted)
chiral differential operators on $G$, which has $V_{-\ka}(\g) \otimes
V_{\ka}(\g)$ as a vertex subalgebra \cite{AG1}. For irrational $\ka$, it
is isomorphic to a direct sum of tensor products of the Weyl modules
over $\ghat$ of levels $-\ka$ and $\ka$:
$$
V^G_\ka(D^{G}_{0,1} \sto D^G_{0,1}) =
\bigoplus_{\la \in P^+(G)} {\mathbb V}_{\la,-\ka} \otimes {\mathbb
  V}_{\la^*,\ka},
$$
where $\la^*=-w_0(\la)$ is the highest weight of the irreducible
$\g$-module dual to the irreducible finite-dimensional $\g$-module
with highest weight $\la$. For irrational $\ka$, we expect that the
localization of this vertex algebra is the push-forward of the
diagonal in $D_{-\ka,\ka}(\Bun_G \times \Bun_G)$. Thus, for irrational
$\ka$, $V^G_\ka(D^{G}_{0,1} \sto D^G_{0,1})$ is the kernel of the
identity functor $D_\ka(\Bun_G) \to D_\ka(\Bun_G)$, the qGL duality
$g=1$.

\section{A family of kernel vertex algebras}    \label{fam ker}

In this section we construct a family of junction vertex algebras
$X_{p,q}(G)$ which we expect to give rise to kernels of specific
quantum Langlands duality functors. The vertex algebra $X_{p,q}(G)$ is
associated to a simple Lie group $G$ and two integers
$$
p \in n(G) \cdot \Z, \quad q \in n(\LG) \cdot \Z,
$$
and it is equipped with the action of the affine Kac--Moody algebras
$\ghat_{-\kappa}$ and $\hlg_{g(\ka)}$, where
$$
g = \wt{T}^{p} S_m \wt{T}^q
$$
and
$$
\wt{T} = S_m T^{-1} S_m = \begin{pmatrix}
1 & 0 \\
m & 1
\end{pmatrix}
$$

When we apply the localization functor to $X_{p,q}(G)$, we obtain a
twisted $D$-module
$$
\Delta_{-\ka,g(\ka)}(X_{p,q}(G))
$$
on $\Bun_G \times \Bun_{\LG}$, with twists $-\ka$ and $g(\ka)$ along
the first and the second factor, respectively
We conjecture that $\Delta_{-\ka,g(\ka)}(X_{p,q}(G))$ is a kernel of the
qGL functor corresponding to $g = \wt{T}^{p} S \wt{T}^q$:
$$
{\mc E}^{G,g}_\ka: D_\ka(\Bun_G) \to D_{g(\ka)}(\Bun_{\LG})
$$
(see formula \eqref{qGLf}; note that $g(G)=\LG$ in this case).

The standard qGL duality, $g=S_m$, corresponds to the case
$p=q=0$. The corresponding junction vertex algebra $X_{0,0}(G)$,
obtained from the chain of junctions $D_{0,1} \to N_{1,0} \to N_{0,1}
\to D_{1,0}$, appears to coincide (for irrational $\ka$) with the
``master vertex algebra'' recently proposed by Gaitsgory
\cite{Ga:Perimeter} as a candidate for a vertex algebra that could be
used to construct a kernel of the qGL duality $S_m$ (this vertex
algebra was also proposed earlier by Feigin; as far as we know, he has
not published anything about it).

However, the vertex algebras $X_{0,0}(G)$ with $p\leq 0$ or $q \leq 0$
have conformal dimensions unbounded from below. This makes it more
difficult to study these vertex algebras and to apply the localization
functor to them (we likely need to take into account higher derived
functors in this case and perhaps apply some regularization procedure,
since the dimensions of the homogeneous components are
infinite). However, we will show that in the case of positive $p$ and
$q$, all conformal dimensions of $X_{p,q}(G)$ are non-negative; and
furthermore, if $p>1$ and $q>0$ or $p>0$ and $q>1$, then only
the vacuum vector has conformal dimension $0$ and conformal dimensions
of all other fields are strictly positive. Furthermore, the dimensions
of all homogeneous components of fields corresponding to a fixed
conformal dimension are then automatically finite.

In this sense, the kernel for the duality
$$
g = \wt{T}^{p} S_m \wt{T}^q = S_m T^{-p} S_m T^{-q} S_m
$$
with, say, $p \geq 1$ and $q>1$ has an advantage over the kernel for
the standard qGL duality $S_m$. And once it is constructed, it can be
used to construct the kernel for $S_m$.

For example, if $G$ is a simply-laced, simply connected group, then we
have $m=1$ and we can take $p=1$, so using the relation $(ST)^3=1$, we
can rewrite $g$ as $S T^{-1} S T^{-q} S = T S T^{1-q} S$. If
$n(\LG)\leq 2$, we can further take $q=2$, to obtain $T S T^{1-q} S =
T^2ST$. But the action of $T$ and $T^{n(\LG)}$ corresponds (up to the
spin and $\theta$-angle subtleties) to tensoring with a line bundle on
$\Bun_G$ and $\Bun_{\LG}$, respectively. Therefore, once we construct
a kernel for the duality $g$, we can construct a kernel for $S=S_1$ as
well.

\subsection{The case of $G=SL_2$, $p=q=0$}

We first construct the kernel for the duality $S=S_1$ for $G=SL_2$
(i.e. $p=q=0$). For that we take the composition of the following
chain of junctions:
$$
D_{0,1} \to N_{1,0} \to N_{0,1} \to D_{1,0}
$$
in the bulk theory ${\mc T}^{PSL_2}_{-1/\ka}$.

The vertex algebra corresponding to the first junction is $\su$ of
level $-1/\ka$. The one corresponding to the second junction is ${\mc
  W}_{1/\ka}(\sw_2) \simeq {\mc W}_\ka(\sw_2)$, i.e. the Virasoro
algebra with the central charge
$$
c(\ka) = 13 - 6\ka - 6\ka^{-1},
$$
because $N_{1,0} \to N_{0,1} = R(N_{0,1} \to N_{1,0})$ and
$-1/\ka=R(1/\ka)$. The one corresponding to the third junction,
$$
V^{PSL_2}_{-1/\ka}(N_{0,1} \sto D_{1,0}) \simeq
V^{SL_2}_{-\ka}(D_{0,1} \sto N_{1,0}),
$$
which is $\su$ of level $-\ka$, because
$$
N^{PSL_2}_{0,1} \to D^{PSL_2}_{1,0} = SR(D^{SL_2}_{0,1} \to N^{SL_2}_{1,0})
$$
and $-1/\ka = SR(-\ka)$ (see formula \eqref{g junction}). Recalling
the procedure of Section \ref{comp junctions} for constructing
compositions of junction vertex algebras, we obtain that for irrational
$\ka$ this vertex algebra is
$$
X_{0,0}(SL_2) = \bigoplus_{r\geq,s\geq 0} {\mathbb V}_{r,-\ka}
\otimes M_{(r,2s),\ka} \otimes {\mathbb V}_{2s,-1/\ka},
$$
where ${\mathbb V}_{m,\ka}$ denotes the Weyl module over $\su$ with
highest weight $m$ and (shifted) level $\ka$, and $M_{(m,n),\ka}$
denotes the irreducible highest weight module over the Virasoro
algebra with the above central charge $c(\ka)$ and conformal dimension
$$
h_{(m,n),\ka} = \frac{1}{4}(m(m+2)\ka^{-1} + n(n+2)\ka) -
\frac{1}{2}(mn + m + n).
$$
Thus, the conformal dimension of the highest weight vector in
${\mathbb V}_{r,-\ka} \otimes M_{(r,2s),-\ka} \otimes {\mathbb
  V}_{2s,-1/\ka}$ is
$$
h_{(r,2s),\ka} - \frac{1}{4}r(r+2)\ka^{-1} -
\frac{1}{4}2s(2s+2)\ka = - \frac{1}{2}(2rs + r + 2s).
$$
We see this vertex algebra has fields with conformal dimensions
unbounded from below, which is a drawback.

However, if we apply the localization functor to $X_{0,0}(SL_2)$
(perhaps, in the derived sense), we obtain a $D$-module on
$\Bun_{SL_2} \times \Bun_{PSL_2}$ with the twisting $-\ka$ along the
first factor and $-1/\ka$ along the second factor. According to our
general conjecture, it should be a kernel of the qGL functor
corresponding to the duality transformation $S$:
$$
{\mc E}^{SL_2,S}_\ka: D_\ka(\Bun_{SL_2}) \to
D_{-1/\ka}(\Bun_{PSL_2}).
$$

\subsection{The case of $G=SL_2$, general $p,q$}

We now generalize this to the duality transformations $g = \wt{T}^{p}
S \wt{T}^q$ with arbitrary integers $p$ and $q$. Consider the
following chain of junctions:
\begin{equation}    \label{junction sl2}
D_{-p,1} \to N_{1,0} \to N_{0,1} \to D_{1,-q}
\end{equation}
in the bulk theory ${\mc T}^{PSL_2}_{-1/\ka}$.

The vertex algebra corresponding to the first junction in
\eqref{junction sl2} is $\su$ of level $1/(p-\ka) =
\wt{T}^{p}(-1/\ka)$, because $D_{-p,1} \to N_{1,0} = \wt{T}^{-p}(D_{0,1}
\to N_{1,0})$. The vertex algebra corresponding to the second junction
is ${\mc W}_{1/\ka}(\sw_2) \simeq {\mc W}_\ka(\sw_2)$, the Virasoro
algebra with the central charge $c(\ka)$ (for the same reason as
explained above). The one corresponding to the third junction,
$$
V^{PSL_2}_{-1/\ka}(N_{0,1} \sto D_{1,-q}) \simeq
V^{SL_2}_{\ka/(q\ka-1)}(D_{0,1} \sto N_{1,0}),
$$
which is $\su$ of level $\ka/(q\ka-1)$, because
$$
N^{PSL_2}_{0,1} \to D^{PSL_2}_{1,-q} =
S\wt{T}^qR(D^{SL_2}_{0,1} \to N^{SL_2}_{1,0})
$$
and $-1/\ka = S\wt{T}^qR(\ka/(q\ka-1))$ (see formula \eqref{g
  junction}). Note that $q$ has to be even, because $\wt{T}^q$ is a
legitimate duality in the bulk theory corresponding to $SL_2$, or
equivalently, $T^q$ is a legitimate duality in the theory
corresponding to $PSL_2$ (ignoring $\theta$-angle subtleties), only if
$q$ is even (indeed, $n(PSL_2)=2$).

In the same way as above, we obtain that the corresponding vertex
algebra is
$$
X_{p,q}(SL_2) = \bigoplus_{r,s\geq 0} {\mathbb V}_{r,\ka/(q\ka-1)}
\otimes M_{(r,2s),\ka} \otimes {\mathbb V}_{2s,1/(p-\ka)},
$$
The conformal dimension of the highest weight vector in ${\mathbb
  V}_{r,\ka/(q\ka-1)} \otimes M_{(r,2s),\ka} \otimes {\mathbb
  V}_{2s,1/(p-\ka)}$ is
\begin{multline*}
h_{(r,2s),\ka} + \frac{1}{4}r(r+2) \frac{q\ka-1}{\ka} +
\frac{1}{4}2s(2s+2)(p-\ka) \\ = \frac{1}{4}(r-2s)^2 +
\frac{1}{4}r(r+2)(q-1) + s(s+1)(p-1)
\end{multline*}
(note that this is always an integer if $q$ is even). Since $r,s \geq
0$, conformal dimensions are non-negative if $p, q \geq 1$, and are
strictly positive, with the exception of the vacuum vector, if in
addition either $p$ or $q$ are positive. But since $q$ has to be even,
this is equivalent to $p\geq 1$ and $q\geq 2$.

Applying the localization functor to $X_{p,q}(SL_2)$, we obtain a
twisted $D$-module on the product $\Bun_{SL_2} \times
\Bun_{PSL_2}$. The twisting is $\ka/(q\ka-1)$ along the first factor
and $1/(p-\ka)$ along the second factor. This should be a kernel of
the qGL functor corresponding to the duality transformation
$\wt{T}^{p} S \wt{T}^q$:
$$
{\mc E}^{SL_2,\wt{T}^{p} S \wt{T}^q}_{\ka/(1-q\ka)}:
D_{\ka/(1-q\ka)}(\Bun_{SL_2}) \to D_{1/(p-\ka)}(\Bun_{PSL_2}).
$$

Note that the junction $D_{-p,1} \to D_{1,-q}$ given by the
composition \eqref{junction sl2} can be transformed by the duality
$\wt{T}^{p}$ to $D_{0,1} \to D_{1-pq,-q} = D_{g(0,1)}$, where $g =
\wt{T}^{p} S \wt{T}^q$. Therefore constructing the junction vertex
algebra for $D_{-p,1} \to D_{1,-q}$ is equivalent to constructing a
junction vertex algebra for $D_{0,1} \to D_{g(0,1)}$. This is why we
have used the same notation $X_{p,q}(SL_2)$ for this vertex algebra as
we used at the beginning of this section. And in fact, if we make a
change of variables $\kappa \mapsto \ka/(q\ka+1)$ in the above
formulas, we obtain the kernel of
$$
{\mc E}^{SL_2,g}_{\ka}: D_\ka(\Bun_{SL_2}) \to
D_{g(\ka)}(\Bun_{PSL_2}),
$$
where $g = \wt{T}^{p} S \wt{T}^q$.

\subsection{General case}    \label{fam gen}

For a general simple Lie group $G$ with lacing number $m$, we consider
the composition of junctions
$$
D_{-pm,1} \to N_{1,0} \to N_{0,1} \to D_{1,-q}
$$
in the bulk theory ${\mc T}^{\LG}_{-1/m\ka}$, where $p \in n(G)\Z, q
\in n(\LG)\Z$.

The vertex algebra corresponding to the first junction is $\hlg$ of
level $1/m(p-\ka) = \wt{T}^{p}(-1/m\ka)$, because $D_{-mp,1} \to
N_{1,0} = \wt{T}^{-p}(D_{0,1} \to N_{1,0})$. The vertex algebra
corresponding to the second junction $N_{1,0} \to N_{0,1} = R(N_{0,1}
\to N_{1,0})$ is ${\mc W}_{1/\ka}(\lg) \simeq {\mc W}_\ka(\g)$. The
vertex algebra corresponding to the third junction,
$$
V^{\LG}_{-1/m\ka}(N_{0,1} \sto D_{1,-q}) \simeq
V^G_{\ka/(mq\ka-1)}(D_{0,1} \sto N_{1,0}),
$$
is $\ghat$ of level $\ka/(mq\ka-1)$.

In the same way as before, we obtain that for irrational $\ka$ the
corresponding vertex algebra is
\begin{equation}   \label{XpqG}
X_{p,q}(G) = \bigoplus_{\la \in P^+,\mu^\vee \in \LP^+} {\mathbb
  V}_{\la^*,\ka/(mq\ka-1)} \otimes M_{(\la,\mu^\vee),\ka} \otimes {\mathbb
  V}_{\mu^{\vee *},1/m(p-\ka)}.
\end{equation}
Here ${\mathbb V}_{\la^*,\ka}$ denotes the Weyl module over $\ghat$
generated from the irreducible finite-dimensional representation
$(V_\la)^*$ of $\g$ which is dual to $V_\la$. Since $(V_\la)^* \simeq
V_{-w_0(\la)}$, we find that $\la^*=-w_0(\la)$. The weight $\mu^{\vee
  *}$ of $\lg$ is defined similarly.

Recall from formula \eqref{Mka} that we denote by
$M_{(\lambda,\mu^\vee),\ka}(\g)$ the ${\mc W}_\ka(\g)$-module
$$
M_{(\lambda,\mu^\vee),\ka}(\g) = H_{\on{DS},\mu^\vee}({\mathbb
  V}_{\la,\ka}).
$$
The conformal dimension of its highest weight vector is (see
\cite{FF})
\begin{equation}    \label{hlamu}
h_{(\la,\mu^\vee),\ka} = \frac{1}{2\ka}(\la,\la+2\rho) +
\frac{\ka}{2}(\mu^\vee,\mu^\vee+2\rho^\vee)- \langle \la,\mu^\vee
\rangle - \langle \rho,\mu^\vee \rangle - \langle \la,\rho^\vee
\rangle.
\end{equation}

We use the invariant inner product $(\cdot,\cdot)_{\h^*}$ on $\h^* = P
\underset{\Z}\otimes \C$ normalized so that the square length of the
maximal root of $\g$ is equal to $2$. This inner product enables us to
identify $\h^*$ with $\h = ({}^L\neg\h)^* = \LP \underset{\Z}\otimes
\C$. We denote the image of $\mu^\vee \in \LP \subset \h$ in $\h^*$
under this identification by the same symbol. In particular, the $i$th
simple root of $\lg$ is identified with $2\al_i/(\al_i,\al_i) \in
\h^*$.

In formula \eqref{hlamu}, $(\mu^\vee,\mu^\vee)$ stands for the square
length $(\mu^\vee,\mu^\vee)_{\h^*}$ of $\mu^\vee \in \h^*$. Note that
it is also equal to $m$ times the square length
$(\mu^\vee,\mu^\vee)_{({}^L\neg\h)^*}$ of this element with respect to
the invariant inner product on $(\cdot,\cdot)_{({}^L\neg\h)^*}$
normalized so that the square length of the maximal root of $\lg$ is
equal to $2$. That's why formula \eqref{hlamu} stays invariant if we
exchange $\g$ and $\lg$ and replace $\ka$ by $1/m\ka$ (and not
$1/\ka$).

We also use the canonical pairing
$$
\langle \cdot,\cdot \rangle: \quad P \times \LP \to \Z,
$$
and the standard notation $\rho$ and $\rho^\vee$ for the elements of
$P$ and $\LP$, respectively, such that $\langle \rho,\al_i^\vee
\rangle = 1$ and $\langle \al_i,\rho^\vee \rangle =1$ for all $i$.

Conformal dimensions of the highest weight vectors of ${\mathbb
  V}_{\la^*,\ka/(mq\ka-1)}$ and ${\mathbb V}_{\mu^{\vee *},1/m(p-\ka)}$
coincide with those of ${\mathbb V}_{\la,\ka/(mq\ka-1)}$ and
${\mathbb V}_{\mu^\vee,1/m(p-\ka)}$, respectively, and are equal to
$$
\frac{mq\ka-1}{2\ka}(\la,\la+2\rho) = \left( -
  \frac{1}{2\ka} + \frac{mq}{2} \right) (\la,\la+2\rho)
$$
and
$$
\frac{m(p-\ka)}{2}(\mu^\vee,\mu^\vee+2\rho^\vee)_{({}^L\neg\h)^*} = \left(
  -\frac{\ka}{2} + \frac{p}{2} \right) (\mu^\vee,\mu^\vee+2\rho^\vee),
$$
respectively. Therefore conformal dimension of the highest weight
vector of the $(\la,\mu^\vee)$-term of the vertex algebra \eqref{XpqG}
is equal to
\begin{multline*}
\frac{1}{2}(\la-\mu^\vee,\la-\mu^\vee) + \frac{mq-1}{2}(\la,\la) +
(mq(\la,\rho) - \langle
\la,\rho^\vee \rangle) \\ + \frac{p-1}{2}(\mu^\vee,\mu^\vee) +
(p(\mu^\vee,\rho^\vee) - \langle \rho,\mu^\vee \rangle).
\end{multline*}
The first term is always non-negative, the second and third terms are
non-negative (resp. strictly positive) if $q\geq 1$ (resp. $q>1$), and
the fourth and fifth terms are non-negative (resp. strictly positive)
if $p\geq 1$ (resp. $p>1$).

We conclude that conformal dimensions in $X_{p,q}(G)$ are bounded from
below (in fact, are non-negative) if and only if $p, q \geq
1$. Furthermore, they are strictly positive, with the exception of the
vacuum vector (which appears in the $\la=\mu=0$ sector) if in addition
either $p$ or $q$ is greater than 1.

Applying the localization functor to $X_{p,q}(G)$, we obtain a
twisted $D$-module on the product $\Bun_{G} \times
\Bun_{\LG}$. The twisting is $\ka/(mq\ka-1)$ along the first factor
and $1/m(p-\ka)$ along the second factor. This should be a kernel of
the qGL functor corresponding to the duality transformation
$\wt{T}^{p} S \wt{T}^q$:
$$
{\mc E}^{G,\wt{T}^{p} S \wt{T}^q}_{\ka/(1-mq\ka)}:
D_{\ka/(1-mq\ka)}(\Bun_G) \to D_{1/m(p-\ka)}(\Bun_{\LG}).
$$
Making a change of variables, we can rewrite it as the kernel of
$$
{\mc E}^{G,g}_{\ka}: D_\ka(\Bun_G) \to
D_{g(\ka)}(\Bun_{\LG}),
$$
where $g = \wt{T}^{p} S \wt{T}^q$ and $g(\ka) =
(mq\ka+1)/m(mpq\ka+p-\ka)$.

\section{Branes, twisted $D$-modules, and compactification
  functors}    \label{stalks}

In this section we discuss in more detail the compactification
functors and their behavior under the quantum Langlands dualities. We
also review the connections between the $D$-modules obtained via
compactification functors and the corresponding branes on the Hitchin
moduli spaces. Much of the material of this section on the links
between branes, $D$-modules, and conformal blocks appeared previously
in \cite{KW,NW,FW,GW3,Ga1,Ga2,BT}. A new ingredient is identifying,
under some mild conditions, the compactification functor with the
localization functor for a junction vertex algebra (up to tensoring
with a line bundle on $\Bun_G$).

\subsection{Branes associated to boundary conditions}    \label{Dmod
  branes}

In their pioneering work \cite{KW}, Kapustin and Witten introduced and
studied the categories of $A$- and $B$-branes on ${\mc M}_H(G,X)$ and
${\mc M}_H(\LG,X)$, and the equivalence (homological mirror symmetry)
between them that is a consequence of the $S$-duality of the 4d gauge
theories ${\mc T}^G_0$ and ${\mc T}^{\LG}_\infty$ (in the notation
introduced in Section \ref{bc}). In particular, they studied the
following branes, which can be viewed as $A$- or $B$-branes in each of
the three complex structures of the Hitchin moduli spaces, relative to
the hyperK\"aler structure described in \cite{KW}:

\medskip

\begin{enumerate}

\item The zero-branes ${\mb B}_{\mc E}$ -- supported at a single
  (smooth) point ${\mc E}$ of ${\mc M}_H(\LG,X)$, each of them has
  type $(B,B,B)$;

\medskip

\item Pairs $({\mb F}_b,\nabla)$, where ${\mb F}_b$ is a smooth fiber 
table of the Hitchin fibration of ${\mc M}_H(G,X)$ and a $\nabla$ is a
  flat $U(1)$-bundle on it -- these are mirror dual to the zero-branes
  ${\mb B}_{\mc E}$ from (i) and have type $(B,A,A)$.

\medskip

\item The canonical coisotropic brane ${\mc B}_{\on{c.c.}}$ -- it
is supported on the entire ${\mc M}_H(G,X)$ and has type $(A,B,A)$;

\medskip

\item The space-filling $(B,B,B)$ brane $\wt{\mc B}$ on ${\mc
    M}_H(\LG,X)$;

\medskip

\item The brane of opers ${\mc B}_{\on{Op}}$ -- it is supported on the
  subspace of $\LG$-opers in ${\mc M}_H(\LG,X)$, has type $(A,B,A)$,
  and is mirror dual to ${\mc B}_{\on{c.c.}}$ from (iii);

\medskip

\item The $(B,A,A)$ brane ${\mc B}_{\on{cl.Op}}$ of ``classical
  opers'' (also known as the Hitchin's section) on ${\mc M}_H(G,X)$ --
  it is mirror dual to the space-filling $(B,B,B)$ brane $\wt{\mc B}$
  from (iv).

\end{enumerate}

\medskip

To make contact with the formalism of the previous section, we note
that to each half-BPS boundary condition $B$ in the physical 4d
theory we can associate a topological boundary condition in the ${\mc
  T}^G_\ka$ theory with $\ka=0$ or $\infty$.  One can also associate
to it two families of branes on the Hitchin moduli space ${\mc
  M}_H(G)$: one corresponds to $\ka=\infty$ and consists of $(B,B,B)$
branes; the other corresponds to $\ka=0$ and consists of $(B,A,A)$
branes. They arise from two different types of compactifications of
the 4d theory on the Riemann surface $X$, preserving two different
$SO(3)$ subgroups of the group $SO(6)$ of $R$-symmetries of the 4d
theory (this is explained in detail in \cite{Ga1,Ga2}).

For a generic $B$, this is the end of the story. But for some special
boundary conditions, one of the two types of branes deform to other
values of $\ka$.  If a brane does deform, then the deformed brane
always has type $(A,B,A)$.

In the above table, the only branes that can be deformed are (iv) and
(vi). The former, the $(B,B,B)$ brane $\wt{\mc B}$, deforms to the
$(A,B,A)$ brane ${\mc B}_{\on{c.c.}}$ from (iii). The latter, the
$(B,A,A)$ brane ${\mc B}_{\on{cl.Op}}$, deforms to the $(A,B,A)$ brane
${\mc B}_{\on{Op}}$ from (v).

Table \ref{bcbr} organizes the branes according to the boundary
conditions to which they are associated (we only consider the basic
ones: Dirichlet, Neumann, and principal Nahm).

\begin{table}[h]
\begin{center}
\begin{tabular}{|c||c||c||c||c||c|}
\hline
{\em Boundary Condition} & $\kappa$ & {\em Brane} & {\em Type} & {\em
  Deforms?} & {\em If yes, $D$-mod} \\
\hline \hline
Dirichlet ($D_{0,1}$) & $\infty$  & ${\mb B}_{\mc E}$ & $(B,B,B)$
& no & \\
\hline
Dirichlet ($D_{0,1}$) & $0$ & ${\mc F}_{\mc P}$ & $(B,A,A)$
& yes, ${\mc F}'_{\mc P}$ & $\delta^\ka_{\mc P}$ \\
\hline
Neumann ($N_{1,0}$) & $\infty$ & $\wt{\mc B}^{x_i,\la_i}$ & $(B,B,B)$ &
yes, ${\mc B}_{\on{c.c.}}^{x_i,\la_i}$ & $D^{x_i,\la_i}_\ka$ \\
\hline
Neumann ($N_{1,0}$) & $0$ & ${\mc B}_0^{x_i,\la_i}$ & $(B,A,A)$ &
no & \\
\hline
Nahm ($N_{0,1}$) & $\infty$ & ${\mc B}_{\on{cl.Op}}^{x_i,\mu^\vee_i}$
& $(B,B,B)$ & yes, ${\mc B}_{\on{Op}}^{x_i,\mu^\vee_i}$ &
$\Psi^{x_i,\mu^\vee_i}_\ka$ \\
\hline
Nahm ($N_{0,1}$) & $0$ & Nahm pole of $\sigma$ & $(B,A,A)$ & no &
\\
\hline
\end{tabular} 
\end{center}
\vspace*{5mm}
\caption{The basic boundary conditions, branes and
  $D$-modules}\label{bcbr}
\end{table}

Note that three of the branes in the middle column, namely, (i), (iv)
and (vi), appeared in the above list; as did two of the branes in the
rightmost column: (iii) and (v). The branes $({\mb F}_b,\nabla)$ from
(ii) correspond to the {\em dual} of the $(B,B,B)$ Dirichlet boundary
conditions from (i) and hence do not appear in our table.

Although the $(B,B,B)$ Dirichlet and their duality images are not the
$\kappa \to q/p$ limits of the $D_{p,q}$ boundary conditions which
exist for irrational $\kappa$, the vertex algebra technology can still
be used to explore their properties. We will comment on this
observation only briefly here, in Section \ref{hecke}, and leave a
more detailed discussion to future work.

Now we give more details on the branes appearing in Table \ref{bcbr}.

\subsubsection{Dirichlet boundary condition} In this case, the
$(B,B,B)$ branes are the zero-branes ${\mb B}_{\mc E}$ which appeared
in \cite{KW}. They cannot be deformed away from $\ka=\infty$ (this can
be seen from the fact that the deformation of ${\mc M}_H(\LG)$ away
from $\ka=\infty$ is non-commutative, so points of ${\mc M}_H(\LG)$ no
longer make sense). Likewise, the branes $({\mb F}_b,\nabla)$,
supported on the fibers of the Hitchin fibration, which are mirror
dual to the zero-branes ${\mb B}_{\mc E}$, as shown in \cite{KW}. They
cannot be deformed away from $\ka=0$.

The $(B,A,A)$ branes associated to the Dirichlet boundary condition
are the branes ${\mc F}_{\mc P}$ on ${\mc M}_H(G)$, where ${\mc P}$ is
a semi-stable $G$-bundle on $X$. Recall that in the complex structure
$I$ the Hitchin moduli space is identified with the moduli space of
semi-stable Higgs bundles on $X$. Hence it contains as an open
subspace the holomorphic cotangent bundle to the moduli space ${\mc
  M}(G,X)$ of semi-stable $G$-bundles. The brane ${\mc F}_{\mc P}$ is
supported on the fiber of this cotangent bundle at ${\mc P}$ (this map
was called ``the second Hitchin fibration'' in \cite{KW} to
distinguish it from the Hitchin map to the Hitchin base). Note that
since these fibers are vector spaces, they do not support any
non-trivial flat connections. Each brane ${\mc F}_{\mc P}$ has type
$(B,A,A)$.

The branes ${\mc F}_{\mc P}$ can be deformed away from $\ka=0$ to the
$(A,B,A)$ branes which we denote by ${\mc F}'_{\mc P}$. These are
constructed in a similar fashion, but using the complex structure $J$,
in which ${\mc M}_H(G,X)$ appears as the moduli space of semi-stable
{\em flat} $G$-bundles on $X$. Then an open part of ${\mc M}_H(G,X)$
also maps to ${\mc M}(G,X)$, and the fibers of this map give rise to a
second family of branes, which we denote by ${\mc F}'_{\mc P}$, where
again ${\mc P} \in {\mc M}(G,X)$ (these are affine spaces, hence they
do not support any non-trivial flat connections). Note that the fibers
of this map are different from the fibers of the previous map, and
hence the branes ${\mc F}'_{\mc P}$ are different from ${\mc F}_{\mc
  P}$.

The $\ka$-twisted $D$-modules on $\Bun_G$ associated to the above
``fiber branes'' are the $\delta$-function $D$-modules
$\delta^\ka_{\mc P}$ discussed in the previous section. More
precisely, ${\mc F}_{\mc P}$ corresponds to $\delta^0_{\mc P}$ for
$\ka=0$ and ${\mc F}'_{\mc P}$ corresponds to $\delta^\ka_{\mc P}$ for
$\ka \neq 0$.

Unlike the branes that are mirror dual to the $(B,B,B)$ zero-branes
${\mb B}_{\mc E}$, which have been described in \cite{KW} in the case
when ${\mc E}$ is a smooth point of ${\mc M}_H(\LG,X)$ (these are the
branes $({\mb F}_b,\nabla)$ from supported on the fibers of the first
Hitchin fibration), the mirror dual branes to the $(B,A,A)$ branes
${\mc F}_{\mc P}$, and to their deformations ${\mc F}'_{\mc P}$, are
rather complicated, and no explicit description for them is presently
known (apart from a small number of cases, including that of abelian
$G$). However, as we argue in this paper, we can construct vertex
algebras which give rise to the twisted $D$-modules corresponding to
these complicated branes using the localization functor to $\Bun_G$.

\subsubsection{Neumann boundary condition} Then the $(B,B,B)$ branes
are the space-filling brane $\wt{\mc B}$ and its generalizations
$\wt{\mc B}^{x_i,\la_i}$ obtained by applying the Wilson line
operators corresponding to the dominant integral weights $\la_i$ of
$G$ at the points $x_i \in X$ to $\wt{\mc B}$. These $(B,B,B)$ branes
deform to the $(A,B,A)$ brane ${\mc B}_{\on{c.c}}$ and its
generalizations ${\mc B}_{\on{c.c}}^{x_i,\la_i}$ on ${\mc M}_H(\LG,X)$
away from $\ka=\infty$. The corresponding twisted $D$-modules on
$\Bun_{\LG}$ are ${\mc D}_\ka$ (whose limit as $\ka \to \infty$ can be
identified with the structure sheaf on $\on{Loc}_{\LG}$, the moduli
stack of flat $\LG$-bundles on $X$) and its generalizations ${\mc
  D}_\ka^{x_i,\la_i}$.

A Neumann-type brane that does not deform is the $(B,A,A)$ brane ${\mc
  B}_0$ which is the zero section of the cotangent bundle $T^* {\mc
  M}(G,X) \subset {\mc M}_H(G,X)$.

\subsubsection{Nahm boundary condition}    \label{nahm branes}

In this case the $(B,A,A)$ branes are the brane ${\mc B}_{\on{cl.Op}}$
of ``classical opers'' on ${\mc M}_H(G,X)$ and its generalizations
${\mc B}_{\on{cl.Op}}^{x_i,\mu^\vee_i}$ obtained by applying the 't
Hooft line operators corresponding to the dominant integral coweights
$\mu^\vee_i$ of $G$ at the points $x_i \in X$ to ${\mc
  B}_{\on{cl.Op}}$. These $(B,A,A)$ brane (for the group $G$) are
mirror dual to the $(B,B,B)$ branes $\wt{\mc B}^{x_i,\mu^\vee_i}$ (for
the group $\LG$), which is consistent with the fact that the category
${\mc C}^G_0(N^G_{0,1})$ is $S$-dual to ${\mc
  C}^{\LG}_\infty(N^G_{1,0})$.

The $(B,A,A)$ brane ${\mc B}_{\on{cl.Op}}$ deforms away from $\ka=0$
to the $(A,B,A)$ brane of opers ${\mc B}_{\on{Op}}$ on ${\mc
  M}_H(G,X)$, and the branes ${\mc B}_{\on{cl.Op}}^{x_i,\mu^\vee_i}$
deform to the $(A,B,A)$ branes ${\mc
  B}_{\on{Op}}^{x_i,\mu^\vee_i}$. Note that for irrational $\kappa$
there are no bulk 't Hooft operators, so ${\mc B}_{\on{Op}}^{x_i,\mu^\vee_i}$ 
cannot be obtained by applying 't Hooft operators to ${\mc B}_{\on{Op}}$.

The $\ka$-twisted $D$-modules on $\Bun_G$ associated to the branes
${\mc B}_{\on{Op}}^{x_i,\mu^\vee_i}$ are the Whittaker sheaves
$\Psi_\ka^{x_i,\mu^\vee_i}$ (see Section \ref{ex fun}, Example (iii)).

This is consistent with the fact that the category ${\mc
  C}^G_\ka(N^G_{0,1}) = \on{Whit}_{-\ka}(G)$ is $S$-dual to ${\mc
  C}^{\LG}_{-1/m\ka}(N^G_{1,0}) = \KL_{-1/m\ka}$. The corresponding
functor
$$
{\mc E}^{G,S_m}_\ka: D_\ka(\Bun_G) \to D_{-1/m\ka}(\Bun_{\LG})
$$
should send the Whittaker sheaves $\Psi_\ka^{x_i,\mu^\vee_i}$ to the
$-1/m\ka$-twisted $D$-modules ${\mc D}^{x_i,\mu^\vee_i}_{-1/m\ka}$ on
$\Bun_{\LG}$ corresponding to the $(A,B,A)$ brane ${\mc
  B}^{x_i,\mu^\vee_i}_{\on{c.c.}}$.

Finally, the Nahm-type brane that does not deform (which is mirror
dual of the above ``zero section'' Neumann-type $(B,A,A)$ brane)
should involve a Nahm pole of additional fields $\sigma$ and
$\ol\sigma$ that appear in the dimensional reduction of the 4d gauge
theory but are not usually included among the degrees of freedom of
the Hitchin moduli space, see \cite{KW,GW1,GW2} (this boundary
condition appears to be related to the Arthur's $SL_2$ in the
classical Langlands correspondence, see Sect. 6.2 of
\cite{F:bourbaki}). It cannot be deformed away from $\ka=\infty$, just
as its mirror dual ``zero section'' Neumann-type $(B,A,A)$ brane
cannot be deformed away from $\ka=0$. This actually provides an
important insight into the expected behavior of the qGL duality
functors ${\mc E}^{G,g}_\ka$, see Remark \ref{rational} below.

\subsection{Branes vs. $D$-modules}    \label{branes vs D}

In \cite{KW}, Kapustin and Witten described a link between the
geometric Langlands correspondence and mirror symmetry of categories
of branes on the Hitchin moduli spaces ${\mc M}_H(G,X)$ and ${\mc
  M}_H(\LG,X)$ in terms of the following triangle of categories:

\smallskip

\begin{equation}    \label{triangle}
\xymatrix{& \boxed{\text{$A$-branes on } {\mc M}_H(G,X)} \ar[dd] \\
\boxed{\text{$B$-branes on } {\mc M}(\LG,X)} \ar[ur] \ar[dr] \\
& \boxed{{\mc D}_0\text{-modules on } \Bun_G}
}
\end{equation}

\medskip

Here $B$-branes on ${\mc M}(\LG,X)$ are considered in the complex
structure $J$, in which ${\mc M}(\LG,X)$ is viewed as the moduli space
${\mc Y}(\LG)$ of semi-stable $\LG$-local systems on $X$. This
category is usually interpreted mathematically as the (derived)
category of coherent sheaves on ${\mc Y}(\LG)$. Hence it is closely
related to the (derived) category of coherent sheaves (or,
equivalently, $\OO$-modules) on $\Loc_{\LG}$, the moduli stack of flat
$\LG$-bundles on $X$. For example, the two categories share some
familiar objects: zero-branes ${\mb B}_{\mc E}$ on ${\mc M}(\LG,X)$
(example (i) from Section \ref{Dmod branes}) correspond to the
skyscraper sheaves on $\Loc_{\LG}$, and the space filling $B$-brane on
${\mc M}(\LG,X)$ (example (iv) from Section \ref{Dmod branes})
corresponds to the structure sheaf on $\Loc_{\LG}$. However, there are
two important differences between the two categories: first,
$\Loc_{\LG}$ is the moduli {\em stack} of flat $\LG$-bundles on $X$,
whereas ${\mc M}_H(\LG,X)$ in complex structure $J$ is the moduli {\em
  space} ${\mc Y}(\LG)$ of semi-stable ones. Second, from the physics
perspective it is more natural to consider coherent sheaves on ${\mc
  Y}(\LG)$ with respect to its complex analytic rather than algebraic
structure, whereas in the traditional formulation of the geometric
Langlands correspondence one considers algebraic $\OO$-modules on
$\Loc_{\LG}$ (see however, \cite{EY}).

The upper arrow in the diagram \eqref{triangle} represents the
(homological) mirror symmetry, while the lower arrow represents the
geometric Langlands correspondence (up to the above two subtleties).

The vertical arrow linking the two should be viewed, according to
\cite{KW}, as an equivalence of two (derived) categories that is
independent from both mirror symmetry and geometric Langlands. Rather,
it is meant to be a special case of a general link between the
(derived) categories of ${\mc D}$-modules on a variety $M$ (twisted by
a square root of the canonical line bundle on $M$) and $A$-branes on
its cotangent bundle $T^*M$ (indeed, ${\mc M}_H(G)$ is closely
connected to $T^* \Bun_G$). We refer the reader to \cite{KW} for more
details.

The picture summarized above corresponds to the case of $\ka=0$ on the
$G$-side and $\check\ka=\infty$ on the $\LG$-side of the geometric
Langlands correspondence. In \cite{KW}, a deformation of this picture
to non-zero values of $\ka$ was considered as well. In this
deformation, one gets a link between the $\ka$-deformed version of the
above categorical mirror symmetry and a quantum deformation of the
geometric Langlands duality.

This is a beautiful idea that has led to important developments in
mathematics aiming at rigorously establishing equivalences of this
kind between categories of $A$-branes and ${\mc D}$-modules (or
constructible sheaves). However, in the context of the above diagram
there still remain several obstacles to making a precise mathematical
statement about such an equivalence. One of them is the difference
between the variety ${\mc M}_H(G)$ and the stack $T^* \Bun_G$. The
other is that in order to have an equivalence between the categories
connected by the lower arrow one needs to modify one of these
categories in a non-trivial way (see \cite{AG,EY2} for details).

Thus, while the diagram \eqref{triangle} offers a useful perspective
on the links between $S$-duality (or mirror symmetry) and geometric
Langlands, the technical difficulties involved in making it
mathematically precise encourages one to look for alternative
proposals.

In this paper we make such an proposal. We show how to use a junction
of a given boundary condition $B$ with the Dirichlet boundary
condition (in the bulk theory ${\mc T}^G_\ka$) to directly construct
the {\em compactification functor} from the category of line defects
associated to $B$ to a category of $\ka$-twisted $D$-modules on
$\Bun_G$, bypassing the categories of branes. Thus, at the outset, we
consider the boundary conditions and the corresponding categories of
line defects separately from one other. {\em A priori}, we only expect
that the groupoid of quantum dualities of the 4d gauge theory acts by
equivalences on the categories of line defects associated to $B$. The
corresponding functors on the categories of twisted $D$-modules on
$\Bun_G$ arise {\em a posteriori}, as functors that should intertwine
the compactification functors. {\em A priori} there is no reason to
expect these functors to be equivalences (and in fact we only expect
that to be true for irrational $\ka$). This understanding removes seeming
inconsistencies between the statements that are made in 4d gauge
theory and in the mathematical theory of quantum geometric Langlands
duality (see Remark \ref{rational} at the end of the next subsection
for more details).

\subsection{From boundary conditions to twisted
  $D$-modules}    \label{bc to D}

The starting point of our proposal is that for each boundary condition
$B$, we have the category of line defects ${\mc C}^G_\ka(B)$
associated to $B$ in the bulk theory ${\mc T}^G_\ka$. At the outset,
we look at these categories separately (rather than considering all
boundary conditions and the corresponding line defects together). For
each $B$, we then construct a compactification functor from ${\mc
  C}^G_\ka(B)$ to $D_\ka(\Bun_G)$, using a junction $D^G_{0,1} \to B$
from the Dirichlet boundary condition to $B$.

The Dirichlet boundary condition plays a fundamental role in 4d gauge
theory. In fact, one could argue that it is only after we specify this
boundary condition that we fix the gauge group $G_c$ of our theory
(and hence its complexification $G$). For the class of twisted gauge
theories we are considering, this means that identifying our 4d gauge
theory with a specific theory ${\mc T}^G_\ka$.

The traditional approach is to first define ${\mc T}^G_\ka$ in the
weak coupling limit (corresponding to $\ka$ close to $0$), and then
argue that it can be analytically continued to other values of
$\ka$. But we find that among the resulting theories, there are many
equivalences given by the duality groupoid (see Section
\ref{symmetries}). Therefore, a more fruitful point of view might be
that each orbit of the duality groupoid gives rise to a single quantum
field theory that has many different Lagrangian descriptions (each of
them corresponding to a particular weak coupling limit, for a suitably
chosen parameter $\ka$). Thus, for example, we see that the gauge
group of the theory is not really a well-defined object in a theory;
rather, it is a {\em pair} of Langlands dual groups that is
well-defined (if we include discrete $\theta$-angles, then we get an
even larger set of groups, though all of them belong to the isogeny
classes of $G$ and $\LG$).

To summarize, given a quantum 4d gauge theory, we have a choice which
boundary condition (among many potential candidates) to call {\em the}
Dirichlet boundary condition $D^G_{0,1}$ of our theory. This choice
implies a particular Lagrangian (weakly coupled) description, and
hence an identification of the theory as ${\mc T}^G_\ka$ for
particular $G$ and $\ka$.

Now, suppose we have declared our choice of $D^G_{0,1}$ (which
includes the choice of the corresponding group $G$). The Dirichlet
boundary condition $D^G_{0,1}$ has a special property: it has a global
{\em symmetry group}, which is nothing but the gauge group $G_c$ of
the theory ${\mc T}^G_\ka$. This allows us to couple this boundary
condition to a background $G$-bundle on the boundary. Suppose that the
boundary is the product of a Riemann surface $X$ and a real line. Then
we can couple this boundary condition to a $G$-bundle ${\mc P}$ on $X$
extended to this product. Given another boundary condition $B$
together with a line defect (i.e. an object ${\mc A}$ of ${\mc
  C}^G_\ka(B)$), we can then define a family of vector spaces ${\mb
  H}^B_{\mc P}({\mc A})$ labeled by ${\mc P} \in \Bun_G$. Furthermore,
one can argue on general grounds of TFT that these vector spaces must
be the fibers of a $\ka$-twisted $D$-module on $\Bun_G$, or more
generally an object of the derived category $D_\ka(\Bun_G)$. For
simplicity, we refer to all of them simply as $\ka$-twisted
$D$-modules on $\Bun_G$. This way we assign to $B$ and ${\mc A}$ a
$\ka$-twisted $D$-module on $\Bun_G$, and this construction gives rise
to the compactification functor from ${\mc C}^G_\ka(B)$ to
$D_\ka(\Bun_G)$.

Let us explain this more precisely. What follows may be viewed a
motivation from the 4d gauge theory perspective. We will give a
rigorous mathematical definition of the compactification functor
(under a few assumptions) in Section \ref{comp and loc} below.

\bigskip

\begin{itemize}

\item[(1)] Let $B$ be a boundary condition in the theory ${\mc
    T}^G_\ka$ and $X$ a Riemann surface. Consider the 4d theory ${\mc
    T}^G_\ka$ on the product of $X$ and an infinite strip, with the
  Dirichlet boundary condition $D^G_{0,1}$ on one side and $B$ on the
  other side. Let ${\mc P}$ be a $G$-bundle on $X$, which we extend to
  the product of $X$ and the real line. Let's couple the Dirichlet
  boundary condition to this $G$-bundle. In the limit in which the
  width of the strip goes to zero, this yields a 3d TFT, whose space
  of states on $X$ we denote by ${\mb H}^B_{\mc P}$. On general
  grounds, as explained in Section B.2 of \cite{Ga1}, we expect the
  spaces ${\mb H}^B_{\mc P}$ for different ${\mc P} \in \Bun_G$ to be
  the fibers of a $\ka$-twisted $D$-module on $\Bun_G$ (the action of
  vector fields comes from a version of the ``Berry connection''
  familiar to physicists). This is the image $F^{G,B}_\ka(I)$ of the
  compactification map.

\bigskip
    
\item[(2)] Consider the same configuration as above, but with a line
  defect corresponding to an object ${\mc A}$ of the category ${\mc
    C}^G_\ka(B)$ on the boundary of the strip on which we have the
  boundary condition $B$, situated at $x \in X$ (see Figure
  \ref{fig1}). Denote the corresponding space of states, in the limit
  when the width of the strip goes to zero, by ${\mb H}^B_{\mc P}({\mc
    A})$. Then, for the same reason as in (1), we expect that ${\mb
    H}^B_{\mc P}({\mc A})$ are the fibers of a $\ka$-twisted
  $D$-module on $\Bun_G$. We denote this $D$-module by
  $F^{G,B}_\ka({\mc A})$.

\bigskip

\item[(3)] Given another object ${\mc A}'$ of ${\mc C}^G_\ka(B)$, and
  a morphism ${\mc A} \to {\mc A}'$, general properties of TFT give
  rise to a morphism $F^{G,B}_\ka({\mc A}) \to F^{G,B}_\ka({\mc A}')$.

Thus we obtain a functor
$$
F^{G,B}_\ka: {\mc C}^G_\ka(B) \to D_\ka(\Bun_G).
$$
This is the {\em compactification functor}. It has a multi-point
generalization compatible with fusion and braiding in ${\mc
  C}^G_\ka(B)$.

\bigskip

\item[(4)] If $g$ is an element of the duality groupoid ${\mc
    G}^G_\ka$, then for irrational $\ka$ we should have an equivalence of
  categories
\begin{equation}    \label{C eq}
{\mc C}^G_\kappa(B) \simeq {\mc C}^{g(G)}_{g(\kappa)}(g(B)).
\end{equation}
Therefore, it is natural to assume the existence of a {\em quantum GL
  duality functor}
\begin{equation}    \label{qGLf}
{\mc E}^{G,g}_\ka: D_\ka(\Bun_G) \to
  D_{g(\ka)}(\Bun_{g(G)})
\end{equation}
intertwining the equivalences \eqref{C eq} via the corresponding
compactification functors $F^{G,B}_\ka$ and $F^{g(G),g(B)}_{g(\ka)}$
in the sense that they fit in the commutative diagram
\eqref{diagram}. Thus, we arrive at the statement of Conjecture
\ref{qGL functors}.

\end{itemize}

\medskip

\begin{rem}    \label{rational}
Note that from the point of view of 4d gauge theory, we can only claim
that the essential images of the compactification functors
$F^{G,B}_\ka$ and $F^{g(G),g(B)}_{g(\ka)}$ for a fixed boundary
condition $B$ are equivalent. {\em A priori}, there is no claim that
the categories $D_\ka(\Bun_G)$ and $D_{g(\ka)}(\Bun_{g(G)})$ are
equivalent. On general grounds, we do expect an equivalence between
the categories of 1d boundary conditions in the compactified 2d theories
${\mc T}^G_\ka[X]$ and ${\mc T}^{g(G)}_{g(\ka)}[X]$, but it is not
clear that the category ${\mc T}^G_\ka[X]$ coincides with
$D_\ka(\Bun_G)$. In general, some of the $D$-modules may have to be
excluded because of various subtleties with the definition of the
theory ${\mc T}^G_\ka[X]$. On the other hand, some 1d boundary
conditions may not correspond to any $D$-modules on $\Bun_G$.

For irrational $\ka$, we expect that $D_\ka(\Bun_G)$ is equivalent to the
category ${\mc T}^G_\ka[X]$. Therefore, for irrational $\ka$ we expect
that each functor ${\mc E}^{G,g}_\ka$ is an equivalence of categories.

For rational values of $\ka$, however, we expect that ${\mc
  T}^G_\ka[X]$ is different from $D_\ka(\Bun_G)$. Perhaps, a part of
it can be described as a certain ``tempered'' subcategory of
$D_\ka(\Bun_G)$. This is motivated by the observation that in the 4d
gauge theory there are additional fields, denoted by $\sigma$ and
$\ol\sigma$ in \cite{KW}, that appear to take care of the parameters
for non-tempered $D$-modules. For example, the ``constant'' $D$-module
in $D_0(\Bun_G)$ corresponds to the principal Nahm pole for the fields
$\sigma$ and $\ol\sigma$ (see Section \ref{nahm branes} and
\cite{F:bourbaki}). But these fields represent additional degrees of
freedom in the dual theory, which suggests that if we only consider
boundary conditions in the dual theory in which these fields are set
to $0$, then we must also exclude the ``constant'' $D$-module on
$\Bun_G$, which is a non-tempered object of the category $D_0(\Bun_G)$
\cite{AG, EY2}.\qed
\end{rem}

\medskip

In general, it is usually difficult to explicitly compute the spaces
${\mb H}^B_{\mc P}({\mc A})$, appearing in part (3) above, purely in
terms of the 4d gauge theory (or the 3d TFT obtained by its
dimensional reduction). And that's where vertex algebras come to the
rescue. In fact, it turns out that the compactification functor for a
boundary condition $B$ can be described rigorously mathematically
(under some assumptions) as the localization functor for modules over
the vertex algebra associated to any non-degenerate junction
$D^G_{0,1} \to B$ (possibly, up to tensoring with a power of the line
bundle ${\mc L}_G$ on $\Bun_G$). In order to explain this, we first
recall the concepts of coinvariants, conformal blocks, and the
localization functor.

\subsection{Generalities on conformal blocks and localization
  functors}    \label{gener}

Here we recall the notions of localization functors, conformal blocks,
and coinvariants (see \cite{FB}, Ch. 18 for details; for a brief
survey aimed at physicists, see Sect. 7 of \cite{F:review}).

Let $V$ be a vertex algebra and ${\mc A}_{g,n}$ the category whose
objects are $(X,(x_i),(M_i))$, where $(X,(x_i)_{i=1,\ldots,n})$ is an
$n$-punctured compact Riemann surface (more generally, a stable
$n$-pointed curve) and $(M_i)_{i=1,\ldots,n}$ is an ordered set of
$V$-modules (we think of each $M_i$ as attached to the point $x_i$),
and morphisms are holomorphic maps. Recall \cite{Segal} that a modular
functor ${\mc H}$ is a functor from the category ${\mc A}_{g,n}$ to
the category of vector spaces:
\begin{equation}    \label{functor}
(X,(x_i),(M_i)) \quad \mapsto \quad H_V(X,(x_i),(M_i)),
\end{equation}
satisfying the well-known ``sewing axiom'' (see \cite{Segal} for
details). The vector space on the right hand side of \eqref{functor}
is the {\em space of coinvariants}, which is dual to the space of
conformal blocks (see \cite{FB,F:review} for the precise definition of
these spaces for any vertex algebra and a given collection
$(X,(x_i),(M_i))$ as above).

Next, suppose that $V$ is a vertex algebra containing as a subalgebra
the vertex algebra $V_\ka(\ghat)$ of an affine Kac--Moody algebra
$\ghat$. In this case, we will say that $V$ has $\ghat$-symmetry of
level $\ka$. Then the space of coinvariants $H_V(X,(x_i),(M_i))$ (as
well as its dual) may be twisted by an arbitrary $G$-bundle ${\mc P}$
on $X$ (see \cite{FB,F:review}).

For instance, let $V = V_\ka(\g)$, the affine Kac--Moody algebra
$\ghat$ of (shifted) level $\ka$. Then the ordinary (untwisted) space
of coinvariants is the quotient
$$
H(X,(x_i),(M_i)) = \otimes_i M_i / \left( \g_{\on{out}} \cdot
  \otimes_i M_i \right),
$$
where
$$
\g_{\on{out}} = \g \otimes \C[X \bs \{ x_1,\ldots,x_n \}]
$$
(it is viewed as a Lie subalgebra in the direct sum of $n$ copies of
$\ghat$ with their centers identified). Its dual space is the ordinary
space of conformal blocks:
$$
{\mc C}(X,(x_i),(M_i)) = \on{Hom}_{\g_{\on{out}}}\left(\otimes_i
  M_i,\C\right),
$$

Now let ${\mc P}$ be a $G$-bundle on $X$. Then we define the space of
${\mc P}$-{\em twisted coinvariants} as
$$
H^{\mc P}(X,(x_i),(M_i)) = \otimes_i M_i / \left( \g^{\mc P}_{\on{out}} \cdot
  \otimes_i M_i \right),
$$
where
$$
\g^{\mc P}_{\on{out}} = H^0(X \bs \{ x_1,\ldots,x_n \},{\mc P}
\underset{G}\times \g).
$$
Its dual space is the space of ${\mc P}$-{\em twisted conformal
  blocks}:
$$
{\mc C}^{\mc P}(X,(x_i),(M_i)) =
\on{Hom}_{\g^{\mc P}_{\on{out}}}\left(\otimes_i M_i,\C\right),
$$

For a general vertex algebra $V$, to obtain the space of coinvariants
(or conformal blocks), we take the quotient (or take invariant
functionals) with respect to a larger Lie algebra, encompassing all
Ward identifies of all fields of the vertex algebra $V$. If $V$ has
$\ghat$-symmetry, then this Lie subalgebra can be twisted by a
$G$-bundle ${\mc P}$ on $X$. This way, we obtain the ${\mc P}$-twisted
spaces of coinvariants and conformal blocks, ${\mc H}_V^{\mc
  P}(X,(x_i),(M_i))$ and ${\mc C}_V^{\mc P}(X,(x_i),(M_i))$, see
\cite{FB}. Thus, we obtain a modular functor ${\mc H}_V^{\mc P}$, and
its conformal blocks version ${\mc C}_V^{\mc P}$, for each $G$-bundle
${\mc P}$. It is natural to call it a ${\mc P}$-twist of ${\mc H}$ (or
${\mc C}$).

As shown in \cite{FB}, if the modules $M_i$ are such that the action
of the Lie subalgebra $\g[[z]]$ of $\ghat$ extends to the group
$G[[z]]$, then the spaces of ${\mc P}$-twisted coinvariants are fibers
(or stalks) of a naturally defined $\ka$-twisted $D$-module on
$\Bun_G$. This $\ka$-twisted $D$-module is obtained by applying the
so-called {\em localization functor} $\Delta_\ka$ to $\otimes_i M_i$.

More generally, suppose that the action of the Lie subalgebra
$z^{m_i}\g[[z]]$ of $\ghat$ extends to the corresponding subgroup of
$G[[z]]$, for some $m_i\geq 0$. Then we obtain a twisted $D$-module on
$\Bun_{G,(x_i),(m_i)}$, the moduli stack classifying the data $({\mc
  P},(\eta_i))$, where ${\mc P}$ is a $G$-bundle on $X$ and $\eta_i$
is a trivialization of ${\mc P}$ on the $(m_i-1)$st formal
neighborhood of the point $x_i$. If $M_i$ is a highest weight
$\ghat$-module (or a module from the category ${\mc O}$), then we can
replace the trivialization of the fiber ${\mc P}_{x_i}$ of ${\mc P}$
at $x_i$ (which corresponds to the case $m_i=1$) by a parabolic
structure, that is a reduction of ${\mc P}_{x_i}$ to a Borel subgroup
of $G$. In this paper, we restrict ourselves to the case $m_i=0$, so
our localization functors takes values in the category of
$\ka$-twisted $D$-modules on $\Bun_G = \Bun_{G,(x_i),(0)}$.

Now recall (see, e.g., \cite{KS,GM}) that for a twisted $D$-module
${\mc R}$ on a smooth manifold $Z$, the stalk of ${\mc D}$ at a point
$p \in Z$ is defined as
$$
i_p^*({\mc R}) = {\mc R} \underset{{\mc O}_Z}\otimes \C_p,
$$
where $\C_p$ is the skyscraper sheaf supported at $p$; that is, the
${\mc O}_Z$-module $\C_p = {\mc O}_Z/{\mathfrak m}_p$, where
${\mathfrak m}_p$ is the sheaf of functions vanishing at $p$.

We can rewrite this as follows:
$$
i_p^*({\mc R}) = {\mc R} \underset{{\mc D}_Z}\otimes \delta_p,
$$

The functor $i^*_p$ is right exact. One can show that the left derived
functor of $i_p^*$ is equivalent to the right derived functor of the
functor $i^!_p$ coupled with cohomological shift by $-\dim Z$, where
$$
i^!_p({\mc R}) = \on{Hom}_{{\mc O}_Z}(\C_p,{\mc R}) = \on{Hom}_{{\mc
    D}_Z}(\delta_p,{\mc R}).
$$
We will ignore this cohomological shift.

If the category of $D$-modules on $Z$ twisted by the square root of the
canonical line bundle on $Z$ is equivalent to the category of
$A$-branes on $T^*Z$, then the latter Hom can be described as a
$\on{Hom}$ in the category of $A$-branes on $T^*Z$ (and similarly, for
the categories of twisted $D$-modules). And that's because under such
an equivalence, the $A$-brane corresponding to the $D$-module
$\delta_p$ to be ${\mc F}_p$, the fiber of $T^*Z$ over $p \in Z$. So,
if ${\mc B}$ is the $A$-brane corresponding to the $D$-module ${\mc
  R}$, then we find that
isomorphic to
\begin{equation}    \label{isom}
\on{Hom}(\delta_p,{\mc R}) \simeq \on{Hom}({\mc F}_p,{\mc B})
\end{equation}
up to a shift of cohomological degrees. (We will ignore this
cohomological shift as well.)

Therefore, for a semi-stable $G$-bundle ${\mc P}$ we can compare the
space of ${\mc P}$-twisted coinvariants with a Hom of $A$ branes (with
respect to the symplectic structure $\omega_K$):
$$
{\mc H}^{\mc P}(X,(x_i),(M_i)) = \on{Hom}(\delta_{\mc P},\Delta\left(
  \otimes_i M_i \right)) \simeq \on{Hom}({\mc F}_{\mc P},{\mc B}),
$$
if ${\mc B}$ is the $A$-brane on ${\mc M}_H(G)$ corresponding to
the $D$-module $\Delta\left( \otimes_i M_i \right)$ obtained by
the localization functor from $\otimes_i M_i$.

This shows that the branes ${\mc F}_{\mc P}$ can be used to
link the spaces of ${\mc P}$-twisted coinvariants and Hom's in the
categories of $D$-modules and $A$-branes.

If our ${\mc D}$-module is holonomic (i.e. all of its fibers are
finite-dimensional), we can express the dual space to the space of
coinvariants, the space of ${\mc P}$-twisted conformal blocks, as
a Hom between $D$-modules in the opposite direction (and without any
cohomological shift):
$$
{\mc C}^{\mc P}_V(X,(x_i),(M_i)) \simeq \on{Hom}(\Delta\left( \otimes_i
  M_i \right),\delta_{\mc P}).
$$
Alternatively, we can write it in terms of the corresponding $A$-brane
${\mc B}$, which in this case should be Lagrangian:
$$
{\mc C}^{\mc P}_V(X,(x_i),(M_i)) \simeq \on{Hom}({\mc B},{\mc
  F}_{\mc P}).
$$
Similarly, we can use the branes ${\mc F}'_{\mc P}$ away from the
critical level.

\subsection{Construction of the compactification functor as a
  localization functor}    \label{comp and loc}

Recall that we denote by $D^G_{0,1}$ the Dirichlet boundary condition
in the 4d gauge theory ${\mc T}^G_\ka$. Let $B$ be another boundary
condition in the same theory. Suppose that we have a specific junction
$D^G_{0,1} \to B$ which is non-degenerate in the sense described in
Section \ref{va j}. Then, according to the general formalism, we have
the junction vertex algebra $V^G_\ka(D^G_{0,1}\sto B)$. From the point
of view of the strip geometry discussed in Section \ref{bc to D}, the
vertex algebra $V^G_\ka(D^G_{0,1}\sto B)$ is a ``boundary chiral
algebra'' of the 3d TFT. On general grounds, we should expect a close
relationship between the spaces of conformal blocks of this boundary
chiral algebra and the spaces of states ${\mb H}^B_{\mc P}({\mc A})$
of the TFT discussed above. In this subsection, we use this idea to
give a mathematically rigorous definition of the compactification
functor $F^{G,B}_\ka$ as a localization functor, under specific
assumptions.

To motivate it, recall that in Section \ref{bc to D} we have argued
that the vector spaces ${\mb H}^B_{\mc P}({\mc A})$ should be fibers
of a $\ka$-twisted $D$-module on $\Bun_G$, which we then took as
$F^{G,B}_\ka({\mc A})$. On the other hand, as explained in Section
\ref{va junction}, we expect that $V^G_\ka(D^G_{0,1}\sto B)$ contains
the affine Kac--Moody algebra $\ghat$ of level $\ka'$ such that $\ka -
\ka' = n(G) \cdot m$ for some $m \in \Z$. Recall also that should be a
functor
\begin{equation}    \label{FGka}
F^G_{\ka,D^G_{0,1}\; \sto \; B}: D_\ka(\Gr_G) \boxtimes {\mc C}^G_\ka(B)
\to V^G_\ka(D^G_{0,1}\sto B)\on{-mod}.
\end{equation}
We now make a list of concrete {\em assumptions} that we expect this
functor to satisfy:

\bigskip

\begin{enumerate}

\item The functor $F^G_{\ka,D^G_{0,1}\; \sto \; B}$ is fully faithful
  (in other words, the junction $D^G_{0,1} \to B$ is non-degenerate).

If this is the case, then slightly abusing notation, we will identify
the image of $I \boxtimes {\mc C}^G_\ka(B)$ under the functor
\eqref{FGka} with ${\mc C}^G_\ka(B)$. Here $I$ is the identity object
of $D_\ka(\Gr_G)$; namely, $I = \delta_{\ol{1}}$, where $\ol{1}$
is the coset of the identity element of $G\zpart$ in
$\Gr_G$. Given an object ${\mc A}$ of the category ${\mc
  C}^G_\ka(B)$, denote the image of $I \otimes {\mc A}$ under the
functor $F^G_{\ka,D^G_{0,1}\; \sto \; B}$ also by ${\mc A}$. The latter ${\mc
  A}$ is thus a $V^G_\ka(D^G_{0,1}\sto B)$-module, and hence a
$\ghat_{\ka'}$-module.

\bigskip

\item The action of the Lie subalgebra $\g[[z]] \subset \ghat_{\ka'}$
  on ${\mc A}$ can be extended to the corresponding group $G[[z]]$. In
  other words, ${\mc A}$ is a $(\ghat_{\ka'},G[[z]])$ Harish-Chandra
  module.

\bigskip

\item The functor $F^G_{\ka,D^G_{0,1}\; \sto \; B}$ can be described as
  follows:
\begin{equation}    \label{conv}
{\mc F} \in D_\ka(\Gr_G), {\mc A} \in {\mc C}^G_\ka(B) \mapsto
{\mc F} \star {\mc A},
\end{equation}
where $\star$ denotes the convolution functor \cite{FG}, which is
well-defined because ${\mc A}$ is a $(\ghat,G[[z]])$ Harish-Chandra
module under the previous assumption (if $\ka\neq\ka'$, we tensor
${\mc F}$ with an appropriate line bundle on $\Gr_G$ to make it into a
$D_{\ka'}$-module before applying the convolution functor). Note that
formula \eqref{conv} coincides with formula \eqref{star} in the case
$B=N^G_{1,0}$.

\end{enumerate}

\bigskip

According to assumption (ii), we have a localization functor
$$
\Delta_{\ka'}: V^G_\ka(D^G_{0,1}\sto B)\on{-mod} \to D_{\ka'}(\Bun_G).
$$
Denote its restriction to the subcategory ${\mc C}^G_\ka(B)$ of
$V^G_\ka(D^G_{0,1}\sto B)$-mod by $\Delta_{\ka'}|_{{\mc
    C}^G_\ka(B)}$.

Under the above assumptions, we now {\em define} the compactification
functor $F^{G,B}_\ka$ from ${\mc C}^G_\ka(B)$ to $D_\ka(\Bun_G)$ in
terms of the localization functor $\Delta_{\ka'}|_{{\mc
    C}^G_\ka(B)}$. Namely, we set
\begin{equation}    \label{F Delta}
F^{G,B}_\ka = \Delta_{\ka'}|_{{\mc C}^G_\ka(B)} \otimes {\mc
  L}_G^{\otimes m},
\end{equation}
where ${\mc L}_G$ is the minimal line bundle on $\Bun_G$ and $m \cdot
n(G) = \ka-\ka'$ (recall that tensoring with ${\mc L}_G$ shifts the
level by $n(G)$).

To motivate the definition \eqref{F Delta}, let us compare the fibers
of the twisted $D$-modules on the left and right hand sides of
\eqref{F Delta} for an object ${\mc A}$ of ${\mc C}^G_\ka(B)$. On the
left hand side, the fiber is
$$
\on{Hom}(\delta_{\mc P},F^{G,B}_\ka({\mc A})).
$$
On the right hand side it is isomorphic to the space of ${\mc
  P}$-twisted coinvariants
$$
H_{V^G_\ka(D^G_{0,1}\to B)}^{\mc P}(X,x,{\mc A}).
$$

In order to show that these two spaces should indeed be isomorphic
under our assumptions (i)-(iii), consider formula \eqref{conv} in the
case that ${\mc F} = \delta_p$, the $\delta$-function twisted
$D$-module corresponding to a point $p \in \Gr_G =
G\zpart/G[[z]]$. Let $\wt{p}$ be a lift of $p$ to $G\zpart$. Then for
any $(\ghat,G[[z]])$ Harish-Chandra module $M$, we have
\begin{equation}    \label{F delta}
\delta_p \star M \simeq \wt{p}^*(M).
\end{equation}
where for a $\ghat$-module $M$ and an element $h \in G\zpart$, we
denote by $h^*(M)$ the $\ghat$-module on which the action of $\ghat$
is modified by $\on{Ad}_h$. Note that if the action of the Lie
subalgebra $\g[[z]] \subset \ghat$ on $M$ extends to an action of the
corresponding group $G[[z]]$, then $h_1^*(M) \simeq h_2^*(M)$ if the
images of $h_1$ and $h_2$ in $\Gr_G$ coincide. That's why the
right hand side of \eqref{F delta} does not depend on the lift of $p$
to $G\zpart$.

By our assumptions (ii) and (iii), the right hand side of \eqref{F
  delta} describes the image of $\delta_p \otimes M$ under the
functor \eqref{FGka}.

Recall that if $G$ is simple, then $\Bun_G \simeq G(X\bs x) \bs
G\zpart/G[[z]]$. Let $p$ be a lift of ${\mc P} \in \Bun_G$ to $\Gr_G =
G\zpart/G[[z]]$, and $\wt{p}$ a lift of $p$ to $G\zpart$. We can think
of $\wt{p}$ as a transition function on the punctured disc around
$x$. If we glue the trivial bundles on the disc and on $X \bs x$ using
this transition function, we obtain the $G$-bundle ${\mc P}$ on $X$.
(As explained in Section \ref{ex fun}, the compactification functor
$F^{G,D_{0,1}}_\ka$ sends $\delta_p, p \in \Gr_G$, to $\delta_{\mc
  P}$, where ${\mc P}$ is the image of $p$ in $\Bun_G = G(X\bs x) \bs
\Gr_G$.)

The definition of twisted coinvariants readily implies that the space
of ${\mc P}$-twisted coinvariants of ${\mc A}$ is the same as the
space of untwisted coinvariants of $\wt{p}^*({\mc A})$, which is
$\delta_p \star {\mc A}$ by formula \eqref{F delta}. Hence
\begin{equation}    \label{coinv tw}
H_{V^G_\ka(D^G_{0,1}\to B)}^{\mc P}(X,x,{\mc
  A}) \simeq H_{V^G_\ka(D^G_{0,1}\to B)}(X,x,\delta_p \star {\mc
  A}).
\end{equation}

Now, according to assumption (ii) and formula \eqref{conv}, Conjecture
\ref{Hom and conf1} states that we have an isomorphism
\begin{equation}    \label{H and Hom}
H_{V^G_\ka(D^G_{0,1}\to
    B)}(X,x,\delta_p \star {\mc A}) \simeq \on{Hom}(\delta_{\mc
    P},F^{G,B}_\ka({\mc A})).
\end{equation}

Combining the isomorphisms \eqref{coinv tw} and \eqref{H and Hom}, we
find that the fibers at ${\mc P}$ of the two sides of \eqref{F
  Delta} should indeed be isomorphic:
\begin{equation}    \label{two fibers}
\on{Hom}(\delta_{\mc P},F^{G,B}_\ka({\mc A})) \simeq 
H_{V^G_\ka(D^G_{0,1}\to B)}^{\mc P}(X,x,{\mc A}).
\end{equation}
Thus, the fibers of the two sides of \eqref{F Delta} are indeed
isomorphic.

Since $\Delta_{\ka'}({\mc A})$ is a $\ka'$-twisted $D$-module on
$\Bun_G$, and $F^{G,B}_\ka({\mc A})$ is a $\ka$-twisted $D$-module,
in light of \eqref{two fibers}, this naturally leads us to stipulate
that
$$
F^{G,B}_\ka({\mc A}) = \Delta_{\ka'}({\mc A}) \otimes {\mc
  L}_G^{\otimes m},
$$
where $m=(\ka-\ka')/n(G) \in \Z$. This motivates the definition
\eqref{F Delta}.

Since there could be multiple junctions $D^G_{0,1} \to B$ with
different vertex algebras associated to them, for this definition to
be correct, we need the following:

\begin{conj}    \label{meta1}
For any two junctions $D^G_{0,1} \to B$ such that the corresponding
functors \eqref{FGka} satisfy the assumptions (i)-(iii), the
functors $\Delta_{\ka'}|_{{\mc C}^G_\ka(B)}$ are naturally isomorphic
(after taking a tensor product with a power of ${\mc L}_G$ if
the levels $\ka'$ corresponding to the two junctions are different).
\end{conj}

This conjecture can be viewed as a generalization of Conjecture
\ref{meta}, according to which the space of coinvariants
$H_{V^G_\ka(D^G_{0,1}\to B)}^{\mc P}(X,x,{\mc A})$ depends (up to an
isomorphism) only on $B$ and the object ${\mc A}$ of the category
${\mc C}^G_\ka(B)$ (as well as ${\mc P} \in \Bun_G$), but is
independent of the junction data between $D^G_{0,1}$ and $B$ used in
the construction of the vertex algebra $V^G_\ka(D^G_{0,1}\sto
B)$. Conjecture \ref{meta1} extends this to an isomorphism of the
corresponding twisted $D$-modules on $\Bun_G$.

\bigskip

\begin{rem}    \label{p and P}
The isomorphism \eqref{coinv tw}, translated
into the language of 4d gauge theory, means that we can obtain the
space of states ${\mb H}^B_{\mc P}({\mc A})$ in the strip geometry
discussed in Section \ref{gener} in two ways: the first is by coupling
the Dirichlet boundary condition to a $G$-bundle ${\mc P}$ on $X$; and
the second is by adding the line defect corresponding to $\delta_p$,
where $p$ is a lifting of ${\mc P} \in \Bun_G$ to $\Gr_G$, at the
boundary. This makes perfect sense because this line defect has the
effect of changing the trivial $G$-bundle on $X$ to a bundle obtained
by gluing the trivial $G$-bundles on the disc around $x$ and $X\bs x$
using as the transition function, an element $\wt{p}$ of $G\zpart$
defined as above.

However, there is a subtle but important difference between these two
procedures of obtaining the space ${\mb H}^B_{\mc P}({\mc A})$: When
we obtain it by coupling the Dirichlet boundary condition to the
$G$-bundle ${\mc P}$, as described in Section \ref{gener}, we
automatically obtain a natural action on ${\mb H}_{\mc P}({\mc A})$ of
the group $\on{Aut}({\mc P})$ of automorphisms of the $G$-bundle ${\mc
  P}$ on $X$. On the other hand, if we instead insert the line defect
$\delta_p$, we obtain the same vector space, but we forget the action
of $\on{Aut}({\mc P})$. If we want these spaces to combine into a
twisted $D$-module on $\Bun_G$, however, the information about the
action of $\on{Aut}({\mc P})$ must be included, so we have to use the
first procedure.

Likewise, the group $\on{Aut}({\mc P})$ acts naturally on the space
$H_{V^G_\ka(D^G_{0,1}\to B)}^{\mc P}(X,x,{\mc A})$ of ${\mc
  P}$-twisted coinvariants (which appears in our definition of the
compactification functor as the localization functor) but doesn't
acts on the space $H_{V^G_\ka(D^G_{0,1}\to B)}(X,x,\delta_p \star {\mc
  A})$.\qed
\end{rem}

\begin{rem}
Roughly speaking, the reason we are able to assign twisted $D$-modules
on $\Bun_G$ to line defects associated to a boundary condition $B$ is
that there is a family of line defects associated to the Dirichlet
boundary condition parametrized by points of $\Bun_G$. What about
the Neumann and Nahm boundary conditions? At first glance, the above
construction can't be generalized to them because their standard line
defects are parametrized by discrete data (integral weights or
coweights of $G$). However, we actually do have a continuous parameter
for those line defects as well; namely, the point $x$ of the Riemann
surface $X$ at which we insert a junction from one of these boundary
conditions to $B$. Unlike the Dirichlet case,
for which varying $x$ does not change the compactification functor, in
the Nahm and Neumann cases it does change. Therefore, using a variant
of the above arguments, we obtain $D$-modules on $X$. More generally, by
considering multiple points, we can obtain $D$-modules on the spaces
of $P^+$- or $\LP^+$-valued divisors on $X$. These $D$-modules can
probably be glued together into $D$-modules on the corresponding Ran
spaces.\qed
\end{rem}

\subsection{A simple test}    \label{test}

Consider the vertex algebra associated to the junction
$$
N^G_{1,0} \to D^G_{0,1} = R(D^G_{0,1} \to N^G_{1,0})
$$
in the bulk theory ${\mc T}^G_{-\ka}$. According to formula \eqref{R
  junction}, this vertex algebra is $V_\ka(\ghat)$. Let us take the
identity object in the category of line defects ${\mc
  C}^G_{-\ka}(N^G_{1,0})$ and the $\delta$-function $D$-modules
$\delta_p, p \in \Gr_G$ in ${\mc C}^G_{-\ka}(D^G_{0,1})$. Applying to
them the compactification functors, we obtain the sheaf of
$\ka$-twisted differential operators ${\mc D}_\ka$ and the
$\delta$-function $D$-modules $\delta_{\mc P}$ on $\Bun_G$. According
to Conjecture \ref{Hom and conf1} and the discussion of Section
\ref{comp and loc}, we expect an isomorphism
\begin{equation}    \label{H as Hom}
H^{\mc P}(X,x,V_\ka(\g)) \simeq \on{Hom}_{{\mc D}_\ka}({\mc
  D}_\ka,\delta_{\mc P}).
\end{equation}

To show that this isomorphism indeed holds, observe that the right
hand side is nothing but the space of sections of $\delta_{\mc P}$,
which we will denote by $\Gamma(\delta_{\mc P})$. On the other hand,
$H_{\mc P}(X,x,V_k(\g))$ is also isomorphic to $\Gamma(\delta_{\mc
  P})$. The proof of this fact (see, e.g., \cite{FB}, Ch. 18) is
straightforward: the tangent space to $\Bun_G$ at ${\mc P}$ is
isomorphic to
\begin{equation}    \label{tangent}
T_{\mc P} \Bun_G \simeq \g^{\mc P}_{\on{out}} \bs \g\zpart/\g[[z]],
\end{equation}
where $z$ is a local coordinate at $x \in X$. By definition,
$$
V_\ka(\g) = U_\ka(\ghat)/U_\ka(\ghat) \cdot \g[[z]],
$$
and $H_{\mc P}(X,x,V_\ka(\g))$ is the quotient of $V_\ka(\g)$ by the
action of the Lie subalgebra $\g^{\mc P}_{\on{out}}$. This gives us an
isomorphism
$$
H_{\mc P}(X,x,V_k(\g)) \simeq \Gamma(\delta_{\mc P}),
$$
which implies \eqref{H as Hom}.

Finally, we can relate the right hand side of \eqref{H as Hom} to a
Hom of branes:
\begin{equation}    \label{Hom1}
\on{Hom}_{{\mc D}_\ka}({\mc
  D}_\ka,\delta_{\mc P}) \simeq \on{Hom}({\mc B}_{\on{c.c.}},{\mc
  F}'_{\mc P}).
\end{equation}
Therefore we obtain
\begin{equation}    \label{Hom2}
H^{\mc P}(X,x,V_\ka(\g)) \simeq \on{Hom}({\mc B}_{\on{c.c.}},{\mc
  F}'_{\mc P}).
\end{equation}
This isomorphism was discussed in \cite{FW}, and an isomorphism
between ``tempered'' versions of these vector spaces has been studied
in \cite{BT} (note that in \cite{BT} this term refers to vectors in
these spaces, and so it has no relation to the notion of tempered
$D$-modules discussed in Remark \ref{rational}).

\subsection{Branes and Hecke eigensheaves}    \label{hecke}

Let $\ka=0$, corresponding to the critical level of $\ghat$. In this
case, Kapustin and Witten defined the $(B,A,A)$ branes $({\mathbf
  F}_b,\nabla)$ supported on the fibers ${\mathbf F}_b$ of the Hitchin
fibration in ${\mc M}_H(G)$ (see item (ii) on the list at the
beginning of Section \ref{Dmod branes}). They are mirror dual to the
$(B,B,B)$ zero-branes ${\mb B}_{\mc E}$ supported at generic points of
${\mc M}_H(\LG)$ (in the complex structure $J$ on ${\mc M}_H(\LG)$
these are flat $\LG$-bundles on $X$ that have no automorphisms other
then those coming from the center of $\LG$). Since ${\mb B}_{\mc E}$
is an eigenbrane of the Wilson operators, we obtain that the branes
$({\mathbf F}_b,\nabla)$ are eigenbranes of the 't Hooft operators
(these are line defect operators corresponding to lines in the bulk of
the 4d gauge theory).

As we discussed in Section \ref{Dmod branes}, the branes $({\mathbf
  F}_b,\nabla)$ do not deform away from $\ka=0$. However, at $\ka=0$
they give rise to twisted $D_0$-modules on $\Bun_G$, which are Hecke
eigensheaves, objects of interest in the geometric Langlands
correspondence. (Recall that $D_0$-modules are $D$-modules twisted by
a square root of the canonical line bundle, so a careful treatment of
the corresponding category includes spin subtleties discussed in the
later sections. But here we choose to ignore these subtleties.)

Based on gauge theory considerations \cite{Ga1, Ga2,CoG}, one can
argue that the Hecke eigensheaves should also computable as conformal
blocks of certain ``kernel'' vertex algebras, which have both a
Kac--Moody subalgebra of level that is a multiple of $n(G)$ and an
extra structure which allows the conformal blocks/coinvariants to be
twisted by a flat $\LG$-bundle.

These $\kappa=0$ kernel vertex algebras can sometimes be built
directly from a microscopic description of the dual to Dirichlet
boundary conditions, involving the three-dimensional SCFT called
$T[G]$ in \cite{GW2}.  The description is only somewhat explicit for
some classical gauge groups and it cannot be deformed to irrational
$\kappa$.

Still, somewhat experimentally, it appears to lead to kernel vertex
algebras $V(T[G])$ which are, in an appropriate sense the $\kappa \to
0$ limits of $V^G_\kappa(D^G_{0,1} \to D^G_{1,0})$: up to some
rescalings, the OPE of $V^G_\kappa(D^G_{0,1} \to D^G_{1,0})$ has a
finite limit which coincides with $V(T[G])$, with the level $\lka$
Kac--Moody currents for $\LG$ replaced in the limit $\lka \to \infty$
by a classical holomorphic connection for $\LG$. This allows conformal
blocks of $V(T[G])$ to be twisted by a flat $\LG$-bundle.
Conjecturally, $V(T[G])$ can be obtained in this fashion even when
$T[G]$ is not directly known.

More generally, vertex algebras of the form $V^G_\kappa(D^G_{1,0}\sto
B)$ tend to have a very nice $\kappa \to 0$ behaviour
\cite{Ga1,Ga2,CG,CoG}: up to some rescalings the OPE have a finite
limit giving rise to a vertex algebra $V^G_0(B)$, with the level
$\lka$ of Kac--Moody currents for $\LG$ replaced by a classical
holomorphic connection for $\LG$. Again, this allows conformal blocks
of $V^G_0(B)$ to be twisted by a flat $\LG$-bundle.  We expect the
$D$-module of coinvariants of $V^G_0(B)$ to be the Geometric Langland
dual of the $D$-module of coinvariants of $V^G_0(D^G_{0,1}\sto B)$.
We hope to explore this in more detail in a future work.

Note that Beilinson and Drinfeld have constructed Hecke
eigensheaves on $\Bun_G$ using the isomorphism of \cite{FF,F:wak}. In
this subsection we discuss a link between these two constructions in
the framework of our general formalism.

Recall that according to \cite{FF,F:wak}, the vacuum module $V_0(\g)$
of $\ghat$ of critical level has a large algebra of endomorphisms,
which is isomorphic to the algebra of functions on the space
$\on{Op}_{\LG}(D_x)$ of opers for the Langlands dual group $\LG$ on
the formal disc $D_x$ at $x$. For $\rho \in
\on{Op}_{\LG}(D_x)$, we can then define a $\ghat$-module ${\mathbb
  V}_\rho$ of critical level as the quotient of $V_0$ by the ideal
$I_\rho$ corresponding to $\rho$:
$$
V_\rho = V_0/I_\rho.
$$
In other words, we set all central elements of the enveloping algebra
of $\ghat$ (such as the quadratic Sugawara elements) equal to the
numeric values prescribed by $\rho$ (see \cite{F:review} for more
details).

Now suppose that $x$ is a point of a Riemann surface $X$ (and so $D_x
\subset X$). The following theorem is due to Beilinson and Drinfeld
\cite{BD}.

\begin{thm} \hfill

  {\em (1)} ${\mc H}^{\mc P}(X,x,V_\rho) = 0$ for all ${\mc
    P} \in \Bun_G$, unless the $\LG$-oper $\rho$ extends from the
  disc $D_x$ to the entire curve $X$.

  {\em (2)} If $\rho$ does extend to an $\LG$-oper on $X$, then the
  critically twisted $D$-module on $\Bun_G$ $\Delta_0(V_\rho)$, whose
  stalks are the spaces of coinvariants ${\mc H}^{\mc
    P}(X,x,V_\rho)$ is a Hecke eigensheaf whose eigenvalue is the
  flat $\LG$-bundle on $X$ defined by this oper.

\end{thm}

The two perspectives on Hecke eigensheaves will agree if 
${\mc H}^{\mc P}(X,x,{\mb V}_\rho)$ coincides with the space of coinvariants for 
$V(T[G])$ twisted by a flat $\LG$-bundle which is also an oper. 
In concrete examples this seems to happen in a very direct fashion: 
coupling to an oper deforms $V(T[G])$ to a vertex algebra equivalent to 
$V_\rho$.

For completeness, we now recall the definition of the Kapustin--Witten $(B,A,A)$ branes on
${\mc M}_H(G)$ that are dual to the zero-branes ${\mb B}_{\mc E}$ on
${\mc M}_H(\LG)$ \cite{KW}. Recall that here ${\mc E}$ is a flat
$\LG$-bundles on $X$ that have no automorphisms other then those
coming from the center of $\LG$, viewed as a point in the Hitchin
moduli space ${\mc M}_H({}^L G)$ with respect to the complex structure
$J$. Now consider ${\mc E}$ as a point of ${\mc M}_H({}^L G)$ with
respect to the complex structure $I$, and let $b = b({\mc E})$ be its
image in the Hitchin base under the (first) Hitchin fibration. Let
${\mb F}_b$ be the dual fiber in ${\mc M}_H(G)$. Then the mirror dual
brane to ${\mb B}_{{\mc E}}$ is the pair $({\mb F}_b,\nabla)$, where
$\nabla=\nabla_{\mc E}$ is a flat unitary line bundle on ${\mb F}_b$
corresponding to ${\mc E}$ under the $T$-duality of Hitchin fibers in
${\mc M}_H(G)$ and ${\mc M}_H(\LG)$.

Suppose that ${\mc E} = {\mc E}(\rho)$ corresponds to an oper $\rho$.
Then we find from formula \eqref{isom} that there should be an
isomorphism between the space of ${\mc P}$-twisted conformal blocks of
the $\ghat$-module ${\mb V}_\rho$ and a Hom in the category of
$A$-branes on ${\mc M}_H(G)$ (with respect to $\omega_K$):
\begin{equation}    \label{crit}
{\mc H}^{\mc P}(X,x,{\mb V}_\rho) \simeq
\on{Hom}({\mc F}_{\mc P},({\mb F}_b,\nabla_{\mc E})).
\end{equation}
This isomorphism has been previously discussed in \cite{FW,BT}.

As Balasubramanian and Teschner argued in \cite{BT}, one can think of
this isomorphism as an expression of a link between the
Beilinson--Drinfeld construction and the Kapustin--Witten construction
of the Geometric Langlands correspondence. Indeed, it shows how to
express the fibers of the Hecke eigensheaf, i.e. a $D_0$-module on
$\Bun_G$ obtained by applying the localization functor $\Delta_0$ to
${\mb V}_\rho$ in terms of Hom's between the fiber-brane ${\mc F}_{\mc
  P}$ discussed in Section \ref{Dmod branes} and the Kapustin--Witten
brane $({\mb F}_b,\nabla_{\mc E})$.

As we explained in \cite{Ga1} and Section \ref{bc to D} above, the
spaces appearing on the right hand side of \eqref{crit} acquire a
natural flat (Berry) connection, which should coincide with the one on
the left hand side, coming from the $D$-module structure on
$\Delta_0({\mb V}_\rho)$.

As a simple test of \eqref{crit}, we can check that for a generic
$G$-bundle ${\mc P}$, the dimensions of the two vector spaces in the
isomorphism \eqref{crit} are the same. According to the left hand
side, this is the generic rank of the $D$-module $\Delta_0({\mb
  V}_\rho)$, which is the multiplicity of the nilpotent cone in the
zero fiber of the (first) Hitchin fibration. It coincides with the
number of points in the intersection of generic fibers ${\mb F}_b$ and
${\mc F}_{\mc P}$ of the first and second Hitchin fibrations. But
the latter is the dimension of the generic space on the right hand
side of \eqref{crit}.

\section{A simple example: $U(1)$ gauge theory}    \label{U1}

The simplest Abelian example -- the 4d $N=4$ supersymmetric $U(1)$
gauge theory -- is already rich enough to demonstrate non-trivial
compositions of junctions.  It is also an excellent example of the
potential spin structure dependence of our constructions and of the
extra subtleties which occur when the spin structure dependence is
lifted.

The notions of Nahm and Dirichlet boundary conditions coincide in
Abelian theories, so we have a single family of boundary conditions
$N^{U(1)}_{p,q}$, which we will denote in this subsection simply by
$N_{p,q}$.

\subsection{Category of boundary lines}
The tensor category ${\mc C}^G_\ka(N_{1,0})$ of line defects
associated to the boundary condition $N_{1,0}$ is
$\KL_\ka(U(1))$.\footnote{If $G_c$ is a compact Lie group and $\g_c$
  its Lie algebra, we will sometimes refer to the affine Kac--Moody
  algebra $\ghat$, where $\g$ is the complexification of $\g_c$, as
  the $\g_c$ (affine) Kac--Moody algebra, and use $\g_c$ rather than
  $\g$ in denoting various categories and functors associated to
  $\ghat$. For example, we will sometimes use the notation $U(1)$
  instead of $GL(1)$, e.g., $\KL_\ka(U(1))$.} It is a semisimple
category with irreducible objects $L_n$ for integer $n$, and the
tensor product (fusion) $L_n \otimes L_{m} \simeq L_{n+m}$. The
twisting (aka topological spin) on $L_n$ is given by $e^{\pi i
  \frac{n^2}{\kappa}}$. We will call its phase factor $\pi
\frac{n^2}{\kappa}$ the twist.

For now, we will consider ${\mc C}^G_\ka(N_{1,0})$ as what we will
call a ``spin-ribbon category.'' This is a modification of the ribbon
category in which the twistings are defined up to multiplication by
$\pm 1$. In other words, the twists are only defined up to addition of
an integer multiple of $\pi$ (rather than $2\pi$). The necessity to
view ${\mc C}^G\ka(N_{1,0})$ as a spin-ribbon category can be
inferred, for example, from the fact that some of the dualities from
the group $SL_2(\Z)$ preserve the spin-ribbon structure but not the
ribbon structure. 

For example, the action of $STS$ maps $\ka \mapsto
(\kappa^{-1} -1)^{-1}$, and hence the twist $\pi \frac{n^2}{\kappa}$
gets mapped to $\pi n^2\left(\kappa^{-1} -1\right)$. The corresponding
categories ${\mc C}^G_\ka(N_{1,0})$ and ${\mc C}^G_{(\kappa^{-1}
  -1)^{-1}}(N_{1,0})$ would only be equivalent if we stipulate that
the twists in them are defined up to addition of integer multiples of
$\pi$.

\subsection{Junction vertex algebras}
The basic junction from $N_{0,1} \to N_{1,0}$ supports the
$\wh{\mathfrak{u}}(1)$ Kac--Moody algebra at level $\ka$ (which is its
own ${\mc W}$-algebra). The standard modules correspond to the
$\wh{\mathfrak{u}}(1)$ vertex operators of charge $n + m \ka$ and
conformal dimension
\begin{equation}
\Delta^{\mathfrak{u}(1)}_{m,n}(\ka) = \frac{n^2}{2 \ka} + n m +
\frac{m^2 \ka}{2}.
\end{equation}
They are the images of the corresponding objects of the category
$\KL_\ka(U(1)) \boxtimes \KL_{\ka^{-1}}(U(1))$ in
$\wh{\mathfrak{u}}(1)_\ka\on{-mod}$ under the functor
$F^{U(1)}_{\ka,N^G_{0,1}\; \sto N^G_{1,0}}$ (we use the notation
introduced in Section \ref{va junction}).

In the same way as in the last example of Section \ref{va junction},
we find that applying the duality transformation $RST^{-1}$ to the
standard junction $N_{0,1} \to N_{1,0}$ we obtain a junction $N_{1,-1}
\to N_{1,0}$ supporting the $\wh{\mathfrak{u}}(1)$ Kac--Moody algebra at
level $\ka^{-1}+1$. On the other hand, applying $T^{-1}$ we obtain
the junction $N_{0,1} \to N_{1,-1}$ supporting the $\wh{\mathfrak{u}}(1)$
Kac--Moody algebra at level $\ka+1$.

The composition of two such junctions, $N_{0,1} \to N_{1,-1}
\to N_{1,0}$, gives an extension of
$\wh{\mathfrak{u}}(1)_{\ka+1} \times \wh{\mathfrak{u}}(1)_{\ka^{-1}+1}$
Kac--Moody by new fields of dimensions
\begin{equation}    \label{lattice conf dim}
\Delta^{\mathfrak{u}(1)}_{0,n}(\ka+1) +
\Delta^{\mathfrak{u}(1)}_{n,0}(\frac{\ka}{\ka+1}) = \frac{n^2}{2}
\frac{1}{\ka+1} + \frac{n^2}{2}\frac{\ka}{\ka+1} = \frac{n^2}{2}.
\end{equation}
The extension can be identified with the product of a
$\wh{\mathfrak{u}}(1)$ Kac--Moody algebra at level $\ka$ and of the
lattice vertex superalgebra corresponding to the lattice $\Z$. The
$\wh{\mathfrak{u}}(1)$ current of the former is
$\frac{\ka}{\ka+1}(J_{\ka+1} - J_{\ka^{-1}+1})$ and the
$\wh{\mathfrak{u}}(1)$ current of the latter is
$\frac{1}{\ka+1}J_{\ka+1}+\frac{1}{\ka^{-1}+1} J_{\ka^{-1}+1}$. The
latter is isomorphic to the free (complex) fermion vertex superalgebra
$\bigwedge$. It is generated by two fermionic fields of conformal
dimensions $1/2$ which correspond to $n = \pm 1$ in formula
\eqref{lattice conf dim}.

More general modules over this vertex algebra combine vertex operators
of dimension
\begin{equation}
\Delta^{\mathfrak{u}(1)}_{m,n}(\ka+1) +
\Delta^{\mathfrak{u}(1)}_{n,e}(\frac{\ka}{\ka+1}) = \frac{(n+e+m)^2}{2}
+ \Delta^{\mathfrak{u}(1)}_{m,-e}(\ka)
\end{equation}

Thus, the vertex algebra $V^{U(1)}_\ka(N_{0,1}\sto N_{1,0})$ obtained
from the composition of junctions $N_{0,1} \to N_{1,-1} \to N_{1,0}$
is the vertex algebra $V^{U(1)}_\ka(N_{0,1}\sto N_{1,0})$ of the basic
junction $N_{0,1} \to N_{1,0}$ tensored with the free fermion vertex
superalgebra. From the point of view of 4d gauge theory, both vertex
algebras are associated to the junction $N_{0,1} \to N_{1,0}$, but
with different junction data.\footnote{Physically, the second junction
  is obtained from the first by ``stacking'' it with a decoupled,
  holomorphic 2d spin-CFT: a free complex fermion. In particular, the
  second junction is intrinsically a ``spin junction''.}

According to Conjecture \ref{meta}, we expect the spaces of
conformal blocks of the two vertex algebras to be isomorphic. And
indeed, this is the case, because the space of conformal blocks of the
free fermion vertex superalgebra $\bigwedge$ is one-dimensional for
any compact Riemann surface $X$. However, because the two generating
fields of $\bigwedge$ have conformal dimensions $1/2$, in order to
define this space of conformal blocks we need to choose a square root
$K^{1/2}$ of the canonical line bundle on $X$. This choice is
equivalent to a choice of spin structure on $X$. This is another
example of how spin structures show up in the twisted TFT.

We can also construct new junctions. For example, we can
compose the basic junction $N_{0,1} \to N_{1,0}$ with the junction
$N_{1,0} \to N_{k,1}$ obtained as $ST^{-k}$ image of the basic
junction $N_{0,1} \to N_{1,0}$. For now, take $k>0$. 

The result is a junction $N_{0,1} \to N_{k,1}$, which is {\it not} a
duality image of the basic junction: there is no combination of $S$ and $T$ operations which could relate it to 
$N_{0,1} \to N_{1,0}$.  The corresponding vertex algebra
is an extension of $\wh{\mathfrak{u}}(1)_{\ka} \times
\wh{\mathfrak{u}}(1)_{k - \ka^{-1}}$ by operators of conformal
dimension $\frac{k}{2} n^2$. It is isomorphic to the product of the
$\wh{\mathfrak{u}}(1)_{\ka - k^{-1}}$ Kac--Moody algebra and the
$u(1)_k$ lattice vertex algebra corresponding to the lattice
$\sqrt{k}\Z$.

More general modules have dimensions
\begin{equation}
\Delta^{\mathfrak{u}(1)}_{m,n}(\ka) +
\Delta^{\mathfrak{u}(1)}_{n,e}(k-\ka^{-1}) = \frac{(k n+e+m)^2}{2 k} +
\frac{m^2 }{2k}(k \ka-1) + \frac{e^2}{2k} \frac{1}{k \ka-1}- \frac{e
  m}{k}
\end{equation}
and combine modules over $u(1)_k$ with modules over
$\wh{\mathfrak{u}}(1)_{\ka - k^{-1}}$ Kac--Moody algebra with integral
magnetic charges (corresponding to $m$) but electric charges of the
form $e/k, e \in \Z$.

For $k=0$, the extension involves modules of conformal dimension $0$
and is isomorphic to the vertex algebra of chiral differential
operators on $GL_1(\C)=\C^\times$. Indeed, by rescaling the first and
the second factors of the $\wh{\mathfrak{u}}(1)_{\ka} \times
\wh{\mathfrak{u}}(1)_{- \ka^{-1}}$ current algebra by $\ka^{-1/2}$ and
$\ka^{1/2}$, respectively, we obtain the $\wh{\mathfrak{u}}(1)_{1}
\times \wh{\mathfrak{u}}(1)_{-1}$ current algebra of the standard
$\beta \gamma$ system in which we make the field $\gamma$
invertible. Note that under this identification, the commuting
$\wh{\mathfrak{u}}(1)$ currents are constructed from the fields
$\beta\gamma$ and $\gamma^{-1} \partial \gamma$, and the extension
vertex operators correspond to $\gamma^n, n \in \Z$.

For $k<0$, the extension involves modules of dimension unbounded from
below and it is less well-behaved as a vertex algebra.

More generally, we would like to construct junctions $N_{0,1} \to
N_{p,q}$ such that in the corresponding vertex algebra all fields,
except the vacuum, have positive conformal dimensions. This is a
favorable property for many reasons, not least the behavior of
conformal blocks.

To achieve that, we need a chain of junctions which support
$\wh{\mathfrak{u}}(1)_{\kappa_i}, i=0,\ldots,n$, with
$$
\frac{1}{\kappa_{i-1}} + \kappa_{i} = k_i,
$$
where each $k_i$ is a positive integer.

Such $\wh{\mathfrak{u}}(1)_{\kappa_i}$ can be found at the junction
$N_{0,1} \to N_{1,0}$ when the bulk coupling is $\kappa_i$.

Applying $ST^{-k_i}$, we find that it can also be found at the junction
$N_{1,0} \to N_{k_i,1}$ when the bulk coupling is
$$
-\frac{1}{\kappa_i - k_i} = \kappa_{i-1}.
$$
We can then compose the $N_{0,1} \to N_{1,0}$ junction and the
$N_{1,0} \to N_{k_i,1}$ junction with the bulk coupling $\kappa_{i-1}$
into a junction $N_{0,1} \to N_{k_i,1}$.

Note that the same vertex algebra will appear when the bulk coupling
is $\kappa_{i-2} = -1/(\kappa_{i-1} - k_{i-1})$ at the
  composition $N_{1,0} \to N_{k_{i-1},1} \to N_{k_i
    k_{i-1}-1,k_i}$. So, we can compose it with $N_{0,1} \to N_{1,0}$
  to get a junction $N_{0,1} \to N_{k_i k_{i-1}-1,k_i}$.

Next, move to the bulk coupling $\kappa_{i-3} = -1/(\kappa_{i-2} -
  k_{i-2})$ by applying $S T^{-k_{i-2}}$ and compose to get $N_{0,1}
  \to N_{k_i k_{i-1} k_{i-2} - k_i -k_{i-2}, k_{i} k_{i-1}-1}$.

  If we run this procedure backwards from $\kappa_n$ to $\kappa_0$, we
  get junctions of the form
\begin{multline*}
N_{0,1}\to N_{1,0} \to ST^{-k_1}(N_{1,0}) \to
ST^{-k_1}ST^{-k_2}(N_{1,0}) \to \\ ST^{-k_1}ST^{-k_2}ST^{-k_3}(N_{1,0})
\to \cdots \to ST^{-k_1}ST^{-k_2} \ldots ST^{-k_n}(N_{1,0})
\end{multline*}
with $k_i>0$, i.e.,
\begin{equation}
N_{0,1}\to N_{1,0} \to N_{k_1,1} \to N_{k_1 k_2 - 1,k_2} \to N_{k_1
  k_2 k_3 - k_1 - k_3,k_2 k_3 - 1} \to \cdots
\end{equation}
If we label the $i$th term after $N_{1,0}$ by $N_{p_i,q_i}$ then
\begin{equation}
\frac{p_i}{q_i} = k_{1} - \frac{1}{k_{2}-}\frac{1}{k_{3}-}\cdots
\frac{1}{k_{i}}
\end{equation}
can be seen as the $i$th {\it convergent}  (i.e., the
truncation after the first $i$ terms) of the continued fraction
\begin{equation}
\frac{p_i}{q_i} = k_{1} - \frac{1}{k_{2}-}\frac{1}{k_{3}-}\cdots
\end{equation}

The junctions support $\wh{\mathfrak{u}}(1)_{\ka_i}$ Kac--Moody with
$\ka_i = k_i - \ka_{i-1}^{-1}$ and $\ka_0 = \ka$. The extension
involves modules of charges $q_{i} \ka_i + q_{i+1}$ for
$\wh{\mathfrak{u}}(1)_{\ka_i}$ (with $q_0=0$) of dimension
$$
\sum_{i=1}^n \frac{k_i}{2} q_i^2 + q_i q_{i+1}.
$$
This vertex algebra is isomorphic to the product of a lattice vertex
algebra of rank $n$ and the $\wh{\mathfrak{u}}(1)$ Kac--Moody algebra
at level
\begin{equation}   \label{kapq}
\ka_{p_n,q_n} = \kappa +\frac{1}{k_{1}-}\frac{1}{k_{2}-}\cdots
\frac{1}{k_{n}}
\end{equation}
with the image of $\KL_{\ka^{-1}}(U(1)) \times \KL_{\ka_n}(U(1))$
given by combinations of modules for the lattice vertex algebra and
modules of integral magnetic charges and fractional electric charges
for the $\wh{\mathfrak{u}}(1)$ Kac--Moody algebra.

All fields of this vertex algebra, except for the vacuum, have
positive conformal dimensions.

This construction can serve as a prototype for building a vertex
algebra satisfying this property corresponding to a junction from
$N_{0,1}$ to any $N_{p,q}$: expand $\frac{p}{q}$ into continued
fractions and use the resulting integers $k_i$ to build the junction
as a composition of basic junctions, as explained above.

\subsection{Compactification functors}

We can now discuss the compactification functors from the categories
${\mc C}^{U(1)}_\ka(B)$ to $D_{\ka}(\Bun_{U(1)})$ (see Section \ref{ex
bc}).

We begin with two important points:

\medskip

\begin{itemize}

\item There is a degree $1$ line bundle ${\mc L}$ on $\Bun_{U(1)}$
  related to the level 1 Chern--Simons action.  It is the line bundle
  of conformal blocks for the $u(1)_1$ lattice vertex algebra (the
  vertex superalgebra $\bigwedge$ of a free complex fermion). Again,
  we are ignoring the spin structure for the moment.

\medskip

\item The category ${\mc C}^{U(1)}_\ka(N_{1,0})$ is $\KL_\ka(U(1))$,
  and the compactification functor $F^{U(1),N_{1,0}}_\ka$ is the
  localization functor $\Delta_\ka$. Applying it to the identity
  object of $\KL_\ka(U(1))$, which is the vertex algebra
  $V_{\ka}(U(1))$, we obtain the sheaf ${\mc D}_\ka$ itself (see
  Section \ref{stalks}). This is a special case of the general
  statement discussed in example (ii) of Section \ref{ex fun}: the
  identity object of ${\mc C}^{U(1)}_\ka(N_{1,0})$ is mapped to ${\mc
    D}_\ka$. In other words, the compactification map sends the
  boundary condition $N_{1,0}$ to ${\mc D}_\ka$.

\end{itemize}

\medskip

The construction of a junction from $N_{0,1} \to N_{1,1}$ as a
composition $N_{0,1} \to N_{1,0} \to N_{1,1}$ of the basic junction
and the junction $N_{1,0} \to N_{1,1}$ discussed in the previous
subsection (the case $k=1$ of the junction $N_{1,0} \to N_{k,1}$) is
instructive. 

It produces a junction vertex algebra isomorphic to $u(1)_1 \otimes
V_{\ka-1}(U(1))$, with the coupling to $U(1)$ bundles implemented by
the diagonal combination of the $\wh{\mathfrak{u}}(1)$ currents of the
two factors. This junction vertex algebra yields a compactification
functor whose image lies directly in $D_{\ka}(\Bun_{U(1)})$. The image
of the compactification functor is the product of the sheaves of
coinvariants of the two vertex subalgebras $u(1)_1 \otimes
V_{\ka-1}(U(1))$, i.e. ${\mc L} \otimes {\mc D}_{\ka-1}$. In other
words, the compactification map sends the boundary condition $N_{1,1}$
to ${\mc L} \otimes {\mc D}_{\ka-1}$.

We could also produce a $N_{0,1} \to N_{1,1}$ junction as a $T$
image of the $N_{0,1} \to N_{1,0}$ junction (up to the spin subtleties
discussed in the next subsection).  Since $N_{1,1} = T(N_{1,0})$,
$$
{\mc C}^{U(1)}_\ka(N_{1,1}) \simeq {\mc C}^{U(1)}_{\ka-1}(N_{1,0}) =
\KL_{\ka-1}(U(1)).
$$
Thus, the relevant localization functor for this second junction is
$\Delta_{\ka-1}$ taking values in $D_{\ka-1}(\Bun_{U(1)})$. When we
apply it to the identity object of ${\mc C}^{U(1)}_\ka(N_{1,1})$
(i.e. the junction vertex algebra itself), we obtain ${\mc
  D}_{\ka-1}$, viewed as a left $(\ka-1)$-twisted $D$-module on
$\Bun_{U(1)}$. As explained in Section \ref{comp and loc}, to get the
corresponding compactification functor $F^{G,N_{1,1}}_\ka$, we should
tensor $\Delta_{\ka-1}$ with the line bundle ${\mc L}$. Under
$F^{G,N_{1,1}}_\ka$, the identity object of ${\mc
  C}^{U(1)}_\ka(N_{1,1})$ therefore goes to ${\mc L} \otimes {\mc
  D}_{\ka-1}$, in agreement with the previous construction.

This example provides a positive test for our prescription for the compactification functor. 
Is also demonstrates how the compactification functors intertwine the action of the
duality $T$ on the categories corresponding to boundary conditions and
the operation of taking a tensor product with ${\mc L}$, in agreement
with Conjecture \ref{qGL functors}.\footnote{From the point of view of 4d gauge
theory, this is clear from the fact that if we have a domain wall such
that gauge theories on the two sides of the wall differ by the duality
symmetry $T$ (i.e. have coupling constants $\ka$ and $\ka+1$), then
the difference between the topological terms in the 4d actions on the
two sides (each of them is a scalar multiple of the first Pontryagin
class, and the two scalars differ by $2\pi$) has to be compensated on
the boundary by the Chern--Simons action of level 1, which is the 4d
gauge theory counterpart of tensoring with the line bundle ${\mc L}$
on $\Bun_{U(1)}$.}

Similar considerations apply to show that the compactification map
sends $N_{1,k}$ to ${\mc L}^k \otimes {\mc D}_{\ka-k}$.

\medskip

Next, we look at $N_{k,1}$. As explained above, the vertex algebra
corresponding to the junction $N_{0,1} \to N_{k,1}$ is $u(1)_k \times
\wh{\mathfrak{u}}(1)_{\ka - k^{-1}}$. If $|k|>1$, we cannot use the
localization functor $\Delta_{\ka - k^{-1}}$ corresponding to the
$\wh{\mathfrak{u}}(1)_{\ka - k^{-1}}$ subalgebra because then $\ka -
k^{-1}$ differs from $\ka$ by a rational number that is is not an
integer. However, this junction vertex algebra has a diagonal
$\wh{\mathfrak{u}}(1)_\ka$ subalgebra generated by the current which
is the linear combination $k^{-1} J_{u(1)_k} + J_{\ka - k^{-1}}$ of
the generating currents of $u(1)_k \times \wh{\mathfrak{u}}(1)_{\ka -
  k^{-1}}$. The compactification functor $F^{U(1),N_{k,1}}_\ka$ can be
obtained from the localization functor $\Delta_\ka$ with
respect to this subalgebra.

In particular, applying the localization functor $\Delta_\ka$ to the
identity object in ${\mc C}^{U(1)}_\ka(N_{k,1})$, we obtain a
$D_\ka$-module on $\Bun_{U(1)}$ which is easy to describe as an ${\mc
  O}$-module: it is the tensor product of the bundles of coinvariants
for $u(1)_k$ and $\wh{\mathfrak{u}}(1)_{\ka - k^{-1}}$, i.e. ${\mc
  V_k} \otimes {\mc D}_{\ka-k^{-1}}$. Here ${\mc V_k}$ is the vector
bundle of coinvariants for $u(1)_k$ which is coupled ``magnetically''
to the line bundles in $\Bun_{U(1)}$ in a non-trivial way: given a
line bundle $\ell$ in $\Bun_{U(1)}$, we represent it as $\ell = {\mc
  O}(D)$, where $D = \sum_i n_i x_i$ is a divisor on $X$, and then
take the space of coinvariants with the insertions of $u(1)_k$ primary
fields of charge $n_i$ at the points $x_i$. (The statement that ${\mc
  V_k} \otimes {\mc D}_{\ka-k^{-1}}$ has the structure of a
$D_\ka$-module is non-trivial. It follows from the fact that it is in
the image of $\Delta_\ka$.)

Thus, we have describe the image of $N_{k,1}$ under the
compactification map. More generally, by using the composition of
junctions above, we can obtain the image of any $N_{p,q}$ as a twisted
$D$-module of the form ${\mc L}^{[q/p]} \otimes {\mc V_{p,q}}
\otimes {\mc D}_{\ka_{p_n,q_n}-[q/p]}$ with ${\mc V_{p,q}}$ being
the bundle of coinvariants of the lattice vertex
algebra built above as an extension of $\otimes_i u(1)_{k_i}$.

It would be nice to prove directly that the qGL dualities ${\mc
  E}^{U(1),g}_\ka$ intertwine these $D$-modules for $(p,q)$ and $g \circ
(p,q)$. The simplest statement corresponds to the fact that
$S(N_{1,-1}) = N_{1,1}$. This implies that the qGL duality $S$ sends
$$
S({\mc L}^{-1} \otimes {\mc D}_{\ka+1}) \mapsto {\mc L} \otimes {\mc
  D}_{-\ka^{-1}-1},
$$
which is a special case of formula \eqref{selfdual}.

It is worth mentioning that all the $N_{p,q}$ boundary conditions can
be given an interpretation as modifications of Neumann boundary
conditions involving coupling to some specific 3d Chern--Simons
theories, i.e. to some physical 3d TFTs. One may consider many more
such boundary conditions, associated to more general 3d TFTs which can
be coupled to $U(1)$ connections.

\subsection{Spin subtleties}    \label{gerbes}
Here we will discuss under which conditions we can dispense 
with the use of spin manifolds for the bulk theory and possibly for 
its boundary conditions and junctions.

Recall that the 4d topological field theory ${\mc T}^G_\ka$ we are
interested in is obtained by a topological twist (the GL twist of
\cite{KW}) from the physical supersymmetric gauge theory. This means
that the action of the Lorentz group on the field content of the
theory has been modified in such a way that the resulting twisted
theory is endowed with a two-dimensional family of supercharges that
can used to define a TFT (the physical theory is not endowed with
supercharges). In particular, while in the physical theory the
fermions transform as spinors, in the twisted TFT they are turned into
$i$-forms with $i=0,1,2$. Thus, while the physical theory can only be
defined on a spin four-manifold $M$ and one needs to make a specific
choice of the spin structure to make the theory well-defined, the
twisted TFT does not require $M$ to be a spin manifold, nor is one
required to make a choice of the spin structure.

However, there is a price to pay: it turns out that the action of the
quantum dualities then has to be refined. If we choose a spin
structure, then every element of the group $PSL_2(\Z)$ gives rise to a
legitimate quantum duality of the theory ${\mc T}^{U(1)}_\ka$ (though
the categories of line defects are then spin-ribbon categories, as
discussed above). However, if we do not have a spin structure, or do
not want to choose a specific one, then we cannot employ the $T$
duality transformation: only $T^2$ and $S$ (and their products) are
legitimate dualities.

Instead of the group $PSL_2(\Z)$ of quantum dualities, we then have a
duality groupoid, obtained by adding two more ``nodes'', which are
certain topological modifications of the standard $U(1)$ twisted gauge
theory ${\mc T}^{U(1)}_\ka$. These three theories are then related to each
other by various duality transformations.

More specifically, we denote the basic gauge theory ${\mc
  T}^{U(1)}_\ka$ by $U(1)_b$. Its $T$-image will be denoted by
$U(1)_t$.\footnote{We should really use the ${\mc
  T}^{U(1)_b}_\ka$, ${\mc
  T}^{U(1)_t}_\ka$, etc. notation, but it is a bit cumbersome.}
  In other words, we define $U(1)_t$ with a coupling $\ka$
to be the same as $U(1)_b$ with coupling $\ka - 1$. Then $T$ maps
$U(1)_b$ theory to $U(1)_t$ theory, and vice versa.

The $S$ image of $U(1)_t$ will be denoted as $U(1)_s$. It is known that 
$U(1)_s$ has an independent definition \cite{M}: it is a gauge theory based on
$\mathrm{Spin}_\C$ connections rather than standard $U(1)$ connections. 
Thus $S$ maps $U(1)_t$ to $U(1)_s$ theories, and vice versa. Finally,
$T$ is a true duality of $U(1)_s$:
\begin{equation}
\begin{matrix} U(1)_b & \longleftrightarrow & U(1)_t &
  \longleftrightarrow & U(1)_s \cr \circlearrowright & T
  &&S&\circlearrowright
\cr S & &&& T\end{matrix}
\end{equation}

The basic boundary conditions $N_{1,0}$ and $N_{0,1}$ for the $U(1)_b$
theory can be defined without a choice of spin structure and are
exchanged by S-duality. On the other hand, in order to define
$N_{1,1}$ in $U(1)_b$, we cannot use $T$.  Instead, we use the direct
definition: $N_{1,q}$ can be defined as a modified Neumann boundary
condition with $q$ extra units of boundary Chern--Simons action. For
odd $q$, this definition is viable but requires a choice of spin
structure at the boundary.

In general, duality transformations tell us that boundary conditions
$N_{p,q}$ do not require a spin structure in $U(1)_b$ as long as $pq$
is even.

Applying $T$, we get by definition the corresponding statement for
$U(1)_t$. In particular, $N_{1,0}$ requires a spin structure in
$U(1)_t$. In $U(1)_s$, we expect then $N_{0,1}$ to require a spin
structure as well. Indeed, Dirichlet boundary conditions for a
$\mathrm{Spin}_\C$ gauge field do not have a canonical choice of
trivial connection, unless a spin structure is selected.

These topological aspects do affect the categories of lines available
at boundary conditions. These will be ribbon categories if the
boundary conditions do not require a spin structure. If they do (as
$N_{1,q}$ with odd $q$ in $U(1)_b$), then they will be spin-ribbon,
i.e. the twists will be defined modulo integer multiples of $\pi$
rather than $2\pi$.

A good example of this phenomenon is the observation that although
$N_{1,1}$ and $N_{1,-1}$ are related by S-duality in $U(1)_b$, the
categories of boundary lines are equivalent only as spin-ribbon
categories: the objects in one category have twists $2 \pi
\frac{n^2}{2 (\kappa+1)}$, while the corresponding objects in the
other categories have spin $-2 \pi \frac{n^2}{2 (\kappa^{-1}+1)}$. The
two differ by possibly half-integer multiples of $2\pi$; namely, by
$2 \pi \frac{n^2}{2}$.

The effects percolate also to vertex algebra calculations. If both
boundary conditions {\it and} the junction itself can be defined with
no reference to spin structure, the vertex algebra will be a true
vertex algebra, with fields of integral dimensions (but possibly a
vertex superalgebra, including odd fields). Otherwise, the vertex
algebra may include fields of half-integer conformal dimensions,
which requires a choice of spin structure on the Riemann surface $X$.

We have already seen this phenomenon implicitly. The standard junction
in $U(1)_b$ from $N_{0,1}$ to $N_{1,0}$ does not require a spin
structure and supports $\wh{\mathfrak{u}}(1)_\kappa$.  On the other
hand, the same junction in $U(1)_t$ must be built as a composition of
junctions leading to $u(1)_1 \times \wh{\mathfrak{u}}(1)_{\kappa-1}$,
a spin-vertex algebra.

These subtleties matter when we build compactification functors to
twisted $D$-modules. This is expressed in the fact that the line
bundle ${\mc L}$ on $\Bun_{U(1)}$ we used above is not defined
canonically unless one selects a spin structure on $X$. It can be
defined canonically as a section of a $\Z_2$ gerbe ${\mc G}$ on
$\Bun_{U(1)}$ which is trivial, but not canonically so.

This is a good moment to discuss some general notions related to
gerbes. Let $\Gamma$ be a finite abelian group, and ${\mc G}$ a
$\Gamma$-gerbe on a manifold $M$. In the Cech definition, this means
that given an open covering of $M$, we have on each open subset $U_i
\subset M$ of the covering a category ${\mc G}(U_i)$ that is a torsor
(in the categorical sense) over the category of $\Gamma$-bundles on
$U_i$. Choosing a trivialization of ${\mc G}(U_i)$, we obtain
$\Gamma$-bundles ${\mc G}_{ij}$ on the overlaps $U_i \cap U_j$ and
hence an element ${\mc G}_{ijk}$ of $\Gamma$ for each triple overlap
$U_i \cap U_j \cap U_k$. These elements should define a Cech
two-cocycle on $M$ with values in $\Gamma$.

A standard example is the following: let ${\mc L}$ be a line bundle on
$M$ and $k$ a positive integer. Let us choose an identification $\Z_k
\simeq \mu_k$, the group of $k$th roots of unity. Define the
$\Z_k$-gerbe ${\mc G}_{{\mc L},k}$ on $M$ by taking as ${\mc G}_{{\mc
    L},k}(U_i)$ the category of $k$th roots of ${\mc L}$ on $U_i$,
i.e. line bundles ${\mc K}$ on $U_i$ such that ${\mc K}^k \simeq {\mc
  L}$. (This is a torsor for the category of $\Z_k$-bundles because
any two such ${\mc K}$ differ by a $k$th root of the trivial line
bundle; or equivalently, a $\Z_k$-bundle.) The gerbe ${\mc G}_{{\mc
    L},k}$ is trivial if and only if there exists on $M$ a $k$th root
of ${\mc L}$ defined on the entire $M$. If ${\mc G}_{{\mc L},k}$ is
trivial, then a trivialization of ${\mc G}_{{\mc L},k}$ is the same as
a choice of such a global $k$th root of ${\mc L}$.

Given a $\Gamma$-gerbe ${\mc G}$ on $M$ and another group $H$ (not
necessarily finite) together with a homomorphism $\Gamma \to H$ whose
image is in the center of $H$, we have the notion of an $H$-bundle
on $M$ modified by ${\mc G}$. In Cech realization, an ordinary
$H$-bundle assigns to each open $U_i$ a torsor over the group
$H(U_i)$. Then on overlaps $U_i \cap U_j$ we obtain $H$-valued
functions which has to satisfy a one-cocycle condition on triple
overlaps. An $H$-bundle on $M$ modified by ${\mc G}$ is defined by the
same data, except that the one-cocycle condition is modified on the
triple overlaps by the images of ${\mc G}_{ijk}$ in $H$.

If $H=\C^\times$, we obtain the notion of a line bundle modified by a
$\Gamma$-gerbe ${\mc G}$ for each homomorphism $\Gamma \to
\C^\times$. We obtain the notions of ${\mc O}$-modules and ${\mc
  D}$-modules on $M$ modified by ${\mc G}$ in a similar way.

The examples we are most interested in are as follows:

\medskip

\begin{itemize}

\item ${\mc L}=K_X$ is the canonical line bundle on a Riemann surface
  $X$ and $k=2$. Then the gerbe ${\mc G}_{K_X,2}$ of square roots of
  $K_X$ is trivial, but not canonically. Its trivialization is the
  same as a choice of spin structure on $X$. Let $H=\C^\times$ and
  $\Z_2 \to \C^\times$ be the embedding with the image $\{ \pm 1
  \}$. We refer the $\C^\times$-bundles on $X$ modified by ${\mc
    G}_{K_X,2}$ as $\on{Spin}_\C$ bundles. More generally, if $\Z_2$
  maps to the center of a group $H$, we refer to the corresponding
  bundles on $X$ as $\on{Spin}_H$ bundles.

\medskip

\item ${\mc L}=K_{\Bun_G}$ is the canonical line bundle on $\Bun_G$
  and $k=2$. For $G=GL_1$, the corresponding gerbe ${\mc
    G}_{K_{\Bun_G},2}$ is the gerbe discussed in this section. For a
  simple Lie group $G$, the gerbe ${\mc G}_{K_{\Bun_G},2}$ has been
  studied in \cite{BD} and, in the setting close to ours, in
  \cite{FW2}. This gerbe is trivial, but not canonically so if $\rho$
  is not an integral weight of $G$. In that case, a trivialization can
  be constructed from a choice of spin structure on $X$.

\medskip

\item ${\mc L}={\mc L}_G$, the minimal line bundle on $\Bun_G$. The
  corresponding gerbes are needed if we want to extend the qGL duality
  groupoid. In 4d gauge theory language, they correspond to the
  theories associated to discrete $\theta$ angles (see Section
  \ref{sec:topo}).

\end{itemize}

\medskip

We can now map the three nodes of our groupoid to three variants of
the usual category of $D$-modules:

\medskip

\begin{itemize}
\item $U(1)_b \to D_{\ka}(\Bun_{U(1)})$, standard twisted $D$-modules.

\medskip

\item $U(1)_t \to {\mc G}-D_{\ka}(\Bun_{U(1)})$, twisted $D$-modules
  modified by the gerbe ${\mc G}$.

\medskip

\item $U(1)_s \to D_{\ka}(\Bun_{\mathrm{Spin}_\C})$, twisted $D$-modules
  on a modified moduli space, that of $\mathrm{Spin}_\C$ bundles on the
  Riemann surface $X$.
\end{itemize}

\medskip

The image of $T$, tensoring with ${\mc L}$, clearly intertwines
between the two first lines. On the other hand, the line bundle ${\mc
  L}$ is defined canonically on $\Bun_{\mathrm{Spin}_\C}$.

The image of the qGL duality $S$ maps the first category
$D_{\ka}(\Bun_{U(1)})$ to itself. We conjecture it exchanges the last
two categories ${\mc G}-D_{\ka}(\Bun_{U(1)})$ and
$D_{\ka}(\Bun_{\mathrm{Spin}_\C})$.

Thus we predict that, taking into account spin subtleties, the qGL
dualities have the form
\begin{equation}
\begin{matrix} D_{\ka}(\Bun_{U(1)}) & \longleftrightarrow & {\mc
    G}-D_{\ka'}(\Bun_{U(1)}) & \longleftrightarrow &
  D_{\ka''}(\Bun_{\mathrm{Spin}_\C}) \cr \circlearrowright & T
  &&S&\circlearrowright
\cr S & &&& T\end{matrix}
\end{equation}

\section{A richer example: $SU(2)$ gauge theory}    \label{SU2}

The vertex algebras appearing at the junctions of boundary conditions
in gauge theories related to the group $SL_2$ (or $SU(2)$) provide a
particularly rich class of examples. The analysis is somewhat
complicated by the choices of global form of the group, i.e. $SU(2)$
or $SO(3)$, and by spin subtleties.

Usually, it is possible to relate gauge theory configurations with
different global forms of the gauge group by topological manipulations
which affect the junction vertex algebras in a relatively minor way,
at most adding/removing some simple auxiliary vertex algebras, such as
the real free fermion vertex algebra $\Ff$ of central charge
$c_{\Ff}=\frac12$. Recall that $\Ff^n = \tso(n)_1$, the simple
quotient of $\mathfrak{so}(n)$ Kac--Moody at level 1. Also, $\Ff^2 =
u(1)_1$.

Because of that observation, we first make some preliminary statements
which are essentially insensitive to the form of the gauge group, and
then refine them to sharper statements involving specific gauge
groups.

\subsection{Major boundary conditions and junction vertex algebras}

At the first, loose level, we can refer to families of boundary conditions $N_{p,q}$ and $D_{p,q}$. We can identify several junctions and corresponding vertex algebras for the 
$\mathfrak{su}(2)$ gauge algebra from string theory constructions which directly produce orthogonal or symplectic groups or by stripping off Abelian contributions from $\mathfrak{u}(2)$
statements. We refer the reader to Sections 8 and 10 of \cite{GR} and Section 4 of \cite{CG} for details.  Up to dualities, we will
encounter five interesting classes of junctions and vertex algebras:

\medskip

\begin{itemize}
\item $N_{0,1} \to N_{1,0}$: $\Vir_\kappa$. This is the Virasoro algebra of central charge $c_{\Vir_\kappa} = 13- 6 \kappa - 6 \kappa^{-1}$. Notice that $\Vir_\kappa \simeq \Vir_{\kappa^{-1}}$, compatible with $RS$ invariance of the junction.  

\medskip

\item $N_{0,1} \to N_{2,1}$: $\sVir_{2 \kappa-1}$. The super-Virasoro algebra $\sVir_\kappa$ has central charge $c_{\sVir_\kappa} = \frac{15}{2}- 3 \kappa - 3 \kappa^{-1}$ and satisfies $\sVir_\kappa \simeq \sVir_{\kappa^{-1}}$, compatible with $RST^2S$ invariance of the junction. 

\medskip

\item $D_{0,1} \to N_{1,0}$: $\wh{\mathfrak{su}}(2)_{\kappa}$. This is the Kac--Moody algebra at critically shifted level $\kappa$. It has central charge $c_{\wh{\mathfrak{su}}(2)_\kappa} = 3-6 \kappa^{-1}$.

\medskip

\item $D_{0,1} \to N_{2,1}$: $\wh{\mathfrak{osp}}(1|2)_{2\kappa-1}$. The super-Kac--Moody algebra $\wh{\mathfrak{osp}}(1|2)_{\kappa}$ has an $\wh{\mathfrak{su}}(2)_{\frac{1+\kappa}{2}}$ sub-algebra. 
It has central charge $c_{\wh{\mathfrak{osp}}(1|2)_\kappa} =1-3 \kappa^{-1}$.

\medskip

\item $D_{0,1} \to D_{1,0}$: $\td(2,1|-\kappa)_1$. This is a quotient of the super-Kac--Moody
  algebra based on the $\mathfrak{d}(2,1|-\kappa)$ exceptional
  superalgebra which has sub-algebra $\wh{\mathfrak{su}}(2)_{\kappa+1}
  \times \wh{\mathfrak{su}}(2)_{\kappa^{-1}+1} \times \tsu(2)_1$, where
  $\tsu(2)_1$ is the simple quotient of $\wh{\mathfrak{su}}(2)$ Kac--Moody
  at level 1. The central charge of this vertex algebra is
  $c_{\td(2,1|-\kappa)_1} = 1$. The algebra is invariant under $\kappa
  \to \kappa^{-1}$, compatibly with $RS$ invariance of the junction. 
\end{itemize}

\begin{table}[h]
\begin{center}
\begin{tabular}{|c||c|}
\hline
{\em Junction} & {\em Vertex Algebra}  \\
\hline \hline
$N_{0,1} \to N_{1,0}$& $\Vir_\kappa$  \\ 
\hline
$N_{0,1} \to N_{2,1}$ & $\sVir_{2 \kappa-1}$  \\
\hline
$D_{0,1} \to N_{1,0}$& $\wh{\mathfrak{su}}(2)_{\kappa}$ \\
\hline
$D_{0,1} \to N_{2,1}$& $\wh{\mathfrak{osp}}(1|2)_{2\kappa-1}$  \\
\hline
$D_{0,1} \to D_{1,0}$& $\td(2,1|-\kappa)_1$  \\
\hline
\end{tabular} 
\end{center}
\vspace*{5mm}
\caption{A brief summary of the vertex algebras which appear at some junction in $SU(2)/SO(3)$ gauge theory.}\label{tab:three}
\end{table}

These vertex algebras are related in interesting ways by the quantum
Drinfeld--Sokolov reductions, which effectively replace $D_{0,1}$ with
$N_{0,1}$ in the junctions:

\medskip

\begin{itemize}
\item The standard $\Vir_\kappa = \DS
  \left[\wh{\mathfrak{su}}(2)_{\kappa}\right]$.\footnote{Indeed
$c_{\Vir_\kappa} = c_{\wh{\mathfrak{su}}(2)_\kappa}-2-6(\kappa-2)$, where
$-2$ is the contribution from the ghost $bc$-system and
$-6(\kappa-2)$ comes from the re-definition of the stress tensor.}

\medskip

\item The relation $\DS \left[\wh{\mathfrak{osp}}(1|2)_{\kappa}\right]=\Ff
  \times \sVir_\kappa$. Here we do the
DS reduction of the $\wh{\mathfrak{su}}(2)_{\frac{1+\kappa}{2}}$
sub-algebra, without stripping off
the extra free fermion originating from the odd current of charge
$\frac12$. This extra fermion is the
decoupled $\Ff$ factor indicated above.\footnote{As a check, $c_{\Ff} +
c_{\sVir_\kappa} =
c_{\wh{\mathfrak{osp}}(1|2)_\kappa}-2-6(\frac{1+\kappa}{2}-2)$.}

\medskip

\item The relation $\DS^{(1)}
  \left[\td(2,1|-\kappa)_1\right]=\tso(4)_1 \times
  \wh{\mathfrak{su}}(2)_{\kappa^{-1}}$. Here we do the DS reduction on the
  $\wh{\mathfrak{su}}(2)_{\kappa+1}$ sub-algebra, without stripping off
the extra four free fermions originating from the odd current of
charge $\frac12$. These extra fermions give the
$\tso(4)_1$ factor indicated above. The residual
$\wh{\mathfrak{su}}(2)_{\kappa^{-1}+1} \times \tsu(2)_1$ in
$\left[\td(2,1|-\kappa)_1\right]$
is embedded in the obvious way in $\tso(4)_1
\times \wh{\mathfrak{su}}(2)_{\kappa^{-1}}$.\footnote{As a check,
  $c_{\td(2,1|-\kappa)_1}-2-6(\kappa+1-2) = 4 c_{\Ff} +
c_{\wh{\mathfrak{su}}(2)_{\kappa^{-1}}}$.} Similarly, $\DS^{(2)}
\left[\td(2,1|-\kappa)_1\right]=\tso(4)_1 \times
\wh{\mathfrak{su}}(2)_{\kappa}$.
\end{itemize}

\medskip

There is also a rich web of coset/extension relations which have a
natural interpretation as compositions of junctions or replacements
$D_{0,1} \mapsto N_{1,k}$:

\medskip

\begin{itemize}
\item $D_{0,1} \to N_{1,-1} \to N_{1,0}$: The GKO-like coset $\Vir_{1+\kappa^{-1}} =
  \frac{\wh{\mathfrak{su}}(2)_{\kappa} \times
    \tsu(2)_1}{\wh{\mathfrak{su}}(2)_{\kappa+1}}$. 

\medskip

\item $D_{0,1} \to N_{1,-2} \to N_{1,0}$:  The super-GKO-like coset $\sVir_{1+2 \kappa^{-1}} =
  \frac{\wh{\mathfrak{su}}(2)_{\kappa} \times
    \tso(3)_1}{\wh{\mathfrak{su}}(2)_{\kappa+2}}$. 

\medskip

\item $N_{0,1} \to N_{1,0} \to N_{2,1}$: $\sVir_{2\kappa-1} \times \Ff$ is an extension of
$\Vir_{\kappa} \times \Vir_{2-\kappa^{-1}}$. 

\medskip

\item $D_{0,1} \to N_{1,0} \to N_{2,1}$: The alternative coset $\Vir_{2-\kappa^{-1}}  =
  \frac{\wh{\mathfrak{osp}}(1|2)_{2\kappa-1}}{\wh{\mathfrak{su}}(2)_{\kappa}}$. 

\medskip

\item $D_{0,1} \to N_{1,-1} \to D_{1,0}$: The $\td(2,1|-\kappa)_1$ VA is an extension of
  $\wh{\mathfrak{su}}(2)_{\kappa+1} \times \wh{\mathfrak{su}}(2)_{\kappa^{-1}+1}
  \times \tsu(2)_1$ 

\medskip

\item $D_{0,1} \to N_{1,-2} \to D_{1,0}$: $\td(2,1|-\kappa)_1$ is also an extension of $\wh{\mathfrak{su}}(2)_{\kappa+2} \times \wh{\mathfrak{osp}}(1|2)_{1
    + 2 \kappa^{-1}}$.
\end{itemize}

\medskip

All these extensions are compatible with each other.\footnote{
There are also compatibility conditions between DS reductions and
extensions. For example,
the $\DS^{(1)}$ reduction of $\td(2,1|-\kappa)_1$, i.e. $\tso(4)_1
\times \wh{\mathfrak{su}}(2)_{\kappa^{-1}}$,
is an extension of $\Vir_{\kappa+1} \times
 \wh{\mathfrak{su}}(2)_{\kappa^{-1}+1} \times \tsu(2)_1$.
 Similarly, the $\DS^{(2)}$ reduction of $\td(2,1|-\kappa)_1$,
i.e. $\tso(4)_1 \times \wh{\mathfrak{su}}(2)_{\kappa}$,
is an extension of $\wh{\mathfrak{su}}(2)_{\kappa+2} \times \DS
\left[\wh{\mathfrak{osp}}(1|2)_{1 + 2 \kappa^{-1}}\right]$.
 Another compatibility relation is the observation that the $\DS$ reduction of 
$\wh{\mathfrak{osp}}(1|2)_{\kappa}$, i.e. $\Ff \times \sVir_\kappa$, is an
extension of $\Vir_{\frac{1+\kappa^{-1}}{2}} \times
\Vir_{\frac{1+\kappa}{2}}$.}

As a preparation to restoring the global form of the gauge group, it
is useful to observe which collections of modules appear in the above
extensions.

\medskip

\begin{itemize}
\item $D_{0,1} \to N_{1,-1} \to N_{1,0}$: $\Vir_{1+\kappa^{-1}} \times \wh{\mathfrak{su}}(2)_{\kappa+1} \subset
\wh{\mathfrak{su}}(2)_{\kappa} \times \tsu(2)_1$. We can expand the right hand sides into even weight 
Weyl modules for $\wh{\mathfrak{su}}(2)_{\kappa+1}$ combined with $\Vir_{1+\kappa^{-1}}$
degenerate modules labelled by the same weight. The extension is based on $\KL_{\kappa+1}(SO(3))$. 

\medskip

\item $D_{0,1} \to N_{1,-2} \to N_{1,0}$:  $\sVir_{1+2 \kappa^{-1}} \times \wh{\mathfrak{su}}(2)_{\kappa+2}\subset
\wh{\mathfrak{su}}(2)_{\kappa} \times \tso(3)_1$. The extension is based on $\KL_{\kappa+2}(SO(3))$. 

\medskip

\item $N_{0,1} \to N_{1,0} \to N_{2,1}$: $\sVir_{2\kappa-1} \times \Ff$ is an extension of
$\Vir_{\kappa} \times \Vir_{2-\kappa^{-1}}$. Here the fields of half-integral/integral spin arise from 
products of modules labelled by odd/even weights. The extension is based on $\KL_{\kappa}(SU(2))$.

\medskip

\item $D_{0,1} \to N_{1,0} \to N_{2,1}$: $\Vir_{2-\kappa^{-1}} \times \wh{\mathfrak{su}}(2)_{\kappa}  \subset
\wh{\mathfrak{osp}}(1|2)_{2\kappa-1}$. The extension is based on $KL_{\kappa}(SU(2))$.

\medskip

\item $D_{0,1} \to N_{1,-1} \to D_{1,0}$: $\wh{\mathfrak{su}}(2)_{\kappa+1} \times \wh{\mathfrak{su}}(2)_{\kappa^{-1}+1}
  \times \tsu(2)_1 \subset \td(2,1|-\kappa)_1$. The right hand side is a sum of products of Weyl modules for the three
  current algebras with the same weight, defined modulo $2$ for $\tsu(2)_1$. We can think about the extension 
  being based on either $\KL_{\kappa^{\pm 1}+1}(SU(2))$. We also have $\wh{\mathfrak{su}}(2)_{\kappa+1} \times \wh{\mathfrak{su}}(2)_{\kappa^{-1}+1} \subset \frac{\td(2,1|-\kappa)_1}{\tsu(2)_1}$, based on 
  $KL_{\kappa^{\pm 1}+1}(SO(3))$.

\medskip

\item $D_{0,1} \to N_{1,-2} \to D_{1,0}$: $\wh{\mathfrak{su}}(2)_{\kappa+2} \times \wh{\mathfrak{osp}}(1|2)_{1
    + 2 \kappa^{-1}} \subset \td(2,1|-\kappa)_1$. The extension is based on $\KL_{\kappa+2}(SO(3))$. 
\end{itemize}

\medskip

The extensions above demonstrate some important facts about the
$KL_\kappa$ categories, seen as spin-ribbon categories:\footnote{The
  algebra objects below are really superalgebra objects, as the
  resulting extensions are super-vertex algebras. Regardless of
  whether we work with ribbon or spin-ribbon categories, all
  categories we use are assumed to be $\Z_2$-graded.}

\medskip

\begin{itemize}
\item There is an algebra object in $KL_{\kappa+1}(SO(3)) \boxtimes KL_{\kappa^{-1}+1}(SO(3))$. Equivalently, 
$KL_{\kappa}(SO(3))$ depends on $\kappa^{-1}$ modulo $1$. 

\medskip

\item There is an algebra object in $KL_{\kappa}(SU(2)) \boxtimes
  KL_{\frac{\kappa}{2 \kappa-1}}(SU(2))$. Equivalently, \linebreak
  $KL_{\kappa}(SU(2))$ depends on $\kappa^{-1}$ modulo $2$. 

\medskip

\item There is an algebra object in $KL_{\kappa+1}(SU(2)) \boxtimes
  KL_{\kappa^{-1}+1}(SU(2)) \boxtimes \tsu(2)_1\mathrm {-mod}$.

  Equivalently, for irrational $\ka$, let us define a new category
  $\widetilde{KL}_{\kappa}(SU(2))$ as a subcategory of
  $$KL_{\kappa}(SU(2)) \boxtimes \tsu(2)_1\mathrm {-mod}$$ whose simple
  objects are tensor products ${\mathbb V}_{n,\ka} \otimes L_{i,1}$
  where $n \in \Z_+, i \in \{ 0,1 \}$ and $n \equiv i \; \on{mod} \;
  2$ (here $L_{i,1}, i \in \{ 0,1 \}$, are the two simple
  $\tsu(2)_1$-modules, $L_{0,1}$ being the vacuum module). In other
  words, we ``dress'' objects in $KL_{\kappa}(SU(2))$ by objects with
  the same weight (mod 2) in $\tsu(2)_1\mathrm {-mod}$. There is a
  similar definition for rational $\ka$ as well. Then
  $\widetilde{KL}_{\kappa}(SU(2))$ to equivalent to
  $KL_{\kappa'}(SU(2))$ with $\kappa^{-1} - (\kappa')^{-1} =1$.
\end{itemize}

\medskip

The first and second facts have a natural interpretation. The category
$KL_{-\kappa^{-1}}(SO(3))$ should be equivalent to the Whittaker
category for $SU(2)$, i.e.  ${\mc C}^{SU(2)}_\ka(N_{0,1})$. The latter
is naturally invariant under $T: \kappa \to \kappa+1$.  On the other
hand, $KL_{-\kappa^{-1}}(SU(2))$ should be equivalent to ${\mc
  C}^{SO(3)}_\ka(N_{0,1})$, which is only invariant under $T^2: \kappa
\to \kappa+2$. Indeed, we have $n(SU(2))=1$ and $n(SO(3))=2$.

The third fact, instead, should be thought of as giving a hint about the nature of 
$N_{1,-1}^{SO(3)}$. Recall that this boundary condition cannot be obtained from 
$N_{1,0}^{SO(3)}$ by a $T$ transformation, as that is not a valid duality. 
Instead, it must be {\it defined} as the $S$ image of $N_{1,1}^{SU(2)}$. 
In particular, we have 
\begin{equation}
{\mc C}^{SO(3)}_\ka(N_{1,-1}) = \widetilde{KL}_{\kappa+1}(SU(2))
\end{equation}

The following analysis requires more physics background than the rest
of the paper. In the $SU(2)$ gauge theory, the boundary conditions
$N^{SU(2)}_{1,q}$ can be given a direct definition as Neumann boundary
conditions modified by a Chern--Simons coupling.  In the $SO(3)$ theory
one would expect a similar direct definition of $N^{SO(3)}_{1,q}$.
However, the required Chern--Simons coupling is only well-defined for
even $q$.

The construction of $\widetilde{KL}_{\kappa+1}(SU(2))$ as a sub-category in 
$KL_{\kappa+1}(SU(2)) \times \tsu(2)_1\mathrm {-mod}$ suggests a direct 
construction of $N^{SO(3)}_{1,2k+1}$ in four steps: 

\medskip

\begin{enumerate}
\item Extend the gauge group from $SO(3)$ to $SU(2)$ at the boundary. 

\medskip

\item Add $2k+1$ units of boundary Chern--Simons coupling for the bulk gauge fields. 
\medskip

\item Add an extra 3d TFT at the boundary: the TFT associated to $U(1)_2$ or 
$SU(2)_1$ Chern--Simons theory. 

\medskip

\item Restrict the boundary lines to the combinations of the original boundary Wilson lines 
and $SU(2)_1$ Wilson lines of the same weight (mod 2). This can be done by ``gauging a 
${\mathbb Z}_2$ one-form symmetry'', the non-anomalous combination of the 
boundary ${\mathbb Z}_2$ one-form symmetry associated to the 
extension of the boundary gauge group and the ${\mathbb Z}_2$ one-form symmetry
of $SU(2)_1$ Chern--Simons theory. 
\end{enumerate}

\medskip

This direct physical construction helps understand the physical origin of some of the junctions we will encounter below. 

\subsection{More details on modules}
Recall that an $\wh{\mathfrak{su}}(2)_{\kappa}$ Weyl module of highest
weight $m$ has a highest weight vector of conformal dimension
\begin{equation}
\Delta^{\wh{\mathfrak{su}}(2)_{\kappa}}_m = \frac{m(m+2)}{4 \kappa}
\end{equation}
This makes it manifest that the topological twist $2 \pi \Delta$ is
invariant for $\kappa^{-1} \to \kappa^{-1}+1$ if the weight is even,
but not if the weight is odd.

The topological twist will be invariant under $\kappa^{-1} \to
\kappa^{-1}+2$ for all weights, but only in the spin sense. As the
dimensions of the highest weight vector of the $\tsu(2)_1$ module
$L_{1,1}$ of highest weight $1$ is $\frac14$, we see how it can be
used to compensate the effect of $\kappa^{-1} \to \kappa^{-1}+1$ on
Weyl modules of odd weight.

The objects of the category $\Vir_\kappa$-mod that are images of the objects of
$$
{\mc C}^{SU(2)}_\ka(N_{0,1}) \boxtimes {\mc
  C}^{SU(2)}_\ka(N_{1,0}) \simeq \KL_{\ka^{-1}}(SO(3))
\boxtimes \KL_\ka(SU(2))
$$
under the functor $F^{SU(2)}_\ka(N_{0,1}\sto N_{1,0})$ should be
given by the fully degenerate modules for Virasoro algebra with
conformal dimensions
\begin{equation}
\Delta^{\Vir}_{m,e} = (- \frac{e}{2b}-\frac{m b}{2})(b+b^{-1} +
\frac{e}{2b}+\frac{m b}{2}) = \frac{e(e+2)}{4 \ka} + \frac{m(m+2)}{4}
\ka - \frac{e m + e + m}{2}.
\end{equation}
with arbitrary $e$ and even $m$. The half-integral shifts compared
with the dimensions of Weyl modules are OK as long as we work with
spin-ribbon categories.

Notice that the full collection of modules, with arbitrary weights $e$
and $m$, does {\it not} give an image of
$\KL_{\ka^{-1}}(SU(2))\boxtimes \KL_\ka(SU(2))$, because of the
half-integral shifts which occur when $e$ and $m$ are both odd.

Finally, because of the negative shifts in the dimensions of Virasoro
modules, the combination of junctions supporting Virasoro algebra is
more likely to give junction vertex algebras with negative conformal
dimensions compared with the combination of junctions supporting
Kac--Moody algebras.

Similar considerations as above apply to categories of modules over
$\wh{\mathfrak{osp}}(1|2)_{2\kappa-1}$ and $\sVir_{2\kappa-1}$. 

The extension construction guarantees the existence of a family of modules over $\wh{\mathfrak{osp}}(1|2)_{2\kappa-1}$ which realize $\KL_{2 - \kappa^{-1}}(SO(3))$. Concretely, we are combining $\Vir_{2 - \kappa^{-1}}$ modules with weights $(2e,m)$ and Weyl modules of $\wh{\mathfrak{su}}(2)_{\kappa}$
with weights $m$. The corresponding dimensions are 
\begin{equation}
\Delta^{\wh{\mathfrak{osp}}(1|2)_{2\kappa-1}}_{e;m} = \frac{e(e+1)\ka}{(2 \ka -1)} + \frac{(m-2e)(m+1)}{2}.
\end{equation}
which is smallest for $m=e$, with conformal dimension of highest weight vector  
\begin{equation}
\Delta^{\wh{\mathfrak{osp}}(1|2)_{2\kappa-1}}_{e;m} = \frac{e(e+1)}{2(2 \ka -1)}.
\end{equation}
These can be identified with Weyl modules of $\wh{\mathfrak{osp}}(1|2)_{2\kappa-1}$.

If we build the extension using only modules in $\KL_\ka(SU(2))$, the
result is the even part $(\wh{\mathfrak{osp}}(1|2)_{2\kappa-1})_e$ of
$\wh{\mathfrak{osp}}(1|2)_{2\kappa-1}$ (in the sense of Lie
superalgebras). Then we can realize the full $\KL_{2 -
  \kappa^{-1}}(SU(2))$ as a category of modules over
$(\wh{\mathfrak{osp}}(1|2)_{2\kappa-1})_e$.

In a similar manner, we can build modules for $\sVir_{2\kappa-1}$
associated to $\KL_{\kappa^{-1}}(SO(3)) \times
\KL_{2-\kappa^{-1}}(SO(3))$. These are well-known degenerate modules
for the super-Virasoro algebra. If we consider the even
subalgebra $(\sVir_{2\kappa-1} \times \Ff)_e$, we can realize the
larger category $\KL_{\kappa^{-1}}(SU(2)) \boxtimes
\KL_{2-\kappa^{-1}}(SU(2))$.

One final observation concerns the embedding $\tsu(2)_1 \boxtimes \tsu(2)_1 \subset \tso(4)_1 =\Ff^4$. It demonstrates the presence of an algebra object in $\tsu(2)_1\mathrm{-mod} \boxtimes \tsu(2)_1\mathrm{-mod}$, seen as spin-ribbon categories. 

There are many holomorphic vertex algebras and spin-vertex algebras
which have a $\tsu(2)_1$ subalgebra, and which thus lead to coset
algebras $V$ such that $V\mathrm{-mod}$ is canonically conjugate to
$\tsu(2)_1\mathrm{-mod}$, as a spin-ribbon or as a ribbon
category.\footnote{A simple class of examples arises from lattice
  vertex algebras associated to (even) unimodular lattices $L$ with a
  vector $v$ of length 2.}

\subsection{Choices of global form}    \label{global}

In this subsection, we give a more detailed analysis, in which we
distinguish between $SU(2)$ and $SO(3)$ while still ignoring the
subtleties related to spin structures.

\subsubsection{Variants of $N_{0,1} \to N_{1,0}$}
The Virasoro algebra $\Vir_\ka$ arises naturally at the junction
$N_{0,1} \to N_{1,0}$ both in $SU(2)$ and in $SO(3)$ gauge
theory. Recall that $\Vir_\ka$ is invariant under $\ka\to
\ka^{-1}$. This transformation corresponds to the duality symmetry
$RS$ which exchanges $SU(2)$ and $SO(3)$.  The difference between the
two choices of gauge group affects the categories of boundary line
defects and hence the corresponding modules $\Vir_\ka$.

An $SU(2)$ gauge theory will have the electric line defects on
$N_{1,0}$ with all possible dominant integral weights and magnetic
line defects on $N_{0,1}$ with even dominant integral weights only.
The corresponding families of modules in $\Vir_\ka$ are mutually local
(indeed, the modules in one family braid trivially with the modules of
the other family and the fusion product of two simple modules of this
kind is again a module of this kind) and so we indeed have a tensor functor
$$
\KL_\ka(SU(2)) \boxtimes \KL_{\ka^{-1}}(SO(3)) \to \Vir_\ka\on{-mod}
$$
The reverse set-up (with electric and magnetic line defects exchanged)
is true for the $SO(3)$ gauge theory, and we have
$$
\KL_\ka(SO(3)) \boxtimes \KL_{\ka^{-1}}(SU(2)) \to \Vir_\ka\on{-mod}
$$
(both as spin-ribbon categories). Notice that modules of odd weights
in the two families braid with a $-1$ sign. In particular, there is no
functor from $\KL_\ka(SU(2)) \boxtimes \KL_{\ka^{-1}}(SU(2))$ to
$\Vir_\ka-\on{mod}$, even as spin-ribbon categories.

The action of $RST^{-1}$ on $N^{SU(2)}_{0,1} \to N^{SU(2)}_{1,0}$ 
gives $N^{SO(3)}_{1,-1} \to N^{SO(3)}_{1,0}$. If the bulk coupling 
for the latter is $\kappa$, then the junction vertex algebra will be 
$\Vir_{1+\ka^{-1}}$. Correspondingly, there is a natural functor 
from $\KL_{-\ka-1}(SO(3)) \boxtimes \KL_{\ka}(SO(3))$ to $\Vir_{1+\ka^{-1}}$-mod.

On the other hand, there is no functor from
$\KL_{-\ka-1}(SU(2)) \boxtimes \KL_{\ka}(SU(2))$ to $\Vir_{1+\ka^{-1}}$-mod.
This fails in two ways: the dimensions (modulo $\frac12$) of families
of modules of $(odd,even)$ or $(even,odd)$ weights differ from the
expected values by $-\frac14$ and furthermore their mutual braiding
has the wrong sign.

This is not surprising as a $N^{SU(2)}_{1,-1} \to N^{SU(2)}_{1,0}$ 
junction cannot be obtained from the
basic junction $N^{SU(2)}_{0,1} \to N^{SU(2)}_{1,0}$ by any legitimate
duality transformation within the $SU(2)$ gauge theory: indeed, $S
T^{\pm 1} S$ is not a legitimate duality symmetry, only $S T^{\pm 2}
S$ and their powers are.

We can seek an appropriate junction in two ways: direct gauge theory construction 
or composition of junctions.

\medskip

\begin{enumerate}
\item The physical construction of the junction requires one to place by hand some extra
degrees of freedom at the junction, in the form of an holomorphic spin vertex algebra
which includes $SU(2)$ currents of level $1$.\footnote{Coupling these to the
gauge fields at the junction, one absorbs the gauge anomaly associated
to the change in the Chern--Simons levels across the junction.} We can take four real fermions as auxiliary degrees of freedom, i.e. $\tso(4)_1$. The junction vertex algebra is thus the coset 
\begin{equation}    \label{Vir su2}
\frac{\wh{\mathfrak{su}}(2)_{\ka} \times
  \tso(4)_1}{\wh{\mathfrak{su}}(2)_{\ka+1}}\simeq \Vir_{1+\ka^{-1}} \times \tsu(2)_1.
\end{equation}

\medskip

\item We can map the problem by $T^2RS$ to the construction of a junction
$N^{SO(3)}_{0,1} \to N^{SO(3)}_{1,1}$ at coupling $2+\ka^{-1}$, and use the composition 
$N^{SO(3)}_{0,1} \to N^{SO(3)}_{1,0} \to N^{SO(3)}_{1,1}$. 
The result is an extension of $\Vir_{2+\ka^{-1}} \times \Vir_{\frac{\ka+1}{2\ka+1}}$.
It can be identified with
\begin{equation}  
\Vir_{1+\ka^{-1}} \times \widetilde{\tsu}(2)_1.
\end{equation}
where $\widetilde{\tsu}(2)_1$ is the same vertex algebra as $\tsu(2)_1$, with a modified choice of 
stress tensor. 
\end{enumerate}

\medskip

The two answers are essentially equivalent. 

Simple modules over $\Vir_{1+\ka^{-1}} \times \tsu(2)_1$  corresponding to a pair of
Weyl modules ${\mathbb V}_{m,\ka}$ and ${\mathbb V}_{n,-\ka-1}$ from
the categories of boundary line defects ${\mc
  C}^{SU(2)}_\ka(N^{SU(2)}_{1,-1})^\vee = \KL_{-\ka-1}(SU(2))$ and
${\mc C}^{SU(2)}_\ka(N^{SU(2)}_{1,0}) = \KL_\ka(SU(2))$ are produced
in the coset description on the right hand side of \eqref{Vir su2}
 as the BRST cohomology of the tensor product ${\mathbb
  V}_{m,\ka} \otimes V \otimes {\mathbb V}_{n,-\ka-1}$ under the
diagonal action of $\wh{\mathfrak{su}}(2)$ of twice the critical
level.

If their weights, $m$ and $n$, have the same parity, then the
coset module coincides with the appropriate $\Vir_\kappa$-module
dressed by the vacuum module of $\tsu(2)_1$ (or $V$). But if the
weights have opposite parity, then the coset module is a
$\Vir_\kappa$-module dressed by the non-trivial module of $\tsu(2)_1$
(or $V$). This gives us the desired functor
$$
\KL_{-\ka-1}(SU(2)) \boxtimes \KL_\ka(SU(2))
 \to \left(\Vir_{1+\ka^{-1}} \times V
\right)\on{-mod}.
$$
Note that the extra dressing cancels the troublesome $-\frac14$ shifts
in conformal dimensions that would result in the undesired $-$ sign
in the mutual braiding.

A small aside: the junctions above are compatible with the 
tentative direct gauge theory description of $N^{SO(3)}_{1,\pm 1}$
involving an auxiliary 3d TFT at the boundary. As we construct a junction 
involving $N^{SO(3)}_{1,\pm 1}$, we need in particular to provide a boundary condition 
for the 3d TFT. 

Such a boundary condition naturally supports a $\tsu(2)_1$ vertex algebra. 

\medskip

\begin{enumerate}
\item In a direct gauge theory construction of $N^{SO(3)}_{1,-1} \to
N^{SO(3)}_{1,0}$, we can use such $\tsu(2)_1$ as auxiliary degrees of freedom at the junction, 
rather than an holomorphic vertex algebra such as $\tso(4)_1$. That gives precisely the GKO construction 
\begin{equation} 
\frac{\wh{\mathfrak{su}}(2)_{\ka} \times
  \tsu(2)_1}{\wh{\mathfrak{su}}(2)_{\ka+1}}\simeq \Vir_{1+\ka^{-1}}.
\end{equation}

\medskip

\item In a direct gauge theory construction of $N^{SO(3)}_{0,1} \to N^{SO(3)}_{1,-1}$,
the $\tsu(2)_1$ vertex algebra goes along for the ride, giving the expected $\Vir_{1+\ka} \times \tsu(2)_1$.
\end{enumerate} 

\begin{table}[h]
\begin{center}
\begin{tabular}{|c||c||c|}
\hline
{\em Junction} & {\em $SU(2)$}& {\em $SO(3)$}  \\
\hline \hline
$N_{0,1} \to N_{1,0}$& $\Vir_\kappa$& $\Vir_\kappa$  \\ 
\hline
$N_{0,1} \to N_{1,-1}$ & $\Vir_{1+\kappa}$& $\Vir_{1+\kappa}\times \tsu(2)_1$  \\
\hline
$N_{1,-1} \to N_{1,0}$& $\Vir_{1+\ka^{-1}}\times \tsu(2)_1$& $\Vir_{1+\ka^{-1}}$ \\
\hline
\end{tabular} 
\end{center}
\vspace*{5mm}
\caption{Global form of the gauge group and junctions, part 1.}\label{tab:four}
\end{table}

\subsubsection{Variants of $N_{0,1}\to N_{2,1}$. }
The simplest junction of this type occurs for $N^{SO(3)}_{1,-2}\to N^{SO(3)}_{1,0}$ 
and its duality images, such as $N^{SU(2)}_{0,1}\to N^{SU(2)}_{2,\pm 1}$, etc. In the former conventions, it supports a $\sVir_{1+2 \kappa^{-1}}$ vertex algebra, though in many constructions 
it emerges dressed by an extra decoupled free fermion. This is the case, 
for example, of the composition $N^{SU(2)}_{0,1}\to N^{SU(2)}_{1,0}\to N^{SU(2)}_{2,1}$.

A direct gauge theory construction of a $N^{SO(3)}_{1,-2}\to N^{SO(3)}_{1,0}$
junction requires a choice of auxiliary holomorphic spin-vertex algebra 
with a $\tsu(2)_2$ sub-algebra.\footnote{This is required in
order to compensate for the shift of boundary Chern--Simons coupling
(note that the level of the $\wh{\mathfrak{su}}(2)$ Kac--Moody algebra
in $\tso(3)_1$ is equal to $2 = n(SO(3))$).} 
The simplest choice is $\tso(3)_1$, which leads to 
a junction vertex algebra given by the 
super-GKO-like coset
$$
\sVir_{1+2 \kappa^{-1}} =
\frac{\wh{\mathfrak{su}}(2)_{\kappa} \times
  \tso(3)_1}{\wh{\mathfrak{su}}(2)_{\kappa+2}}
$$
The lines of even weight on both boundaries end on
appropriate coset modules, which are the degenerate modules of 
$\sVir$.

Another interesting junction is $N^{SO(3)}_{0,1}\to N^{SO(3)}_{2,1}$.
Junction composition $N^{SO(3)}_{0,1}\to N^{SO(3)}_{1,0} \to N^{SO(3)}_{2,1}$
now produces the even subalgebra $(\sVir_{2\kappa-1} \times \Ff)_e$. 

Notice that the duality orbit of this junction does ${\it not}$ include 
$N^{SO(3)}_{0,1}\to N^{SO(3)}_{2,-1}$, which will require an alternative construction. 
Instead, it includes $N^{SU(2)}_{1,2}\to N^{SU(2)}_{1,0}$, $N^{SU(2)}_{1,1}\to N^{SU(2)}_{1,-1}$
and $N^{SO(3)}_{1,-1}\to N^{SO(3)}_{1,1}$.

The latter provides another perspective on the junction: the boundary conditions for the 
3d TFTs involved in $N^{SO(3)}_{1,-1}$ and $N^{SO(3)}_{1,1}$ can provide 
$\tsu(2)_1 \times \tsu(2)_1$ auxiliary degrees of freedom, resulting in the coset description
$$
( \sVir_{1+2 \kappa^{-1}}  \times \Ff)_e=
\frac{\wh{\mathfrak{su}}(2)_{\kappa} \times
 \tsu(2)_1 \times \tsu(2)_1}{\wh{\mathfrak{su}}(2)_{\kappa+2}}
$$

In order to study the final duality orbit, including $N^{SO(3)}_{0,1}\to N^{SO(3)}_{2,-1}$, 
we can look at $N^{SU(2)}_{1,-2}\to N^{SU(2)}_{1,0}$. A direct gauge theory construction 
requires an auxiliary holomorphic vertex algebra which has a $\tsu(2)_2$ sub-algebra, 
but also includes fields of odd weight, so that all boundary lines can end at the 
junction and we can find a fully faithful functor from the boundary categories into 
the junction vertex algebra modules. 

There are many choice of such a vertex algebra, and none is obviously canonical.
We can realize $\tsu(2)_2$ as the diagonal combination of two $\tsu(2)_1$'s, 
each included in a separate $\tso(4)_1$ factor. 

We can describe the resulting junction vertex algebra as a coset
\begin{equation}
\frac{\wh{\mathfrak{su}}(2)_{\kappa} \times
  \tso(8)_1}{\wh{\mathfrak{su}}(2)_{\kappa+2}}
\end{equation}
where $\wh{\mathfrak{su}}(2)$ is embedded in the $\tso(8)_1$ in a
non-trivial way, acting separately on two blocks of four fermions each.
With a bit of care, we can simplify that to
\begin{equation}
( \sVir_{1+2 \kappa^{-1}}  \times \tso(5))_e
\end{equation}

We can obtain the same result by a composition
$N^{SU(2)}_{1,-2}\to N^{SU(2)}_{1,-1}\to N^{SU(2)}_{1,0}$ leading to
an extension of $\Vir \times \tsu(2)_1 \times \Vir \times \tsu(2)_1$. 

\begin{table}[h]
\begin{center}
\begin{tabular}{|c||c||c|}
\hline
{\em Junction} & {\em $SU(2)$}& {\em $SO(3)$}  \\
\hline \hline
$N_{1,-2}\to N_{1,0}$& $( \sVir_{1+2 \kappa^{-1}}  \times \tso(5)_1)_e$ & $\sVir_{1+2 \kappa^{-1}}$  \\ 
\hline
$N_{1,2}\to N_{1,0}$& $( \sVir_{-1+2 \kappa^{-1}}  \times \Ff)_e$& $\sVir_{-1+2 \kappa^{-1}}$  \\ 
\hline
$N_{1,-1} \to N_{1,1}$ & $( \sVir_{\frac{\ka-1}{\ka+1}} \times \tso(5)_1)_e$& $( \sVir_{\frac{\ka-1}{\ka+1}}  \times \Ff)_e$  \\
\hline
\end{tabular} 
\end{center}
\vspace*{5mm}
\caption{Global form of the gauge group and junctions, part 2.}\label{tab:five}
\end{table}

\subsubsection{Variants of $D_{0,1} \to N_{1,0}$}

We know that both $D^{SU(2)}_{0,1} \to N^{SU(2)}_{1,0}$ and $D^{SO(3)}_{0,1} \to N^{SO(3)}_{1,0}$
support the Kac--Moody algebra $\wh{\mathfrak{su}}(2)_{\kappa}$. Upon compactification, 
these junctions produce the expected ${\mc D}_{\ka}$ in $D_{\ka}(\Bun_{SL_2})$ or in $D_{\ka}(\Bun_{PSL_2})$.

Very similar considerations apply to $D^{SU(2)}_{0,1} \to N^{SU(2)}_{1,k}$. These junctions support 
$\wh{\mathfrak{su}}(2)_{\kappa-k}$. Upon compactification, 
these junctions produce ${\mc D}_{\ka-k} \otimes {\mc L}^k$ in $D_{\ka}(\Bun_{SL_2})$.

The same is true for $D^{SO(3)}_{0,1} \to N^{SO(3)}_{1,2 k}$. These junctions support 
$\wh{\mathfrak{su}}(2)_{\kappa-2k}$. Upon compactification, 
these junctions produce ${\mc D}_{\ka-2k} \otimes {\mc L}^{2k}$ in $D_{\ka}(\Bun_{PSL_2})$.

A separate class of interesting junctions is $D^{SO(3)}_{0,1} \to N^{SO(3)}_{1,2 k-1}$,
which cannot be produced by the action of $T^2$ on the standard junction. Instead, we can 
consider the extension $D^{SO(3)}_{0,1} \to N^{SO(3)}_{1,0}\to N^{SO(3)}_{1,1}$, 
leading to $\wh{\mathfrak{su}}(2)_{\kappa-1} \otimes \tsu(2)_1$, coupled to $PSL_2$ bundles 
by the diagonal combination of the currents. 

Upon compactification, this junction produces an interesting object in $D_{\ka}(\Bun_{PSL_2})$,
which should be qGL dual to ${\mc D}_{-\ka^{-1}+1} \otimes {\mc L}^{-1}$ in $D_{-\ka^{-1}}(\Bun_{SL_2})$.

\subsubsection{Variants of $D_{0,1} \to N_{2,1}$}

First, we can look at $D^{SU(2)}_{0,1} \to N^{SU(2)}_{2,\pm 1}$. This
should support $\wh{\mathfrak{osp}}(1|2)_{2\ka \mp 1}$. Upon compactification, 
this junction should produce a rather non-trivial object: the restriction of ${\mc
  D}_{2\ka-1}[\Bun_{OSp_{1|2}}]$ to $\Bun_{SL_2}$.

We thus expect this object to be qGL dual to ${\mc D}_{-\ka^{-1}\pm 2}
\otimes {\mc L}^{\mp 2}$ in $D_{-\ka^{-1}}(\Bun_{PSL_2})$.

For $SO(3)$ gauge group, we should distinguish two junctions not related by dualities:
\begin{enumerate}
\item $D^{SO(3)}_{0,1} \to N^{SO(3)}_{2,1}$: the composition of junctions involving $N^{SO(3)}_{1,0}$
only gives the (Grassmann) even part $(\wh{\mathfrak{osp}}(1|2)_{2\ka - 1})_e$.
\item $D^{SO(3)}_{0,1} \to N^{SO(3)}_{2,-1}$: the composition of junctions involving $N^{SO(3)}_{1,-1}$
gives instead $(\wh{\mathfrak{osp}}(1|2)_{2\ka +1} \times \tso(4)_1)_e$.
\end{enumerate}

\subsubsection{Kernel junctions}
Now we have the information needed to study a junction
$D^{SU(2)}_{1,0} \to D^{SU(2)}_{1,0}$ or, equivalently,
$D^{SO(3)}_{0,1} \to D^{SO(3)}_{1,0}$. We can use the composition
$D_{0,1}^{SU(2)} \to N_{1,-1}^{SU(2)} \to D_{1,0}^{SU(2)}$. Now the
second junction has an extra $\tsu(2)_1$ factor and the composition
gives precisely $\td(2,1|-\ka)_1$.

Crucially, the coupling to $SL_2$ bundles associated to
$D_{0,1}^{SU(2)}$ employs the $\wh{\mathfrak{su}}(2)_{\ka+1}$
sub-algebra, but the coupling to $PSL_2$ bundles associated to
$D_{1,0}^{SU(2)}$ employs the diagonal sub-algebra of the
$\wh{\mathfrak{su}}(2)_{\ka^{-1}+1} \times \tsu(2)_1$ sub-algebra.

This observation is very important: the currents in $\td(2,1|-\ka)_1$
have either (even, even, even) or (odd, odd,odd) weights for the three
subalgebras. Under the diagonal sub-algebra of the
$\wh{\mathfrak{su}}(2)_{\ka^{-1}+1} \times \tsu(2)_1$, though, they
have even weights and thus the coupling to $PSL_2$ bundles is
possible.

The sheaf of coinvariants of $\td(2,1|-\ka)_1$ is an object in
$D_{\ka+1}(\Bun_{SL_2}) \otimes D_{\ka^{-1}+2}(\Bun_{PSL_2})$ which can be
mapped canonically (up to spin subtleties) to an object in
$D_{\ka}(\Bun_{SL_2}) \otimes D_{\ka^{-1}}(\Bun_{PSL_2})$, which is the
conjectural qGL kernel.

An interesting variant is the junction $D^{SO(3)}_{0,1} \to
D^{SO(3)}_{1,1}$. We can use the composition $D_{0,1}^{SO(3)} \to
N_{1,0}^{SO(3)} \to D_{1,1}^{SO(3)}$. The second junction can be
dualized to $N_{0,1}^{SU(2)} \to D_{1,-1}^{SU(2)}$ and then to
$N_{0,1}^{SU(2)} \to D_{1,0}^{SU(2)}$ and thus $D_{0,1}^{SO(3)} \to
N_{1,0}^{SO(3)}$. The composition gives a smaller vertex algebra: the
coset $\frac{\td(2,1|-\ka)_1}{\tsu(2)_1}$.

\subsection{Spin subtleties}    \label{spin sub}

Finally, we include spin structures in our analysis. In the absence of
a spin structure, the $SU(2)$ gauge theory is still $T$ invariant. It
comes in two variants, though, as we can define a standard theory
$SU(2)_b$ based on an $SU(2)$ connection and a twisted theory
$SU(2)_s$ based on an $\mathrm{Spin}_\C$-like $SU(2)$ connections.

On the other hand, $SO(3)$ gauge theory is invariant under $T^4$
rather than $T^2$. We can denote the corresponding two versions of
$SO(3)$ gauge theory as $SO(3)_b$ and $SO(3)_t$, related by $T^2$.

Perhaps surprisingly, it is natural to take $SO(3)_t$ and $SO(3)_b$
to be respectively the $S$-dual images of $SU(2)_b$ and $SU(2)_s$. 
We thus have the following duality groupoid:
 
\begin{equation}
\begin{matrix} SU(2)_b & \longleftrightarrow & SO(3)_t & \longleftrightarrow &
SO(3)_b & \longleftrightarrow & SU(2)_s \cr \circlearrowright & S &&T^2&&S&\circlearrowright 
\cr T & &&&&& T\end{matrix}
\end{equation}

In this convention, Neumann boundary condition $N_{1,0}$ does not require
a spin structure for $SO(3)_b$ but does require it for
$SO(3)_t$. Indeed, the Nahm pole boundary condition for $SU(2)_b$
requires a choice of spin structure, but does not for $SU(2)_s$.

On the other hand, Neumann boundary condition $N_{1,0}$ does not
require a choice of spin structure for either $SU(2)$ theory, nor does
Nahm for either $SO(3)$ theory.

Acting with $T$, we find that $N_{1,\pm 1}$ do not require a choice of
spin structure for either $SU(2)$ theory and thus should not require a
spin structure for either $SO(3)$ theory as well.

On the other hand, acting with $T^2$ we find that $N_{1,\pm 2}$
require a spin structure for $SO(3)_b$ but not for $SO(3)_t$.  Acting
with $S$, we find that $N_{2,\pm 1}$ require a
spin structure for $SU(2)_s$ but not for $SU(2)_b$.  This makes sense:
the boundary hypermultiplets are twisted into fields of integral spin,
but transform as a doublet of $SU(2)$.  They are not naturally
sections of $\mathrm{Spin}_\C$-like $SU(2)$ bundle. On the other hand,
the $N_{2,\pm 1}$ should not require a spin structure for either
$SO(3)$ theory.

Standard Dirichlet boundary conditions only require a spin structure
for $SU(2)_s$, and even then it is better to think about coupling to a
background $\mathrm{Spin}_\C$-like $SU(2)$ bundle at the boundary.

Upon compactification on a Riemann surface, we can map the four nodes
of our groupoid to four variants of the usual category of $D$-modules:

\medskip

\begin{itemize}
\item $SU(2)_b \to D_{\ka}(\Bun_{SL_2})$, standard twisted $D$-modules.

\medskip

\item $SO(3)_b \to D_{\ka}(\Bun_{PSL_2})$, standard twisted $D$-modules.

\medskip

\item $SO(3)_t \to {\mc G}-D_{\ka}(\Bun_{PSL_2})$, twisted $D$-modules
  modified by the gerbe ${\mc G}$.

\medskip

\item $SU(2)_s \to D_{\ka}(\Bun_{\mathrm{Spin}-SL_2})$, twisted $D$-modules on a
  modified moduli stack, that of $\mathrm{Spin}_{SL_2}$ bundles
  on the Riemann surface $X$ (we use notation from Section
  \ref{gerbes}).
\end{itemize}

\medskip

This agrees with the duality statements in \cite{FW2}.

As an example of non-spin junctions, the $N^{SO(3)_b}_{0,1} \to
N^{SO(3)_b}_{1,0}$ or $N^{SU(2)_s}_{0,1} \to N^{SU(2)_s}_{1,0}$
junctions can be defined with no reference to spin structure and
indeed support the standard $\Vir_\ka$ algebra.

It is straightforward, if tedious, to build more examples of junctions
which do not require a spin structure.

Instead, we will just make a couple of observations about the duality
kernel vertex algebra.

\medskip

\begin{itemize}
\item The conformal blocks of $\td(2,1|-\ka)_1$ will give an object in
  $D_{\ka+1}(\Bun_{SL_2}) \otimes D_{\ka^{-1}+2}(\Bun_{PSL_2})$.
Because ${\mc L}^2$ is a section of the gerbe ${\mc G}$, this maps to an object in $D_{\ka}(\Bun_{SL_2}) \otimes {\mc G}-D_{\ka^{-1}}(\Bun_{PSL_2})$,
as needed. 

\medskip

\item Consider the subalgebra $(\td(2,1|-\ka)_1 \times \tso(3)_1)_e$. This vertex algebra can be coupled to $\mathrm{Spin}_{SL_2}$ bundles. If we couple the second factor to the $PSL_2$ bundles, the conformal blocks will give an object in $D_{\ka+1}(\Bun_{s-SL_2}) \otimes D_{\ka^{-1}+4}(\Bun_{PSL_2})$.
This maps canonically to an object in $D_{\ka}(\Bun_{s-SL_2}) \otimes D_{\ka^{-1}}(\Bun_{PSL_2})$, as needed. 
\end{itemize}

\section{General gauge groups and discrete $\theta$ angles} \label{sec:topo}

Until now, we have employed a somewhat traditional perspective on the
S-duality groupoid, in which we focus on the standard 4d
supersymmetric gauge theories for a simple compact Lie group $G_c$
and its Langlands dual $\LG_c$.

The lift of the duality group from spin-TFTs to TFTs for the groups
$U(1)$ and $SU(2)/SO(3)$ forced us to consider some generalizations of
that structure, either involving $\mathrm{Spin}_{\C}$-like
modifications of the gauge group or the images of the standard gauge
theories under the duality transformation $T$ (we will call them
$T$-images).

Even if we remain in the realm of spin-TFTs, it turns out that
extending the duality groupoid by introducing $T$-images of standard
gauge theories, however useful, is still incomplete \cite{GMN,AST}. It turns out that
the definition of 4d gauge theories can be modified by a variety of
``discrete $\theta$ angles'' and some of the resulting modifications
are {\it not} simply the $T^k$-images of standard gauge
theories. Furthermore, there is a full duality groupoid generated by
$T$ and $S_m$, whose nodes can be identified with a variety of thus
modified gauge theories.

The full duality groupoid organizes all theories associated to the Lie
algebras $\mathfrak{g}$ and $\lg$ into a complicated pattern of
orbits, whose structure depends sensitively on the specific gauge Lie
algebra.

When such modified 4d gauge theories are compactified on a Riemann
surface $\Sigma$, we obtain as the result certain modifications of the
sigma model with the target ${\mc M}_H(G)$, the Hitchin moduli space
associated to $\Sigma$ and $G$. As a result, the corresponding
categories of twisted $D$-modules on $\Bun_G$ get modified in such a
way that each connected component of $\Bun_G$ gets equipped with an
appropriate discrete $B$-field and the corresponding categories of
$D_\ka$-modules get twisted by appropriate gerbes. As a result, we
obtain a variety of duality relations involving these gerbe-twisted
categories of $D_\ka$-modules.

This overall structure is reasonably well understood for the physical
gauge theories, but has to be further refined for the topologically
twisted gauge theories, because of spin subtleties. 

There is some recent mathematical work \cite{GL,gaitsP} which
describes modifications of the standard Geometric Langlands
correspondence that appear to be closely related to the modifications
due to the discrete $\theta$ angles. It would be very
interesting to make the dictionary explicit and compare the expected
actions of S-duality on these data.

\subsection{Topological actions and lattice of line defects}

In order to consider topological modifications of gauge theory actions
(a.k.a. ``discrete $\theta$-angles'') in full generality, one has to
study certain generalized cohomology theories of $BG$.

In the standard physical gauge theory, these generalized cohomology
theories will be associated to cobordisms of spin manifolds equipped
with a $G$ bundle and perhaps an $SU(4)_R$-bundle.

Alternatively, as the spinors in the physical theory transform in the
fundamental representation of the $SU(4)_R\simeq Spin(6)_R$ R-symmetry
group, while fields of integral spin transform in the representations
of $SO(6)_R$, it should be possible to couple the theory to
$\mathrm{Spin}_{SU(4)_R}$-bundles, i.e. bundles on a four-manifold $M$
with the structure group $(Spin(4) \times SU(4))/\Z_2$ (which is a
$\Z_2$-extension of the structure group $SO(4) \times (SU(4)/\Z_2) =
SO(4) \times SO(6)$), such that the corresponding $SO(4) \times
(SU(4)/\Z_2)$-bundle has the frame bundle of $M$ along the first
factor. These should be available on all manifolds.  Topological
actions for this variant of the physical theories will be classified
by some cohomology theories associated to cobordisms of manifolds
equipped with a $G$-bundle and an $\mathrm{Spin}_{SU(4)_R}$-bundle.

The latter choice is probably better in preparation for a topological
twist. It suggests that the spin-refined duality action, which we will
explore later on, may already be probed within the physical theory.

Upon topological twist, the classification of discrete $\theta$-angles
should simplify a bit, as the cohomology theories will be associated
to cobordisms of manifolds equipped with a $G$-bundle only and should
coincide with some standard version of the group cohomology
$H^4(BG,U(1))$.

The group cohomology description of the topological action is
well-suited for understanding the compactification to the 2d sigma
model on a Riemann surface $\Sigma$ with the target Hitchin moduli
space ${\mc M}_H(G)$. Indeed, there is an obvious map $\Sigma \times
\Bun_G(\Sigma) \to BG$ and the pull-back of the class in
$H^4(BG,U(1))$ along that map can be integrated over $\Sigma$ to give
a class in $H^2(\Bun_G(\Sigma),U(1))$, i.e. a discrete $B$-field, or
equivalently a gerbe, of the sigma model.\footnote{Notice that this
  statement holds true for whatever generalized cohomology theory is
  relevant to the problem at hand: a topological action for the gauge
  theory will descend to a topological action for the 2d sigma model.}

Let us denote a choice of the theory as a pair $(G, \omega)$ with
$\omega$ being a choice of topological action added on top of the
usual gauge theory action. An important open question is to describe
explicitly the action of the duality groupoid on pairs $(G,
\omega)$. This has not yet been done, as far as we
know.\footnote{While some of the spin subtleties are expressed in
  $\omega$, to get a complete description of the possible theories one
  needs to allow $\mathrm{Spin}_{SU(4)_R}$-bundles in addition to
  $G$-bundles, as we explained above in the case of $G=U(1)$ and
  $SU(2)/SO(3)$.} What is known is the action of the duality groupoid
there on a simpler piece of data obtained from pairs $(G, \omega)$. It
seems to capture the information hidden in the pair $(G, \omega)$ that
we need, up to the spin subtleties.

This piece of data is the lattice $\Lambda$ of allowed electric and
magnetic charges for the line defects in the theory. We will use this
data even though these line defects are actually not available in
general in the bulk of the topologically twisted theory (only at the
boundary), as a formal proxy for the underlying topological action
$\omega$ of the gauge theory. If we further add to $\Lambda$ the data
of the topological spin associated to the line defects, we should be
able to keep track of spin subtleties.

Recall that the charges of line defects in a gauge theory with the Lie
algebra $\mathfrak{g}$ are pairs $({\mb m},{\mb e})$ where ${\mb m}$
is in the magnetic weight lattice $\LP$ (that is, the coweight lattice
of $\mathfrak{g}$) and ${\mb e}$ is in the weight lattice $P$ of $\g$.

There is a natural (Dirac) inner product on $\LP \times P$:
\begin{equation}
\langle ({\mb m},{\mb e}), ({\mb m}',{\mb e}') \rangle = {\mb m} \cdot
{\mb e}' - {\mb m}' \cdot {\mb e}.
\end{equation}
By definition, a lattice $\Lambda$ of allowed line defects is a
maximal local sublattice of $\LP \times P$. Here ``local'' means that
the restriction of the above inner product to $\Lambda$ should take
integer values, and ``maximal'' means that it is not be possible to
add any other elements to $\Lambda$ without violating locality
property. In particular, it follows that $\Lambda$ necessarily
contains $\LQ \times Q$, where $Q \subset P$ and $\LQ \subset \LP$ are
the lattices of roots and coroots, respectively. Hence such $\lambda$
is completely determined by its projection onto $(\LP/\LQ) \times
P/Q$.

The duality groupoid acts on $\Lambda$ by transforming the elements in
the obvious way:
\begin{equation}
  S_m: ({\mb m},{\mb e}) \to ({\mb e},-{\mb m}) \qquad \qquad T: ({\mb
    m},{\mb e}) \to ({\mb m},{\mb e} + {\mb m}).
\end{equation}

In order to make contact with the cohomology class $\omega$ discussed
above, we rearrange the data in $\Lambda$ a bit.  First of all, we can
read off the global form of the gauge group $G$ by looking at the
sublattice of Wilson lines $({\mb 0},{\mb e})$ inside $\Lambda$,
i.e. the intersection $\Lambda \cap P$. This should be the weight
lattice of the gauge group $G$, which we denote by
$\Lambda_G$. Likewise, the intersection $\Lambda \cap \LP$ should be
the coweight lattice of $G$, which we denote by $\Lambda^M_G$. For
each ${\mb m}$ the values of allowed electric charges ${\mb e}$ in our
theory (i.e. such that $({\mb m},{\mb e}) \in \Lambda$) then form a
torsor for $\Lambda_G$. Picking a specific representative ${\mb e}$ of
this coset, we define a pairing $\Lambda^M_G \times \Lambda^M_G \to
U(1)$ by the formula
$$
\langle {\mb m},{\mb m}' \rangle = e^{2\pi i \; {\mb m}' \cdot {\mb
    e}},
$$
which is symmetric because of the locality property. If we identify
$\Lambda^M_G$ with the possible magnetic fluxes on a sphere, this
pairing is nothing but the evaluation of the corresponding cohomology
class $\omega$ on the four-manifold which is the product of two
spheres.

In the next subsection, we will consider examples of lattices
$\Lambda$ in the case of $SL_n$. We will consider the projections of
the lattices onto $(\LP/\LQ) \times P/Q$.

\subsection{Example: $SL_2$, revisited}

In this case, $(\LP/\LQ) \times (P/Q) = \Z_2 \times \Z_2$ and there
are three possible lattices of charges whose projections onto
$(\Z_2,\Z_2)$ contain $(0,0)$ and either of the three $(0,1)$, $(1,0)$
or $(1,1)$.

The first choice is $SU(2)$ gauge theory, the second is the standard
$SO(3)$ gauge theory, which we denote by $SO(3)_+$ and the third is
the $T$-image of $SO(3)_+$, which we denote by $SO(3)_-$. The duality
transformation $S$ exchanges $SU(2)$ and $SO(3)_+$ and maps $SO(3)_-$
to itself.

The spin-enrichment of the duality groupoid keeps track of four
distinct variants of $SO(3)$, corresponding to the basic $SO(3)_0$
theory and to the $T^k$ images $SO(3)_k$. Now $S$ relates $SO(3)_0$
and $\on{Spin-}SU(2)$, $SO(3)_1$ and $SO(3)_3$, $SO(3)_2$ and
$SU(2)$.

Upon compactification on a Riemann surface, $SO(3)_k$ gives
$D_k$-modules twisted by the gerbe associated to the $k$th power of
the naive line bundle ${\mc L}$.

\subsection{Example: $SL_N$}

Ignoring spin subtleties, the number of distinct variants of
$\mathfrak{sl}_N$ gauge theories is equal to the sum of the divisors
of $N$.

As the basic $SU(N)$ gauge theory is invariant under $\Gamma_0(N)$,
the duality groupoid images of $SU(N)$ are counted by the index of
$\Gamma_0(N)$ in $PSL_2(\Z)$, which is
\begin{equation}
N \prod_{p|N} \left(1+\frac{1}{p}\right)
\end{equation}
Hence if $N$ is square-free, all variants of $\mathfrak{sl}_N$ gauge
theories can be found as images of $SU(N)$ gauge theory under some
gauge transformation. If not, there are separate duality orbits.

\subsubsection{Example: $SL_3$}

The Lie algebra $\mathfrak{sl}_3$ gives an example with no
subtleties. The standard $SU(3)$ gauge theory, the standard $PSU(3)_0$
gauge theory and the two $T$-images $PSU(3)_{1,2}$ exhaust the
possible variants, even keeping track of the spin structure
dependence. They are all in the same duality orbit, with $S$
exchanging $PSU(3)_{1}$ and $PSU(3)_{2}$.

\subsubsection{Example: $SL_4$}

This is a much more intricate example. Ignoring spin subtleties, there
are $7$ variants of the gauge theory, but the main duality orbit only
includes $6$ of them.

Starting from $SU(4)$, which has line defects of charges $(0,e), e \in
\Z_4$ and is $T$-invariant, $S$-duality gives $(SU(4)/\Z_4)_0$, which
has line defects of charges $(m,0), m \in \Z_4$. There are three more
variants produced by the action of $T$, $(SU(4)/\Z_4)_{i}$ with
$i=1,2,3$. The $(SU(4)/\Z_4)_{1,3}$ are exchanged by $S$, but
$(SU(4)/\Z_4)_{2}$ is mapped to a modified version of $SU(4)/\Z_2$,
which we denote by $(SU(4)/\Z_2)_{1}=SO(6)_1$.  On the other hand, the
standard $(SU(4)/\Z_2)_{0}= SO(6)_0$ is invariant under both $S$ and
$T$.

If we keep track of spin subtleties, the situation should become even
richer. The Nahm pole for $SU(4)$ requires a spin structure,
suggesting that its $S$-dual gauge theory's Neumann boundary condition
should require a spin structure as well. Indeed, the periodicity of
the $\theta$ angle in $SU(4)/\Z_4$ is twice as long on non-spin manifolds, and
$(SU(4)/\Z_4)_{4}$ is a good candidate for the S-dual of $SU(4)$. We
expect the $S$-dual of $(SU(4)/\Z_4)_{0}$ to be $\on{Spin-}SU(4)$.

Because of the identity $STS = T^{-1} S T^{-1}$ applied to $SU(4)$, we
get that $S$ maps $(SU(4)/\Z_4)_{3}$ to $(SU(4)/\Z_4)_{5}$.
Similarly, $S$ maps $(SU(4)/\Z_4)_{1}$ to $(SU(4)/\Z_4)_{7}$.

The $T$ operation does not fix $SU(4)/\Z_2=SO(6)$ anymore, but $T^2$
does. We also have $\on{Spin-}SO(6)$ versions of the gauge
theory. It would be nice to fill in the remaining structure of the
duality groupoid.

\section{Other duality kernels}    \label{other}

In this section, we discuss another family of kernel vertex algebras
$Y_\ka(G)$ associated to the $D_{0,1} \to D_{1,0}$ junction and thus to
the standard qGL duality, in which conformal dimensions away from the
vacuum are strictly positive.

In the case when $G$ is a reductive Lie group that is Langlands
self-dual (for instance, $E_8$ or $GL_n$) the vertex algebras
$Y_\ka(G)$ were introduced in \cite{Ga1} and \cite{CG}, as the result
of a $D_{0,1} \to N_{1,-1} \to D_{1,0}$ composition of junctions. The
objective of this section is to generalize that construction for an
arbitrary simple Lie group $G$ and discuss the corresponding qGL
duality functors.

The generalization has two levels of complexity. For simply-laced
groups, it turns out that the simplest generalization of the above
construction involves the extended duality groupoid: we can build
analogous $D_{0,1} \to D_{1,0}$ junctions for the generalized gauge
theory $G_{-1}$ which is the $T^{-1}$ image of the gauge theory
associated to the adjoint form of the gauge group $G$.

Notice that the $G_{-1}$ theory is mapped by S to the $T$ image of
$G$, aka $G_1$.  Indeed, $S \circ G_{-1} = ST^{-1} \circ G = T (S T S)
\circ G = G_{1}$. On the other hand, the $RS$ transformation maps the
$G_{-1}$ theory to itself.

The construction of $Y_\ka(G)$ is simpler than the one given in the
Section \ref{fam ker} in that for general $\ka$ there is only one
direct sum over dominant integral weights (rather than two). The
vertex algebra $Y_\ka(G)$ comes equipped with the action of two copies
of $\ghat = \wh{\lg}$, of levels $\ka+1$ and $\ka^{-1}+1$.

In the case that $G$ is self-dual and simply-connected (which for
simple Lie groups means $G=E_8$), we have already described this
vertex algebra and the corresponding kernel the qGL duality $TST$ in
Section \ref{building}. Tensoring this kernel with the line bundle
${\mc L}_G^{-1}$ on each factor $\Bun_G$, we should then obtain the
kernel of the main qGL duality $S$. For other simply-laced simple Lie
groups the construction is identical, but tensoring with ${\mc
  L}_G^{-1}$ maps the kernel to a gerbe-twisted version of $\Bun_G$.

The action on $S$ on other nodes of the duality groupoid requires
appropriate modifications of the vertex algebra $Y_\ka(G)$. Based on
the examples we treat below in detail, we expect that the
modifications are always finite index extensions of $Y_\ka(G) \otimes
A$ for some rational vertex algebra $A$ that depends sensitively on
the choice of gauge groups and discrete $\theta$-angles.

We then generalize $Y_\ka(G)$ to non-simply laced simple Lie
groups. In this case, $Y_\ka(G)$ is larger: it involves two
summations over dominant weights for groups with the lacing number
$m=2$ and four for $G_2$, which has lacing number $m=3$.

\subsection{Construction of $Y_\ka(G)$ for simply-laced
  $G$}    \label{YkaG}

Let's recall and extend the construction in the self-dual case (see
\cite{Ga1,CG} and Section \ref{building} above). Take $G$ to be the
adjoint form of the group, so that $\LG$ is simply-connected.

Consider the following composition of junctions
\begin{equation}    \label{E8}
D_{0,1} \to N_{1,-1} \to D_{1,0}
\end{equation}
in the bulk theory ${\mc T}^{G_{-1}}_{\ka} \equiv T^{-1} \circ {\mc
  T}^{G}_{\ka+1}$.

The first junction is the $T^{-1}$ image of $D_{0,1} \to N_{1,0}$ in
${\mc T}^{G}_{\ka+1}$.  The corresponding junction vertex algebra is
$\ghat$ of level $\ka+1$.

The second junction is the $RST^{-1}$ image of $D_{0,1} \to N_{1,0}$
in ${\mc T}^{G}_{\ka^{-1}+1}$. The corresponding junction vertex
algebra is $\ghat$ of level $\ka^{-1}+1$.

The composition of the two junctions gives an extension 
$$
Y_\ka(G_{-1}) = \bigoplus_{\la \in P^+} {\mathbb V}_{\la,\ka+1}
\otimes {\mathbb V}_{\la^*,\ka^{-1}+1},
$$
where $\la^*$ was defined in Section \ref{fam gen}. Notice that $\la$
and $\la^*$ belong to the root lattice and the vertex algebra has two
$\ghat$- and $G[[z]]$-actions.
 
Applying the localization functor to it, we therefore obtain a twisted
$D$-module $\Delta^{TST}_\ka(G)$ on $\Bun_G \times \Bun_G$, with the
twists $\ka+1$ and $\ka^{-1}+1$ along the two factors, which we expect
to be a kernel of the functor
$$
{\mc E}^{G,TST}_{-\ka-1}: D_{-\ka-1}(\Bun_G) \to
D_{\ka^{-1}+1}(\Bun_G)
$$
corresponding to the qGL duality $TST=TS_1T$. Since the action of $T$
on the category of $\ka$-twisted $D$-modules corresponds to tensoring
with the line bundle ${\mc L}_G$, the kernel of the qGL duality $S$
can now also be constructed: we simply take the tensor product
$$
\Delta^{TST}_\ka(G) \otimes ({\mc L}_G^{-1} \boxtimes {\mc
  L}_G^{-1}).
$$

This a now gerbe-twisted $D$-module on $\Bun_G \times \Bun_G$ with the
twists $\ka$ and $\ka^{-1}$ along the two factors, which should give
rise to the kernel
$$
{\mc E}^{G,S}_{-\ka}: {\mc G}_{1}-D_{-\ka}(\Bun_G) \to {\mc
  G}_{-1}-D_{\ka^{-1}}(\Bun_G).
$$
where ${\mc G}_{\pm 1}$ is the $\Z_{n(G)}$-gerbe of $n(G)$th roots of
${\mc L}_G^{\otimes \pm 1}$.

For other choices of generalized gauge group, different from
$G^{\mathrm{Adj}}_{-1}$, the procedure above can still be implemented,
but we need to identify appropriate junctions $D_{0,1} \to N_{1,-1}$
for $G$ and $\LG$.

Based on general gauge theory considerations and past examples, we
expect the existence of junctions to support vertex algebras which
include two copies of $\ghat$ as well as an auxiliary rational vertex
algebras, which are needed to implement correct functors from ${\mc
  C}^G_\ka(N_{1,-1})$.  The composition of such junctions should give
a finite index extension of $Y_\ka(G^{\mathrm{Adj}}_{-1}) \otimes A$
for some rational vertex algebra $A$.

It is far from obvious, though, that our strategy of composing simpler
junctions in order to produce, say, a $D_{0,1} \to N_{1,-1}$ junction
will give vertex algebras of this form, as we can see from the example
in this remark.

\begin{rem}
Suppose that we start with the composition of junctions
\begin{equation}    \label{E8-1}
D_{0,1} \to N_{1,-1} \to D_{1,0}
\end{equation}
in the bulk theory ${\mc T}^G_{\ka}$ rather than ${\mc
  T}^{G_{-1}}_{\ka}$ as we did before. Then we can represent the
second junction $N_{1,-1} \to D_{1,0}$ as the $SRT^{-1}$ image of
junction $D_{0,1} \to N_{1,0}$ in ${\mc T}^{\LG}_{\ka^{-1}+1}$
because $T^{-1}$ is a legitimate duality of ${\mc
  T}^{\LG}_{\ka^{-1}+1}$ (recall that $\LG$ is assumed to be
simply-connected). But we can no longer represent the first junction
as the $T^{-1}$ image of $D_{0,1} \to N_{1,0}$ in ${\mc T}^G_{\ka+1}$
because $T^{-1}$ is not a legitimate duality there.

Instead, we can try to build it as a composition of two junctions. The
simplest one is
\begin{equation}    \label{DN}
D_{0,1} \to N_{1,0} \to N_{1,-1}
\end{equation}
(again, in the bulk theory ${\mc T}^G_\ka$), where the second junction
is the $ST$ image of the basic junction $N_{0,1} \to N_{1,0}$ in the
bulk theory ${\mc T}^{\LG}_{-(\ka+1)/\ka}$. The vertex algebra
corresponding to \eqref{DN} is therefore, for irrational $\ka$,
$$
\bigoplus_{\la \in P^+_{\on{ad}}} {\mathbb V}_{\la,\ka} \otimes
M_{(\la^*,0),-\ka/(\ka+1)},
$$
where we use the notation \eqref{Mka}. Hence the vertex algebra of the
junction \eqref{E8-1} is, for irrational $\ka$,
\begin{equation}    \label{tildeYka}
\bigoplus_{\la \in P^+_{\on{ad}},\mu^\vee \in P^+}
{\mathbb V}_{\la,\ka} \otimes M_{(\la^*,\mu^\vee),-\ka/(\ka+1)}
\otimes {\mathbb V}_{\mu^{\vee *},\ka^{-1}+1}.
\end{equation}
Unfortunately, this vertex algebra has unbounded conformal
dimensions. That's why we prefer $Y_\ka(G)$ which does not suffer
from this defect.\qed
\end{rem}

\subsection{Kernels for $SL_2$}

In Section \ref{SU2} we encountered the vertex algebra
$\td(2,1,-\kappa)_1$, which enters in slightly different guises in the
construction of duality kernels for a variety of qGL dualities
involving $SU(2)$ and $SO(3)$. According to the definition above, 
$Y_\ka(SO(3)_{1}) \equiv \frac{\td(2,1,-\kappa)_1}{\tsu(2)_1}$.

Ignoring spin subtleties, but keeping
track of the full groupoid with nodes $SU(2)$, $SO(3)_0$ and
$SO(3)_1 \equiv SO(3)_{-1}$, we summarize the appearances of $\td(2,1,-\kappa)_1$:

\begin{enumerate}
\item The affine subalgebra $\hsl(2)_{\kappa+1} \times
  \hsl(2)_{\kappa^{-1}+2}$ of $\td(2,1,-\kappa)_1$ can be used to
  associate (via the localization functor) to the vertex
  algebra $\td(2,1,-\kappa)_1$ a twisted $D$-module on $\Bun_{SL_2}
  \times \Bun_{PSL_2}$ with the twists $\ka+1$ along the first factor
  and $\kappa^{-1}+2$ along the second factor. Then, tensoring with
  the inverse of the generating line bundle ${\mc L}_{SL_2}$ along the
  first factor and the inverse of the generating line bundle ${\mc
    L}_{PSL_2}$ along the second factor, we obtain a
  $(\ka,\ka^{-1})$-twisted $D$-module on $\Bun_{SL_2} \times
  \Bun_{PSL_2}$. This is our candidate for the kernel of the standard
  qGL duality associated to the $S$ operation $SU(2) \to SO(3)_0$.

\item The coset $\frac{\td(2,1,-\kappa)_1}{\tsu(2)_1}$ allows
  conformal blocks to be defined as twisted $D$-modules on
  $\Bun_{PSL_2} \times \Bun_{PSL_2}$ with the twist $\ka+1$ along the
  first factor and the $\ka^{-1}+1$ along the second factor. This is
  our candidate for the kernel of a modified qGL duality associated
  to the $S$ operation $SO(3)_1 \to SO(3)_1$.
\end{enumerate}

In particular, we can naturally define $Y_\ka(SO(3)_0) = Y_\ka(SU(2))=\td(2,1,-\kappa)_1$,
with appropriate choices of currents to couple to $SL(2)$ and $PSL(2)$ bundles. 

Spin subtleties manifest themselves in a rather straight-forward
manner. The spin-extended groupoid has nodes $SU(2)_b$, $SU(2)_s$, $SO(3)_0$,$SO(3)_1$,$SO(3)_2$ and
$SO(3)_3\equiv SO(3)_{-1}$.
\begin{enumerate}
\item In the above statement (i), tensoring with ${\mc
    L}^{-1}_{PSL_2}$ actually maps $(\ka^{-1}+2)$-twisted $D$-modules
  on $\Bun_{PSL_2}$ to $\ka^{-1}$-twisted and ${\mathcal G}$-twisted
  $D$-modules on $\Bun_{PSL_2}$. Thus, the resulting kernel should
  correspond to the qGL duality $S: SU(2)_b \to SO(3)_2$.

\item A different vertex algebra is required for the qGL duality
  $SU(2)_s \to SO(3)_0$. A likely candidate is the even part of
  $\td(2,1,-\kappa)_1 \times \tso(3)_1$. This vertex algebra has an
  affine $\hsl(2)_{\kappa+1} \times \hsl(2)_{\kappa^{-1}+4}$ subalgebra
  and the spin and $\hsl(2)_{\kappa+1}$ weight of fields are tied
  together. This means that its localization functor yields
  $\ka$-twisted $D$-modules along $\Bun_{\mathrm{Spin}-SL_2}$ and
  $\ka^{-1}$-twisted $D$-modules along $\Bun_{PSL_2}$.

\item The $\frac{\td(2,1,-\kappa)_1}{\tsu(2)_1}$ vertex algebra is the natural candidate
  for the qGL duality $SO(3)_3 \to SO(3)_1$.

\item Likewise, a kernel for the qGL duality $SO(3)_1 \to SO(3)_3$
  would require a vertex algebra with an $\hsl(2)_{\kappa+3} \times
  \hsl(2)_{\kappa^{-1}+3}$ sub-algebra. A possible solution is to
  combine $\frac{\td(2,1,-\kappa)_1}{\tsu(2)_1}$ with the even part of
  $\tso(3)_1 \times \tso(3)_1$.

\end{enumerate}

\subsection{Kernels for $SL_3$.}

This is a particularly straightforward example. It has no spin subtleties, but it illustrates 
the new challenges which arise when we go beyond $SL_2$. 

We have learned above how to build a $D_{0,1} \to D_{1,0}$ junction in 
the $PSU(3)_2 \equiv PSU(3)_{-1}$ generalized gauge theory. It supports the extension of 
$\hsl(3)_{\kappa+1} \times \hsl(3)_{\kappa^{-1}+1}$ we denote as $Y_\ka(PSU(3)_2)$. 

The next simplest junction we can produce is a $D_{0,1} \to D_{1,0}$ junction in 
the $PSL(3)_1$ generalized gauge theory, the $T$ image of the standard 
$PSL(3)$ theory. 

As an intermediate step, we can decompose the junction as 
$D_{0,1} \to N_{1,-2} \to D_{1,0}$ and associate $D_{0,1} \to N_{1,-2}$
to a $\hsl(3)_{\kappa+2}$ Kac--Moody algebra. We are left with the problem 
of building $N_{1,-2} \to D_{1,0}$. 

A similar, $RS$ dual decomposition $N_{1,-2} \to N_{2,-1}\to D_{1,0}$
gives an $\hsl(3)_{\kappa^{-1}+2}$ factor, but still leaves a non-trivial problem: 
build an $N_{1,-2} \to N_{2,-1}$ junction in the $PSU(3)_1$ theory. 

We can solve the problem with one final step: the composition 
$N_{1,-2} \to N_{1,-1} \to N_{2,-1}$ in the $PSU(3)_1$ theory.
Indeed, both partial junctions can be dualized to  
$N_{1,0} \to N_{1,1}$ in the standard $PSU(3)$ theory,
i.e. $N_{0,1} \to N_{1,-1}$ in the $SU(3)$ theory. 

Keeping track of couplings, that gives a ${\mc W}_{\frac{\ka+1}{\ka+2}}$ and a ${\mc W}_{\frac{\ka+1}{2 \ka+1}}$
algebras. We can then go back to the original junction and associate differently. 
The $D_{0,1} \to N_{1,-2}\to N_{1,-1}$ composition in the $PSL(3)_1$ theory
can be recognized by the GKO coset construction of ${\mc W}$ as supporting 
$\hsl(3)_{\ka+1} \times \tsu(3)_1$, possibly coupled to $PSL(3)$ 
bundles through the total level $\ka+2$ currents. 

Overall, we obtained a junction $D_{0,1} \to D_{1,0}$ in 
the $PSU(3)_1$ gauge theory supporting an extension of 
$$\hsl(3)_{\kappa+1} \times \hsl(3)_{\kappa^{-1}+1}\times \tsu(3)_1 \times \tsu(3)_1$$
we denote as $Y_\ka(PSU(3)_1)$. The extension involves the category of lines 
on $N_{1,-1}$ in the $PSU(3)_1$ theory, i.e. the category of lines on 
on $N_{1,-1}$ in the standard $SU(3)$ gauge theory: $\KL_{\frac{\ka+1}{\ka+2}}({\mathfrak sl}_3)$.
In particular, it involves a sum over all highest weights. 

Conformal blocks of such vertex algebra 
give a potential $S$ kernel for the category of 
twisted $D$-modules associated to the $PSL(3)_1$ theory.

It is very tempting to assume that the same vertex algebra may also occur at a junction 
$D_{0,1} \to D_{1,0}$ in the standard $SU(3)$ or $PSU(3)$ gauge theories. 
Indeed, it has an $\hsl(3)_{\kappa+1} \times \hsl(3)_{\kappa^{-1}+3}$ sub-algebra 
and it may be possible to couple it to $SL_3 \times PSL_3$ bundles to produce 
$D$-modules with the correct twists. 

We cannot easily confirm this naive expectation by composing basic junctions. With a small extra conjecture, 
we can get a somewhat similar statement, but involving an extra $\tsu(3)$ spectator factor. We have derived 
$\hsl(3)_{\ka+1} \times \tsu(3)_1$ at $D_{0,1} \to N_{1,-2}\to N_{1,-1}$ in the $PSL(3)_1$ theory.
We can look as a similar composition: $N_{0,1} \to N_{1,-2}\to N_{1,-1}$ in the $PSL(3)_1$ theory.

In analogy with the $SU(2)$ case, we conjecture that, up to a re-definition of the stress tensor, 
there is an $N_{0,1} \to N_{1,-1}$ junction 
in the $PSL(3)_1$ theory which supports ${\mc W}_{\ka+1} \times \tsu(3)_1$.
Equivalently, we conjecture that there is an $N_{0,1} \to N_{1,0}$ junction 
in the $PSL(3)_2$ theory which supports ${\mc W}_{\ka} \times \tsu(3)_1$.

If this conjecture is correct, then a direct construction of $D_{0,1} \to D_{1,0}$ in the standard $PSU(3)_0$ gauge theory 
can proceed as follows. We can first decompose it to $D_{0,1} \to N_{1,-2} \to N_{1,-1} \to D_{1,0}$.
The $D_{0,1} \to N_{1,-2}$ junction can be produced from $D_{0,1} \to N_{1,-3}\to N_{1,-2}$:
$N_{1,-3}\to N_{1,-2}$ in $PSU(3)_0$ is dual to $N_{1,0} \to N_{1,1}$ in $PSU(3)_0$
and thus to $N_{0,1} \to N_{1,-1}$ in $SU(3)$ and supports ${\mc W}_{\frac{\ka+2}{\ka+3}}$.
It can combine with $\hsl(3)_{\ka+3}$ to give $\hsl(3)_{\ka+2} \times \tsu(3)_1$ at the $D_{0,1} \to N_{1,-2}$ junction.

Next, we map $N_{1,-2} \to N_{1,-1}$ in $PSU(3)_0$ to $N_{1,-1} \to N_{1,0}$ in $PSU(3)_1$,
to $N_{0,1} \to N_{1,-1}$ in $PSU(3)_1$, to $N_{0,1} \to N_{1,0}$ in $PSU(3)_2$. 
Then the junction should support ${\mc W}_{\frac{\ka+2}{\ka+1}} \times \tsu(3)_1$.

Assembling the pieces, we get an extension of $\hsl(3)_{\ka+2} \times \tsu(3)_1 \times {\mc W}_{\frac{\ka+2}{\ka+1}} \times \tsu(3)_1$
at $D_{0,1} \to N_{1,-1}$ in $PSU(3)_0$, which we can presumably be identified with something like 
 $\hsl(3)_{\ka+1} \times \tsu(3)_1 \times \tsu(3)_1 \times \tsu(3)_1$. If this is correct, the result will be some extension of 
 $$\hsl(3)_{\kappa+1} \times \hsl(3)_{\kappa^{-1}+1}\times \tsu(3)_1 \times \tsu(3)_1\times \tsu(3)_1$$.
 
 It would be interesting to test this conjecture further. We leave that to future work. 
 
\subsection{$Sp(N)$ and $SO(2N+1)$ and other non-simply laced groups.}

The basic challenge here is that, unlike the simply-laced case,
$N_{1,-1}$ is not self-dual under the basic duality $RS_m$. Rather, it
is mapped by $RS_m$ to $N_{m,-1} \neq N_{1,-1}$ if the lacing number
$m>1$. This means that if we want to use the composition $D_{0,1} \to
N_{1,-1} \to D_{1,0}$ for a non-simply laced group $G$, we need to
describe the vertex algebra for the junction $N_{1,-1} \to D_{1,0}$
for the coupling $\ka$ and group $G$. Hence, we have to describe its
$RS_m$-dual: $D_{0,1} \to N_{m,-1}$ for the coupling $(m\kappa)^{-1}$
and group $\LG$. 

In what follows, we will pretend that $T$ is a
legitimate duality for both $G$ and $\LG$.
We will simply ignore here subtleties concerning the global form of the group. 

\bigskip

Consider first the case $m=2$. Then we can decompose further the
latter junction as $D_{0,1} \to N_{1,-1} \to N_{2,-1}$.  The $N_{1,-1}
\to N_{2,-1}$ junction for the coupling $(2\kappa)^{-1}$ and group
$\LG$ is the same as $N_{1,0} \to N_{2,1}$ at $(2\kappa)^{-1}+1$ and
group $\LG$, which is the same as $N_{0,1} \to N_{1,-1}$ at
$-\frac{\ka}{1+2\ka}$ and the original group $G$. In turn, that is the
same as $N_{0,1} \to N_{1,0}$ at $\frac{\ka+1}{2\ka+1}$ and group $G$.

On the other hand, $D_{0,1} \to N_{1,-1}$ at $(2\kappa)^{-1}$ and
group $\LG$ is the same as $D_{0,1} \to N_{1,0}$ at $(2\kappa)^{-1}+1$
and group $\LG$.

Thus, the resulting kernel vertex algebra $Y_\ka(G)$ is a double
extension of $\mathfrak{g}_{\ka+1} \times {\mc
  W}_{\frac{\ka+1}{2\ka+1}}(\g) \times
\wh{\lg}_{(2\ka)^{-1}+1}$. For irrational $\ka$, it can be written
explicitly as a double direct sum, as in Section \ref{fam ker}. One
can show that this vertex algebra has non-negative conformal
dimensions. When we apply the localization functor to the vertex
algebra $Y_\ka(G)$, we should obtain a kernel in a suitable version of
the category of twisted $D$-modules on $\Bun_G \times \Bun_{\LG}$ with
the twists $\ka+1$ along the first factor and $(2\ka)^{-1}+1$ along
the second factor. This $D$-module should give rise to the qGL duality
$T S T$.

When $\mathfrak{g} = \mathfrak{so}(2n+1)$, string theory suggests that
the extension ${\mc W}_{\frac{\ka+1}{2\ka+1}}(\mathfrak{so}(2n+1))
\times \wh{\mathfrak{sp}}(n)_{(2\ka)^{-1}+1}$ should be identified
with $\wh{\mathfrak{osp}}(1|n)_{\ka^{-1}+1}$ and hence the kernel
vertex algebra is an extension of $\wh{\mathfrak{so}}(2n+1)_{\ka+1}
\times \wh{\mathfrak{osp}}(1|n)_{\ka^{-1}+1}$.

With a bit more care, and perhaps extra factors of $\tso(2n+1)$ and
$\mathrm{sp}(2n)$ WZW models at level $1$, we should be able to
manufacture kernels adapted to specific pairs of gauge groups and
discrete $\theta$-angles. Similar considerations should apply to other
groups of lacing number $m=2$.

\bigskip

For lacing number $m=3$, we need to work harder. 

If we attempt a decomposition $D_{0,1} \to N_{2,-1} \to N_{3,-1}$, the
$N_{2,-1} \to N_{3,-1}$ junction is easy: it is dual to $N_{1,-1} \to
N_{3,-2}$, which is dual to $N_{1,0} \to N_{3,1}$, dual to $N_{0,1}
\to N_{1,1}$ and hence finally to $N_{0,1} \to N_{1,0}$.

On the other hand, $D_{0,1} \to N_{2,-1}$ requires a further
decomposition, such as $D_{0,1} \to N_{1,-1} \to N_{2,-1}$. 

Still, $N_{1,-1} \to N_{2,-1}$ is not an elementary junction. 
We can attempt a further decomposition $N_{1,-1} \to N_{3,-2}\to N_{2,-1}$.
The first half is dual to $N_{1,0} \to N_{3,1}$ and thus to $N_{0,1} \to N_{1,-1}$,
which is elementary. The second half is dual to $N_{3,1}\to N_{2,1}$
and then to $N_{1,-1} \to N_{3,-2}$, which we just analyzed. 

Thus a potential kernel vertex algebra for lacing number $m=3$ will be a
quadruple extension of the rough form $\ghat_{\ka+1} \times {\mc
  W}_{\ka'}(\g) \times {\mc W}_{\ka''}(\g) \times {\mc W}_{\ka'''}(\g) \times
\ghat_{(3\ka)^{-1}+1}$.

We leave to future work checks of positivity of conformal dimensions
for this extension.

\section{Some future directions}    \label{future}

This work leaves open a variety of questions in gauge theory, vertex
algebras and the Geometric Langlands Program.  Here we list some of
the most important ones.

\medskip

\begin{itemize}
\item {\em Rational $\kappa$:} What are the junctions associated to
  the extra boundary conditions which only exist at rational values of
  $\ka$?

Do sheaves of coinvariants of the corresponding junction
  vertex algebras give rise to dual pairs of objects?

What is the best
  way to describe the limit of the kernel vertex algebras constructed
  in this paper when $\kappa$ tends to a rational value?

What are the limits of the corresponding qGL
  duality functors?

Do issues of temperedness affect equally all
  possible kernels obtained from the junctions, or different kernels
  have a different domain of effectiveness?

\medskip

\item Same questions at the critical level. Furthermore, the theory of
  coinvariants twisted by flat bundles needs to be developed
  mathematically.

\medskip

\item What is the full spin-refined duality groupoid? 

\medskip

\item Can auxiliary rational vertex algebras be found, which can
  combine with ${\mc W}$- or Kac--Moody algebras to give elementary
  junctions for all possible global forms of the gauge group?

\medskip

\item Is there some simple criterion which guarantees or forbids the
  existence of junction vertex algebras with positive conformal
  dimensions?
\end{itemize}

\end{document}